\documentclass[useAMS,usenatbib] {mn2e}
\usepackage{graphicx}
\usepackage{times}
\usepackage{natbib}

\def\pc{{\rm pc}} %.........parsecs
\def\kpc{{\rm k}\pc} %......kiloparsecs
\def\Mpc{{\rm M}\pc} %......megaparsecs
\begin{document}

%\title[A Comparison of Cosmological Codes: Thermal, Shocks and Accelerated Cosmic Rays Properties] {A Comparison of Cosmological Codes: Thermal, Shocks and Accelerated Cosmic Rays Properties}
\title[A Comparison of Cosmological Codes] {A Comparison of Cosmological Codes: Properties of Thermal Gas and Shock Waves in Large Scale Structures}
\author[F.Vazza, K.Dolag, D.Ryu, G.Brunetti, C.Gheller, H.Kang, C.Pfrommer] {F.Vazza$^{1,2}$\thanks{
E-mail: f.vazza@jacobs-university.de}, K.Dolag $^{3,4}$, D.Ryu$^{5}$, G.Brunetti$^{2}$, C.Gheller$^{6}$, H.Kang$^{7}$, C.Pfrommer$^{8}$\\
%EndAName
$^{1}$ Jacobs University Bremen, Campus Ring 1, 28759 Germany \\
$^{2}$ INAF-Istituto di Radioastronomia, via Gobetti 101, I-40129 Bologna,
Italy\\
$^{3}$ University Observatory Munich, Scheinerstr. 1, D-81679 Munich, Germany \\
$^{4}$ Max-Planck Institut fur Astrophisik, P.O. Box 1317,D-85741,Garching,Germany\\
$^{5}$ Department of Astronomy and Space Science, Chungnam National University, Daejeon 305-764, Korea\\
Italy\\
$^{6}$ CINECA, High Performance System Division, Casalecchio di 
Reno--Bologna, Italy\\
$^{7}$ Department of Earth Sciences, Pusan National University, Busan 609-735, Korea\\
$^{8}$ Heidelberg Institute for Theoretical Studies, Schloss-Wolfsbrunnenweg 35, D-69118 Heidelberg, Germany }
\date{Accepted ???. Received ???; in original form ???}
\maketitle

\begin{abstract}

  Cosmological hydrodynamical simulations are a valuable tool for
  understanding the growth of large scale structure and the observables
  connected with this. Yet, comparably little attention has been given to
  validation studies of the properties of shocks and of the resulting thermal gas
  between different numerical methods -- something of immediate importance as
  gravitational shocks are responsible for generating most of the entropy of the
  large scale structure in the Universe.  Here, we present results for the
  statistics of thermal gas and the shock wave properties for a large volume
  simulated with three different cosmological numerical codes: the Eulerian
  total variations diminishing code TVD, the Eulerian piecewise parabolic
  method-based code ENZO, and the Lagrangian smoothed-particle hydrodynamics
  code GADGET.  Starting from a shared set of initial conditions, we present
  convergence tests for a cosmological volume of side-length $100 \Mpc/h$,
  studying in detail the morphological and statistical properties of the thermal
  gas as a function of mass and spatial resolution in all codes. By applying
  shock finding methods to each code, we measure the statistics of shock waves 
  and the related cosmic ray acceleration efficiencies, within the sample of
  simulations and for the results of the different approaches.  We discuss the regimes of
  uncertainties and disagreement among codes, with a particular
  focus on the results at the scale of galaxy clusters.  Even if
  the bulk of thermal and shock properties are reasonably in agreement among the
  three codes, yet some significant differences exist (especially between Eulerian
  methods and smoothed particle hydrodynamics). In particular, we report:
a) differences of huge factors ($\sim 10-100$) in the values of average gas density,
temperature, entropy, Mach number and shock thermal energy flux in the most 
rarefied regions of the simulations ($\rho/\rho_{\rm cr} < 1$) between grid and SPH methods; b) the hint of an
entropy core inside clusters simulated  in grid codes; c) significantly different 
phase diagrams of shocked cells in grid codes compared to SPH; d) sizable differences in the morphologies
of accretion shocks between grid and SPH methods. 

\end{abstract}

%% ======================================================================= %%

\label{firstpage} 

\vskip 0.4cm \fontsize{11pt}{11pt} \selectfont  
%% ======================================================================= %%
%% REPORT REPORT REPORT REPORT REPORT REPORT REPORT REPORT REPORT REPORT   %%
%% ======================================================================= %%

\section{Introduction}

\label{sec:intro}

Cosmological numerical simulations are a powerful tool to investigate
the properties of the Universe at the largest scales.
From galaxy formation to the precise measurement of cosmological parameters, from
the propagation of ultra high cosmic rays to the growth of the non-thermal energy components of the intra cluster medium (e.g. magnetic field, relativistic particles), cosmological simulations represent an
effective complement to
theoretical models and observations (e.g. Borgani et al.  2008; Borgani \& Kravtsov 2009 and Norman 2010 for recent reviews).
In order to model the evolution of cosmic structures in the most reliable way,
numerical methods must follow the non-linear dynamics of the gas and dark matter (DM) assembly across a very large dynamical range (e.g. from 
scales of $\sim (10^{2}-10^{3}) \Mpc$ to $\sim (1-10) \kpc$),
 over the age of the Universe.

To accomplish this task, a number of finite difference methods
have been developed in the past, which can be broadly
divided into 2 classes (e.g. Dolag et al.  2008 for 
a modern review). ``Lagrangian'' methods
discretize baryon gas by mass, using a finite 
number of particles, and the equation of fluid-dynamics
are solved with the approach of smoothed particle hydrodynamics
(SPH, see Price 2008 and Springel 2010 for recent reviews). Further
details of the SPH method investigated in this project
will be discussed in Sec.\ref{subsec:sph}.

Contrarily, ``Eulerian'' methods discretize space, by dividing
the computational domain into regular cells (with fixed or
variable size), and the gas-dynamics is
evolved by solving cell-to-cell interactions (e.g. Le Veque 1990
for a review).
A variety of numerical schemes can be applied for
the reconstruction of the gas velocity, density, and pressure fields for a given
number of neighbors (e.g. piecewise linear method, Colella \& Glaz 1985; piecewise parabolic method, Colella \& Woodward 1984), 
as well as for the time integration of the fluxes across the cells 
(e.g. ROE method, Powell et al.  1999; HLL/HLLE method, Harten 
et al.  1983; HLLC method, Li 2005). Further details of the grid
methods employed in this project will be presented
in Sections~\ref{subsec:enzo_cp} and \ref{subsec:tvdhydro}.

\bigskip

Despite the enormous progresses made since their first applications (e.g. Peebles 1978; Efstathiou \& Eastwood 1981; Davis et al.  1985; Efstathiou 1985;), the mutual convergence of the results of 
cosmological numerical methods is still matter of debate and research.
This is true also for the most simple physical modeling of 
large scale structures, where no  forces other than gravity
and pressure are taken into account.

A few comparison works in the literature (e.g. Kang et al. 1994, Frenk et
al.1999, O'Shea et al. 2005, Heitmann et al. 2008) have provided
evidences that most of the relevant quantities involved in large scale structure
dynamics are generally reproduced with similar accuracy by most codes
on the market. The general findings suggest that the simplest clustering properties of DM, and their dependencies
on assumed cosmological and numerical parameters are fairly well understood (e.g. Heitmann et al.  2008).

A less satisfactory agreement is generally found
when the properties of gas in different methods are compared,
even when simple non-radiative numerical setups are considered.
In simulations of galaxy clusters, for instance, the entropy profile, the baryon fraction and the X-ray
luminosities are affected by the larger uncertainties among codes
reaching differences up to a factor of a few (e.g. Frenk et al. 1999;
O'Shea et al.  2005; Voit et al.  2005; Kravtsov et al.  2005; Ettori et al.  2006; Vazza et al.  2010).

More recent works aiming at comparing different numerical methods
in more idealized test cases (e.g. shock tubes, blast waves, halo
profile stability, ram pressure stripping of substructure) produced additional insights in the ways in which 
the numerical implementations of different codes work
(e.g. Agertz et al.  2007, Tasker et al. 2008, Mitchell
et al. 2009; Springel 2010; Robertson et al.  2010; Hess \& Springel 2010; Merlin et al.  2010). One of the reported key findings is that the effective numerical viscosity 
acting within each code has a sizable impact on the overall evolution of quantities tightly linked to ram pressure stripping, turbulence and shocks in the gas medium.

\bigskip

Cosmological simulations also  proved to be important
tools to study the acceleration and evolution of cosmic ray particles (CR) in the
Universe, and their connection to the observed statistics of non-thermal
emission from galaxy clusters (e.g. Dolag et al.  2008 for 
a review).
Several mechanisms related to cluster
mergers and to the accretion of matter can act as sources of non-thermal components in the ICM. 
The most important mechanism  during cluster
formation is likely diffusive shock acceleration (DSA):
the thermal particles in the high-energy tail of the Maxwellian distribution
  function are able to experience multiple scatterings across the shock surface
  which can be modeled as a diffusion process. This leads to an exponential
  gain of energy and an exponential loss of the number of particles which
  results in a power-law distribution in particle momentum extending into the
  relativistic regime and giving rise to so-called cosmic rays
(e.g. Bell 1978; Blandford \& Ostriker 1978; Drury \& V\"{o}lk 1981; see also Kang \& Ryu 2010 and Caprioli et al.  2010 for recent reviews). 

Energetic shocks generated by mergers are believed to accelerate supra-thermal
electrons from the thermal pool and explain the origin of radio relics (Ensslin
et al. 1998; R\"{o}ttiger et al.  1999; Pfrommer et al.  2008; Pfrommer 2008; Hoeft
\& Br\"{u}ggen 2007; Battaglia et al.  2009; Skillman et al.  2011), while
high-energy electrons accelerated at these shocks can produce X-rays and
gamma-rays via inverse Compton scattering off CMB photons (e.g., Sarazin 1999;
Loeb \& Waxmann 2000; Blasi 2001; Miniati 2003; Pfrommer et al.  2008; Pfrommer
2008).  Relativistic hadrons accelerated at shocks can be advected in galaxy
clusters and efficiently accumulated there (V{\"o}lk, Aharonian, \&
Breitschwerdt 1996; Berezinsky, Blasi, \& Ptuskin 1997), possibly leading to a sizable
non-thermal component which could be detected by gamma-ray observations (e.g.,
Pfrommer \& Ensslin 2004; Blasi, Gabici \& Brunetti 2007; Pfrommer et al.  2007;
Pfrommer 2008; Pinzke \& Pfrommer 2011).  The re-acceleration of relativistic electrons by MHD turbulence can be responsible for the episodic
diffuse radio emission  observed in the form of radio halos (e.g., Petrosian \& Bykov 2008; Brunetti et al. 2008; Brunetti \& Lazarian 2011); in addition
secondary particles injected in the ICM via proton--proton collisions may also
produce detectable synchrotron radiation (e.g., Blasi \& Colafrancesco 1999;
Dolag \& Enssil 2000; Miniati et al.  2001; Pfrommer et al. 2008; En{\ss}lin et
al.2011).

The occurrence of shock waves in large scale structures has been studied in
detail with cosmological numerical simulations (e.g., Miniati et al.  2001; Ryu
et al. 2003; Pfrommer et al. 2006; Pfrommer et al. 2007; Kang et al. 2007; Hoeft et
al. 2008; Skillman et al. 2008; Vazza, Brunetti \& Gheller 2009; Molnar et
al.2009) or indirectly by the action of shock waves on radio plasma
  bubbles, employing a novel method of combining radio observations and
  analytical insight that is supported by idealized hydrodynamic simulations
  (e.g., En{\ss}lin et al.  2001; Pfrommer \& Jones 2011).  Most of these
numerical works agree on the fact that the bulk of the energy in the Universe is
dissipated at relatively weak shocks, $M \sim 2-3$ (where $M$ is the Mach
number), internal to clusters, while strong and larger shocks are found outside
of large scale structures, $M \sim 10-100$, at the boundary layers between the
``collapsing'' and the ``expanding'' universe.  However when the properties of
CR injection by DSA are compared across the different simulations, differences
up to $1-2$ orders of magnitude in various quantities are found, including (but
not limited to) quantities such as the ratio of energies of CR and thermal gas,
the spectral energy distribution (e.g., Miniati et al.  2002; Ryu et al.  2003;
Pfrommer et al.  2006,2007,2008; Pfrommer 2008; Kang et al.  2007; Hoeft et
al. 2008; Vazza et al.  2009,2010; Skillman et al.  2008,2011; Pinzke \& Pfrommer
2011).  This limits our present understanding of the main mechanism for the
enrichment of CRs in the intra cluster medium.

\bigskip

In this work, we explicitly aim at comparing three independent
numerical approaches for cosmological simulations, applied to
the evolution of  a large volume of the Universe: the 
smoothed particle hydrodynamics code GADGET (Springel, Yoshida \& White 2001; Springel 2005); the
total variation diminishing code developed by D. Ryu and collaborators (Ryu 
et al.  1993; Ryu et al.  2003) and the parabolic piecewise method ENZO, developed by G.Bryan
and collaborators (e.g. O'Shea et al.  2004; Norman et al.  2007).

We adopted a set of shared {\it identical} initial conditions generated
at different resolution, and we re-simulated them with the three codes; the output of
all runs were  then compared in detail, looking at the convergence of several
thermal and non-thermal properties across the various codes and for different numerical resolutions.
We chose the simplest possible physical setup for this project, and include
only non-radiative physics (i.e., no radiative cooling, no UV radiation 
background from primordial stars, no magnetic fields, etc.).

This approach helps us to understand which differences are due to the numerical
methods (e.g. ``Lagrangian'' versus ``Eulerian'' method for gas dynamics) and which
are due to the post-processing (e.g. temperature-based method to detect shocks
versus velocity based methods).  Also, this approach helps in assessing some of
the more robust findings of present cosmological simulations, and determines the
minimum resolution requirements needed to achieve a good convergence independent
of the particular adopted numerical method.

  The  paper is organized as follows. In Sec.\ref{sec:codes} we give a
  brief description on the underlying numerical schemes of these
  codes. Based on the simulations using different resolutions we 
  present a
  comparison of the general distribution statistics for dark matter
  and thermal gas in sections \ref{sec:dm} and \ref{sec:thermal}. In
  particular, we focus on the galaxy clusters properties according to
  the various codes in section \ref{subsec:clusters_cp} and present an
  exploratory test showing important differences between the
  underlying numerical, hydrodynamical schemes (specially between PPM
  and SPH) in the matter accretion pattern  inside halos in section
  \ref{subsec:tracers}. We then apply different shock detecting
  schemes in section \ref{subsec:basic}  to the various re-simulations
  and we present results for the characterization of shock waves in
  all codes in sections
  \ref{subsec:dn_comp}--\ref{subsec:phas_comp}. We particularly focus
  on shocks in galaxy clusters, their properties and their role in the
  acceleration of CR predicted according to the different, underlying
  numerical schemes in section \ref{subsec:comp_CR_clust}.  

\begin{table}
\label{tab:tab_cp}
\caption{Details of the simulations run for this comparison 
project. First column: name of the run; second column: initial redshift
of the simulation; third column: mass resolution for Dark Matter particles; 
fourth column: softening length (for SPH runs) or uniform mesh spacing (for
ENZO and TVD) employed in the runs.}
\centering \tabcolsep 3pt 
\begin{tabular}{c|c|c|c}

{\small GADGET} \\ 
Run & $z_{\rm in}$ & $M_{dm}$ [$M_{\odot}/h$] &        $R_{soft}$
[kpc/$h$] \\ \hline
64     & 34.63 &                   $2.4\cdot 10^{11}$ &        31.0 \\
128     &  44.77     &                   $3.0\cdot 10^{10}$  &        15.75 \\
256     & 55.92      &                  $3.76\cdot 10^{9}$  &        7.875 \\ \hline
{\small ENZO} \\
Run  &  $z_{\rm in}$ & $M_{dm}$ [$M_{\odot}/h$] & $\Delta x$
[kpc/$h$] \\ \hline
64  &   34.63 &                    $2.4 \cdot 10^{11}$ &        1562.5 \\
128 &   44.77 &                   $3.0 \cdot 10^{10}$        &        781.25 \\
256 &  55.92 &     $3.76 \cdot 10^{9}$ & 390.625 \\  
512 &  67.99 & $4.7 \cdot 10^{8}$ & 195.31 \\ \hline
{\small TVD}\\
Run & $z_{\rm in}$ & $M_{dm}$ [$M_{\odot}/h$] &        $\Delta
x$ [kpc/$h$] \\ \hline
64-32  &  34.63 &                     $3..0 \cdot 10^{12}$ &        1562.5 \\
128-64 &  44.77  &                $2.4 \cdot 10^{11}$        &        781.25 \\
256-128 &  55.92  &          $3.0 \cdot 10^{10}$ & 390.625 \\  
512-256 &  67.99 &                          $3.76 \cdot 10^{9}$ & 195.31         \\
\end{tabular}
\end{table}

\section{Numerical Codes}
\label{sec:codes}

\subsection{Eulerian method: ENZO PPM}
\label{subsec:enzo_cp}

ENZO is an adaptive mesh refinement (AMR) cosmological hybrid code
originally written by Greg Bryan and Michael Norman
(Bryan \& Norman 1997, 1998; O'Shea et al.  2004; Norman et al. 2007). It couples
a particle-mesh solver with an adaptive mesh method for ideal 
fluid-dynamics (Berger \& Colella, 1989).

ENZO uses a particle-mesh N-body method (PM) to follow
the dynamics of collision-less systems.  This method computes trajectories 
of a representative sample of individual DM particles and it is much more 
efficient than a direct solution of the Boltzmann equation in most 
astrophysical situations. 

DM particles are distributed onto a regular grid
using the cloud-in-cell (CIC) interpolation technique, forming a spatially
discretized DM density field.  After sampling dark matter density onto the
grid and adding baryon density (calculated in the hydro method of the code), 
the gravitational potential is calculated 
on the periodic root
grid using Fast Fourier Transform algorithms, 
and finally solving the elliptic Poisson's equation.

The effective force
resolution of a PM calculation is approximately twice
as coarse as the grid spacing at a given level of resolution.  The 
potential is solved in each grid cell;
however, the quantity of interest, namely the acceleration, is the gradient
of the potential, and hence two potential values are required to calculate
this.

In the case of ENZO simulations employing AMR, the potential is recursively
computed within sub-grids at a higher resolution and the boundary conditions
are interpolated from the potential values of the parent grid. Then a multi-grid
relaxation technique is adopted to compute the gravitational force for each cell
within sub-grids (e.g. O'Shea et al. 2006). This enables the use of a gravitational
softening of the order of the highest resolution available in the simulation; however
in this project we did not use AMR capabilities of ENZO, and thus the maximum available
softening is the fixed resolution of the adopted mesh.

As hydrodynamical solver, ENZO adopts the Eulerian
Piecewise Parabolic 
Method (PPM, Woodward \& Colella, 1984). 
The PPM algorithm belongs to a class
of schemes in which an accurate representation of flow discontinuities is made
possible by building into the numerical method the calculation of the
propagation and interaction of non--linear waves. 
It is a higher order extension of Godunov's shock capturing
method (Godunov 1959). It is at least second--order accurate in space (up
to the fourth--order, in the case of smooth flows and small time-steps) and
second--order accurate in time. This leads to an
optimal treatment of energy
conversion processes, to the
minimization of errors due to the finite size of the cells of the grid and
to a spatial resolution close to the nominal one. 

In order to treat more accurately  bulk hypersonic
motions, where the kinetic energy of the gas can dominate the internal
energy by many orders of magnitude, both the gas internal energy
equation and total energy equation are solved everywhere on the grid
at all times.  This {\it dual energy formulation} ensures that the
method produces the correct entropy jump at strong shocks and also
yields accurate pressures and temperatures in cosmological hypersonic
flows.

This works uses the public 1.0.1 version of ENZO{\footnote
  {http://lca.ucsd.edu/software/enzo/v1.0.1/download/}}.  To simplify the
comparison with the other codes of this project, this work employs a fixed grid
only instead of the adaptive multilevel grids and additional physics (e.g. star
formation, re-ionization and cooling processes) which are powerful tools in ENZO.

%%%%%%%%%%%%%%%%%%%%%%%%%%%%%%%%%%%%%%%%%%%%%%%%%%%%%%%%%%%%%%%%%%%%%%%%%

\subsection{Eulerian method: Cosmological TVD}
\label{subsec:tvdhydro}

The cosmological code created by Ryu et al. (1993) is based on the Harten (1983) 
Total Variation Diminishing (TVD) scheme. It is a 
flux-based Eulerian code with second-order
accuracy in space and time. It captures shocks within two to three cells without generating
oscillations, but limiting the numerical flux according to the TVD scheme instead of
adding a simple artificial viscosity.
Several important improvements were made while incorporating the TVD scheme into
the cosmological code.  
The numerical artificial heating around the extremely supersonic flows where the bulk
kinetic energy is much greater than the thermal energy is reduced; this was achieved
by following the adiabatic changes of the thermal energy using a modified -entropy
equation instead of using the total energy equation. 
The leakage of the gravitational energy into the thermal energy in regions of supersonic
flows was prevented by including the effects of the gravitational force only to
the momentum and kinetic energy and keeping the thermal energy rather than solving
the conservation of the total energy.
Also, a correction due to the mass diffusion under the gravitational field has been
added in the gravitational force term in order to obtain better conservation of the
total energy and to satisfy the cosmic energy equation.  
Additional details can be found in Ryu et al.  (1993) and Ryu et al.  (2003).

The treatment of gravity and DM particle dynamics follows the
Particle Mesh approach on a fixed
resolution grid (see Sec.\ref{subsec:enzo_cp}).
Additionally, in this code there is the possibility of 
using a number of DM particles smaller
than the total number of cells in the grid, in order to spare 
memory usage. This is motivated by the fact that, as stressed in Sec.\ref{subsec:enzo_cp}, in 
the PM scheme the effective force
resolution is approximately twice
as coarse as the mesh spacing. Therefore, adopting a number of DM particles
which is $(N/2)^3$ for a $N^{3}$ grid, has a very little or negligible
difference in the final accuracy of the resulting potential and accelerations.

%%%%%%%%%%%%%%%%%%%%%%%%%%%%%%%%%%%%%%%%%%%%%%%%%%%%%%%%%%%%%%%%%%%%%%

\subsection{Smoothed Particle Hydrodynamics: GADGET3}
\label{subsec:sph}

We compare Eulerian methods with the parallel TreeSPH
code GADGET3 (Springel 2005), which combines smoothed particle
hydrodynamics with a hierarchical TreePM algorithm for gravitational
forces. SPH uses a set of tracer particles to discretize mass elements 
of the fluid. Continuous fluid quantities are estimated by a
kernel interpolation technique (e.g. Monaghan 1992). The equation of
motion for these tracer particles can be derived (by applying the variational 
principle) from the Lagrangian of such system.
The thermodynamic state of each fluid element may either be
defined in terms of its thermal energy per unit mass, $u_i$, or in
terms of the entropy per unit mass, $s_i$. The latter is used as the
independent thermodynamic variable evolved in SPH, as discussed in
full detail by Springel \& Hernquist (2002). The adaptive smoothing 
lengths $h_i$ of each SPH particle are defined such that their kernel 
volumes contain a constant mass for the estimated density (e.g. corresponding 
to the mass of $N=64$ particles is a common choices). Accounting for
the fact that then the adaptive smoothing lengths $h_i$ are a function of
density allows SPH to be formulated so that both
energy and entropy are manifestly conserved. Provided there are no shocks 
and no external sources of heat, the derivation of equations for the 
reversible fluid dynamics in SPH is straightforward (see Price 2008 and Springel 2010 
for recent reviews on SPH).
However, flows of ideal gases can easily develop discontinuities where
entropy must be generated by micro-physics. Such shocks need to be
captured by an artificial viscosity in SPH, which is active
only when fluid elements approach one another in space, preventing
particle interpenetration and transforming kinetic energy irreversibly
into heat (e.g. Monaghan \& Gingold 1983). Modern schemes like GADGET3
make also use of an artificial viscosity based on an analogy with 
Riemann solutions of compressible gas dynamics, as proposed by Monaghan 1997; 
additional viscosity-limiters are also introduced in GADGET3 in the presence of
strong shear flows to alleviate spurious angular momentum transport (Steinmetz 1996).

Both the collision-less dark matter and the gaseous fluid
are represented by particles, allowing the self-gravity of both
components to be computed with gravitational N-body methods. GADGET3
allows the pure tree algorithm to be replaced by a hybrid method
consisting of a synthesis of the particle-mesh method and the tree
algorithm, with significant reduction of the computational effort.

The effective force resolution is controlled by the gravitational
softening $R_{\rm soft}$ used in the tree part as listed in the 
last column of table 1 for the different simulations and the 
particles are allowed to have individual time steps, based on different
time stepping criteria (see Springel 2005 for details).

\begin{figure*}
\begin{center}
\includegraphics[width=0.9\textwidth]{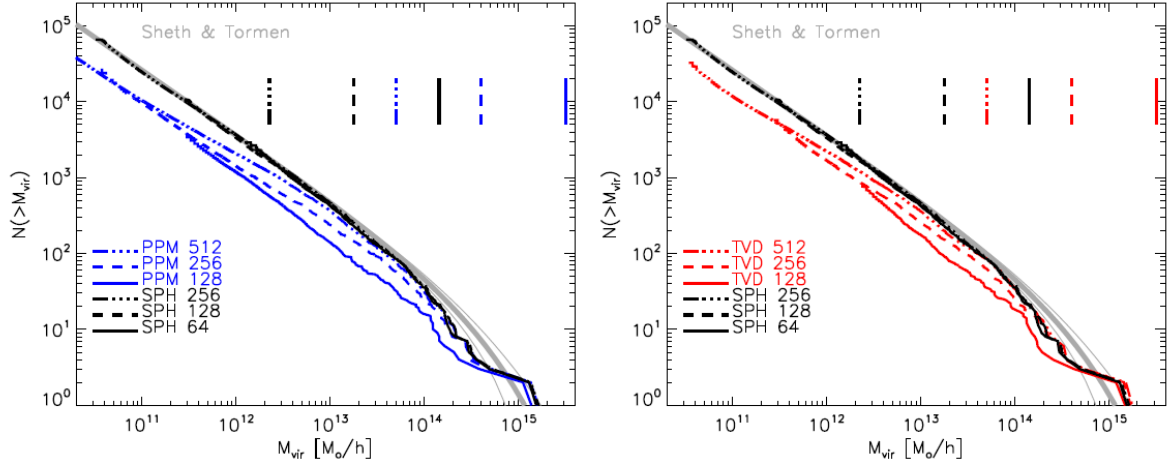}
\caption{Cumulative mass functions of the virialised halos in the various runs. In both panels the GADGET results
are reported in black for the various resolutions, while the left
panel reports the mass functions from ENZO runs (in blue) and the right
panel reports the mass functions from TVD  runs (in red). 
The Sheth \& Tormen (1999) mass function is shown for reference in bold 
(grey lines), with the thin lines showing the Poisson errors.
The vertical lines indicate the minimum mass resolution for
each cluster run, as outlined in Sec.\ref{sec:dm}.}
\label{fig:mf1}
\end{center}
\end{figure*}

\begin{figure*}
\begin{center}
\includegraphics[width=0.9\textwidth]{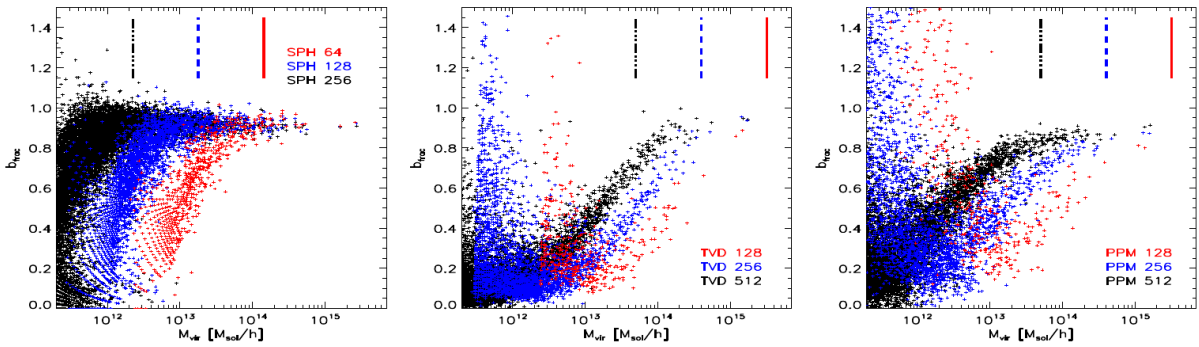}
\caption{Baryon fraction for all halos in the three codes, at all resolutions. The vertical lines
mark the minimum mass resolution criterion outlined in Sec.\ref{sec:dm}.}
\label{fig:bf}
\end{center}
\end{figure*}

\begin{figure*}
\begin{center}
\includegraphics[width=0.31\textwidth,height=0.29\textwidth]{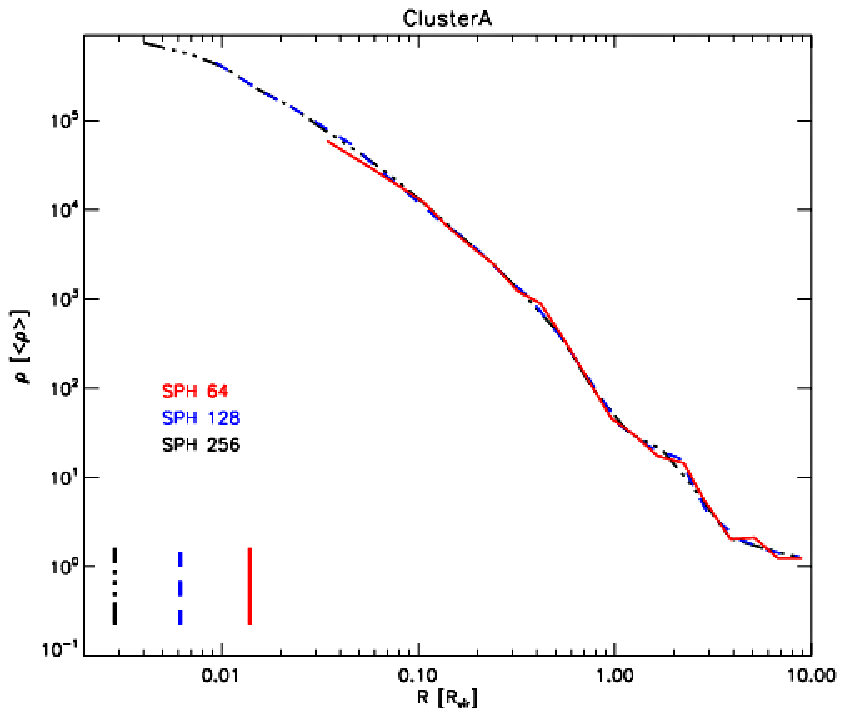}
\includegraphics[width=0.31\textwidth,height=0.29\textwidth]{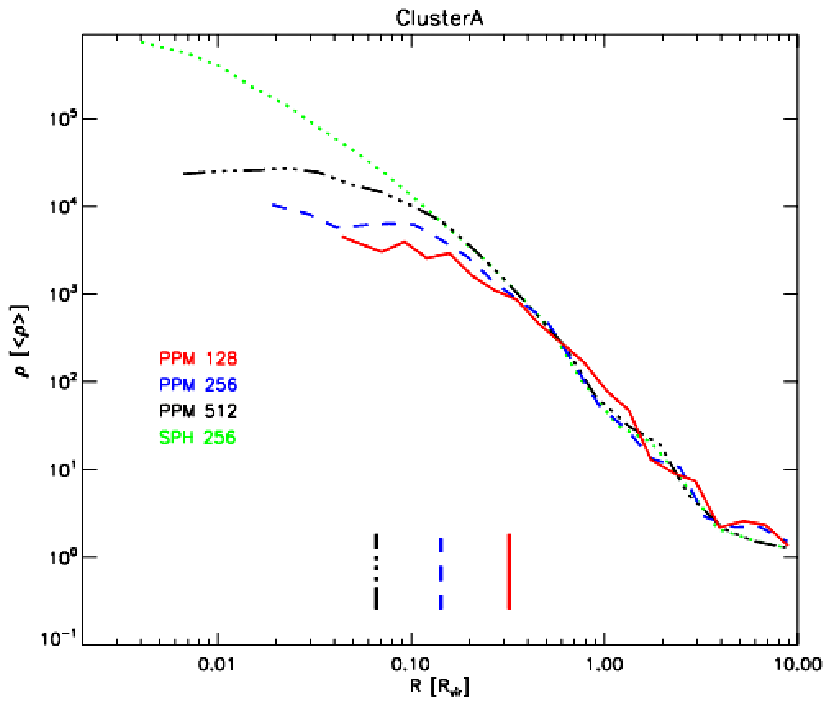}
\includegraphics[width=0.31\textwidth,height=0.29\textwidth]{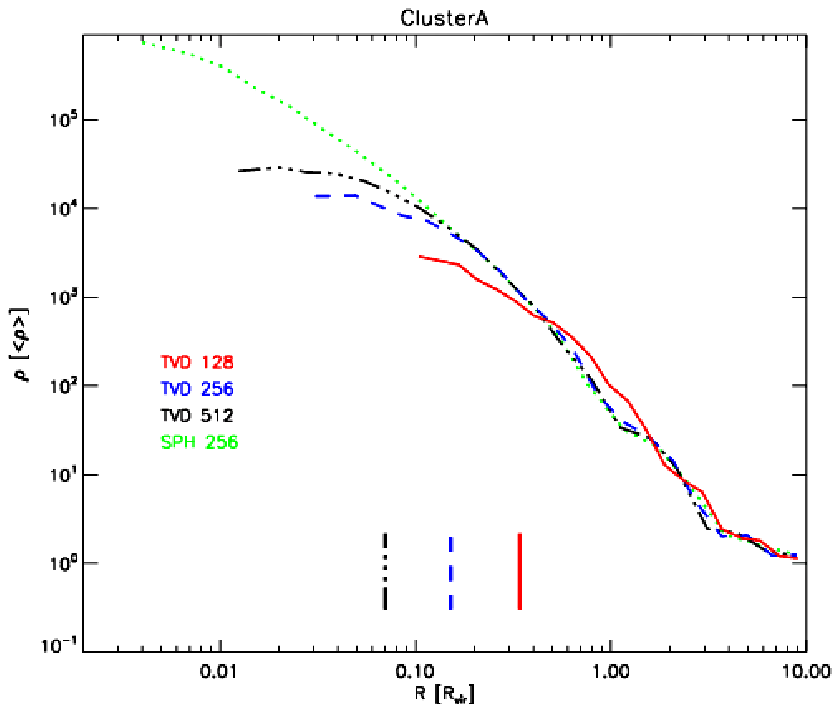}
\caption{Radial profile of DM density for the most massive galaxy cluster in our volume, for the three codes; the profiles of the 256 run in GADGET3 are reported for comparison in the
last two panels. The vertical lines in each panels show the 
value of gravitational softening for each run.}
\label{fig:dm_profile}
\end{center}
\end{figure*}

\section{Initial Conditions}
\label{sec:sim}

We have assumed a ``concordance'' model, with
density parameters $\Omega_\mathrm{curv} = 0$, $\Omega_{b} = 0.043$, $\Omega_{DM} =
0.227$, $\Omega_{\Lambda} = 0.73$, Hubble parameter $h = 0.70$, 
a power spectrum with slope $n=1$ and a normalization of the primordial matter power spectrum  
$\sigma_{8} = 1.2$. 
The $\sigma_{8}$ parameter is intentionally set
to a larger value compared to recent estimate from 
CMB data (e.g. Spergel et al. 2007) in order to enhance the probability
of forming massive halos within the simulated volume of side $100 \Mpc/h$.
Any modeling of cooling, radiative and heating processes for the gas component is neglected, and therefore the thermal history of cosmic gas here is mainly driven by shock waves induced by gravity.
Table \ref{tab:tab_cp} lists the main parameters of all simulations
run for the project. 

\bigskip

The initial displacements and 
velocities of DM particles were identical for all codes; 
the numbers of DM particles adopted are $512^{3}$, $256^{3}$, $128^{3}$
and $64^{3}$.
 The GADGET3 simulations preliminary looked remarkably converged with resolution
already at $256^{3}$, and therefore we choose to skip the production of the $512^{3}$ case in SPH, in order to spare
computational resources time.
The initial redshift of simulations were computed in order
to reach the same growth rate at $z=0$ for
the smallest available density perturbations: 
$z_{in}=67.99$, $z_{in}=55.92$, $z_{in}=44.77$ and $z_{in}=34.63$
for the different resolutions, respectively\footnote {
The initial conditions used in this Project 
are public and accessible at: http://canopus.cnu.ac.kr/shocks/case0/.}.

Usually in SPH cosmological runs both the DM and the gas particle
distributions are perturbed in their initial positions and velocities  according to the Zel'Dovich approximation (e.g. Dolag et al. 2008
for a review). In grid runs, on the other hand, the initial gas
distribution is at rest compared to the DM initial velocities.
Since computing exactly the same initial perturbation in velocity
for SPH particles and cells is not a trivial issue, for the sake
of simplicity in this work we neglected 
initial perturbations in velocities also for the SPH distribution. 

In the following we will refer to a given run according to the number
 of its gas particles or
gas cells; in the case of the TVD code, the number of DM particles is kept
8 times smaller than the number of gas cells (see Sec.\ref{subsec:tvdhydro}).
In what follows, we will typically refer to ``{\it self}--convergence'' meaning the convergence of a code with
respect to increasing resolution, and to ``{\it cross}--convergence'' meaning the convergence between
different codes, at a given resolution.

\begin{figure*}
\includegraphics[width=0.95\textwidth]{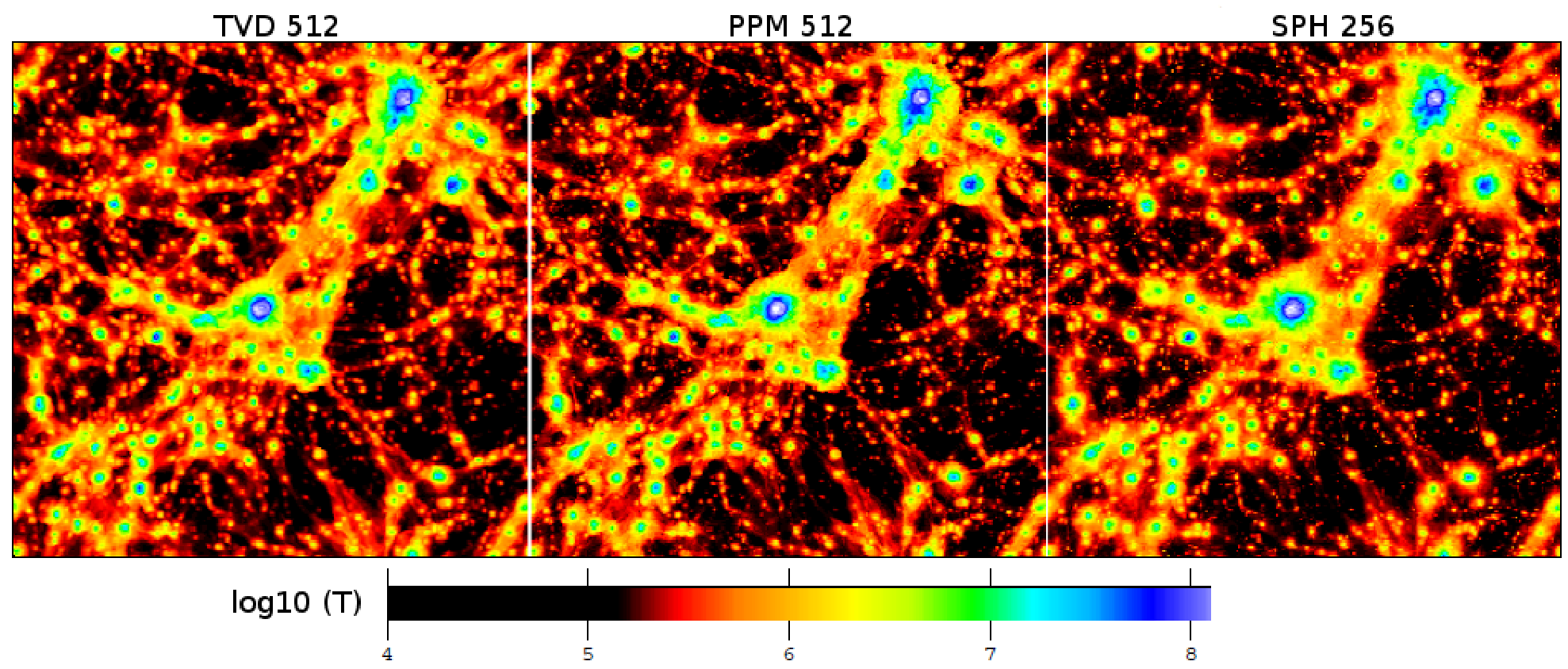}
\caption{Maps of projected mass-weighted temperature through the full
cosmological volume of $100 \Mpc/h$, for the three most resolved runs
of our project.}
\label{fig:maps}
\end{figure*}

\begin{figure*}
\includegraphics[width=0.95\textwidth]{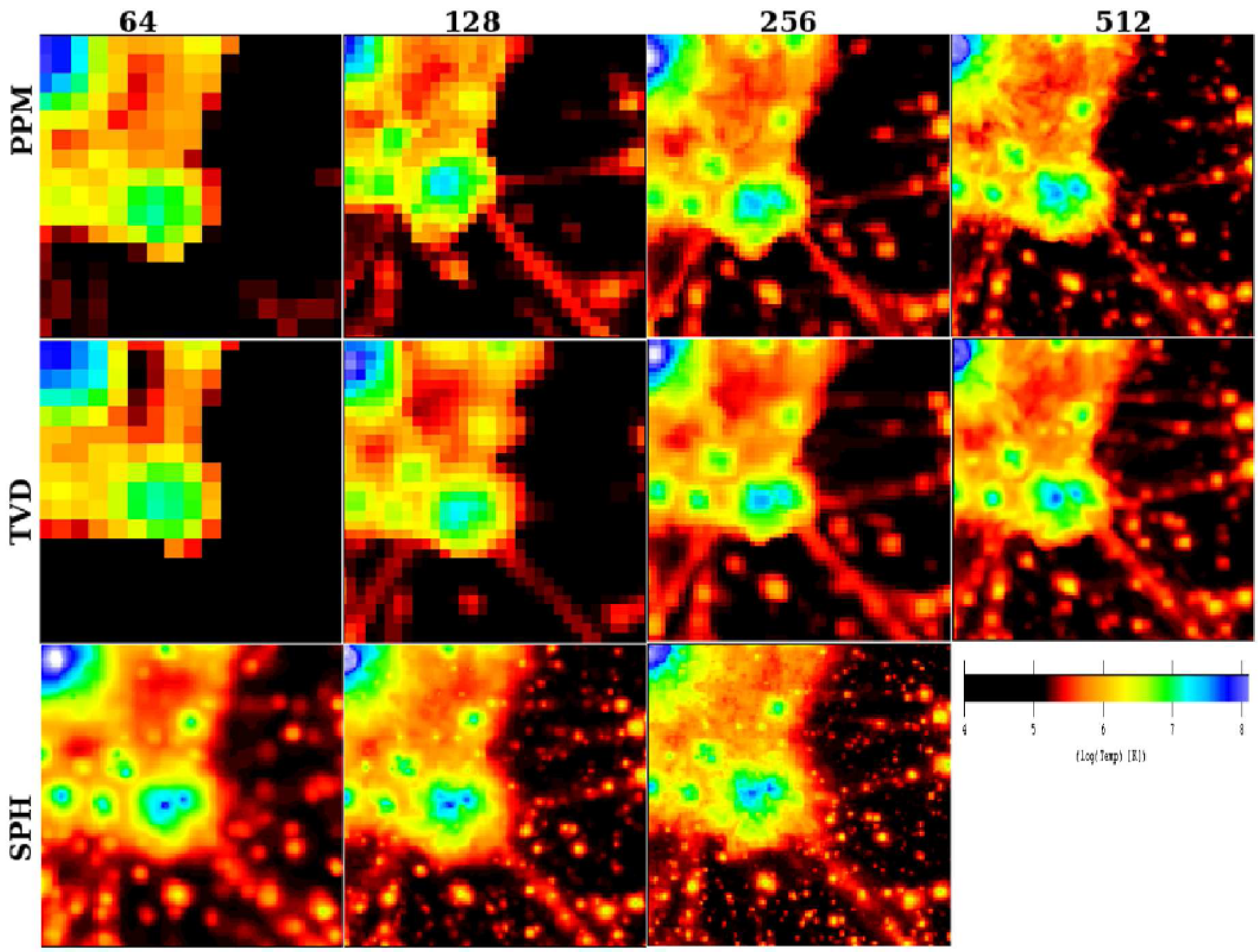}
\caption{Maps of projected mass-weighted temperature for a sub region with the side of  $40 \Mpc/h$ for all codes and
resolutions.}
\label{fig:maps_zoom}
\end{figure*}

\section{Dark matter properties}
\label{sec:dm}

A number of works in the literature have shown that 
present day numerical codes at their best achieve an 
agreement within $\approx 5-10$ per cent on the mass
functions of halos (e.g. Frenk et al. 1999;
O'Shea et al. 2005; Heitmann et al. 2007). However, 
subtle differences
in the adopted numerical methods should be responsible for
the exact shape of the inner DM profiles
(e.g. Bullock et al. 2001; Warren et al. 2006).

We compared the properties of the DM component
for all resolutions and codes in order to ensure that
the distribution of DM in our simulated large scale
structures is characterized by a similar
degree of intrinsic ``scatter'' reported in the
literature.

The most important statistics related to DM is the mass
function of halos, for which analytical solutions as a function
of cosmological parameters are available (e.g. Press \& Schecter
1974; Sheth \& Tormen 1999).
We report in Fig.~\ref{fig:mf1} the cumulative mass functions (DM plus gas) 
for all runs in the project. 
The virial mass, $M_{\rm vir}$, is customarily defined as the spherical
over-density of gas+DM, enclosing a mean over-density of $\approx 109 \rho_{\rm cr}$, where $\rho_{\rm cr} \approx 9.31 \cdot 10^{-30} g/cm^{3}$ is the critical density of the universe (e.g. Eke et al.  1998). 
The virial halo masses are computed using the same 
halo finder in all codes, based on the gas+DM spherical over-density. 
In the case of grid runs, the cells distributions have been converted into a distribution of particles, in order to
apply  exactly the same procedure used to analyse GADGET runs.

In order to compare different codes and resolution, it is useful to assign
a ``formal'' resolution to each run. This allows us to understand which  halos in our simulations are suitable for
 ``convergence'' studies and which are not, because of  under-sampling problems at a given cluster size.
Even if in GADGET runs the mass functions are resolved down to the smallest halos (with $<20$ particles), Power et al.  (2003) showed that 
convergence in the inner dynamical structures of halos is achieved with 
at least
$\sim 500$ particles inside $R_{\rm vir}${\footnote {We notice, however, that tests with radiative
runs have shown that a larger number of particles, $N \sim 1000-5000$,
may required to achieve a good convergence in the X-ray luminosities
of clusters (e.g. Valdarnini, Ghizzardi \& Bonometto 1999; Valdarnini
2002).}}.

We preliminary consider that the resolution limit in GADGET is 
achieved with 500 DM particles
within the virial radius.
For grid runs, we apply the following empirical
approach: we consider only halos whose virial radius is resolved with at
least 500 cells, and we assign a {\it formal} minimum mass  
to have halos ``suitable for convergence'' taking the corresponding
virial mass, extracted from the theoretical $R_{\rm vir}$ versus
 $M_{\rm vir}$ relation.
 
The corresponding minimum masses for all codes and resolutions are shown
as vertical lines in Fig.~\ref{fig:mf1}.
Although this methods is rather artificial, we find it predicts rather
well the convergence observed for halos in grid codes, which takes place at
larger masses compared to corresponding GADGET runs at the same DM mass
resolution. For instance, GADGET run 256 shows a halos mass function
which is converged down
to masses of $\sim 2 \cdot 10^{12}M_{\odot}$, while run 256 in ENZO and TVD
achieve convergence only for halos with masses larger than
$5 \sim  10^{14} M_{\odot}$. 

Therefore, we would expect to see cross-convergence of the virial 
parameter for none but the largest halos in grid results, while we expect good self-convergence
across a larger range of masses in GADGET.

A similar trend is also observed in the baryon fraction of halos 
in the various run, as reported in Fig.~\ref{fig:bf}.
The baryon fraction in GADGET is rather perfectly converged at all 
resolutions for $M > 10^{14} M_{\odot}/h$, with a value 
of $f_{b}\approx 0.9 f_{\rm cos}$, where $f_{\rm cos}=\Omega_{\rm b}/(\Omega_{\rm b}+\Omega_{\rm dm})=0.159$ is the cosmic baryon
fraction in our runs)
. In grid codes, 
the convergence to a slightly larger baryon fraction, $f_{\rm b} \sim 0.95 f_{\rm cos}$, 
seems to be reached only for masses larger than $M> 10^{15} M_{\odot}/h$
(as for the halos mass functions, ENZO shows the a slower rate of convergence
compared to TVD).

The radial profiles of DM mass density for the most massive galaxy cluster in our sample
are shown in Fig.~\ref{fig:dm_profile} for various resolutions. All profiles in GADGET3
runs are remarkably self converged, while the profiles of DM in both grid methods present a slower rate of convergence. At the best available resolution, the grid codes agree at the 
percent level with the reference profile of GADGET3 runs, with sizable differences only
in the core region of the cluster, $<0.1 R_{\rm vir}$, due to the well known lack of force 
resolution in the PM method (the softening length for the gravity force in the $512^{3}$ runs is $293 \kpc$).

Overall, the trend found are in line with those reported by
O'Shea et al. (2005) and Heitmann et al. (2008). Based on our results, we suggest
that the representation of the underlying DM distribution is 
similar to what can be found in the recent literature, and that the 
bulk of differences that will be reported in the next Sections are
mostly connected with a different modeling of hydrodynamics
in the various methods.

\begin{figure}
\begin{center}
\includegraphics[width=0.49\textwidth,height=0.45\textwidth]{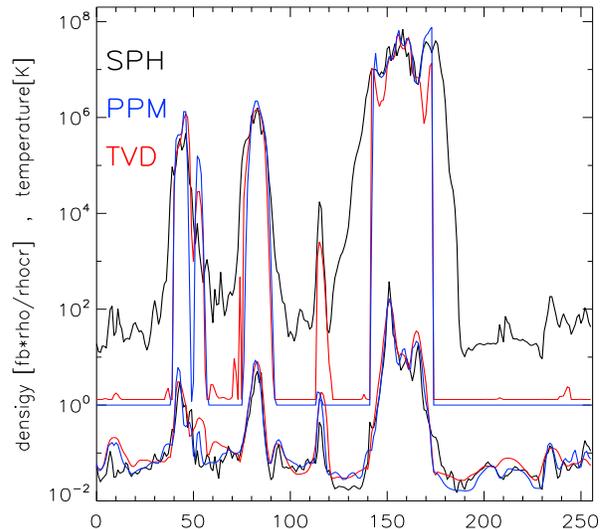}
\caption{One dimensional distribution of gas mass density ({\it lower lines}) and volume-weighted temperature 
({\it upper lines}) for a line crossing our simulated volume,
for all $256^{3}$ runs.}
\label{fig:1d_line}
\end{center}
\end{figure}

\section{Baryonic Matter Properties}
\label{sec:thermal}

In the following Sections we compare the
distributions of several gas thermodynamical variables in all runs
as a function of numerical resolution. The final
goal is to identify which are the cosmic environments and minimum 
resolution requirements necessary to achieve a good 
convergence in the estimates provided by the different methods.

\subsection{Maps}
\label{subsec:map_cp}

A preliminary inspection of the morphological distribution
of baryon gas in the cosmic structures captured by all 
methods ensure that at a zero order all simulations correctly
sample a cosmological volume with identical density
fluctuations.
In Figure~\ref{fig:1d_line} we report the one-dimensional behavior of gas density and gas temperature along 
a line crossing the position of the most massive galaxy
cluster in the volume, for all $256^{3}$ runs. The spatial distribution of gas density
is well matched in all codes, and in particular the positions
of the gas density peaks associated with halos and filaments agree within a $1-2$ cells accuracy (i.e. $\sim 400-800 \kpc/h$ at this 
resolution). The one-dimensional gas temperature profiles show very similar maxima near the gas density peaks, 
but sizable differences can be found 
in the outer regions. The bulk of the difference here is
however a simple effect of the variable smoothing length in GADGET3,
which provide a coarser resolution compared to grid codes
for the regions outside of clusters.

In the panels in Fig.~\ref{fig:maps} we report the maps of projected mass-weighted temperature across the simulated volume,  for the most resolved runs of the project (run 256 for GADGET3 and runs 512 for ENZO and TVD).
The trend with resolution of the projected mass-weighted
temperature, at all resolutions, is reported in Fig.~\ref{fig:maps_zoom} for a sub-volume of $40 \Mpc/h$
inside the cosmological box. 

To readily compare Lagrangian and Eulerian data at the
same spatial resolution, the gas fields of
GADGET3 runs have been interpolated
onto a regular grid, with resolution equal to that of the corresponding grid  runs, using the same 
SPH kernel employed during the simulation for each gas particle.

In GADGET3, over dense non-linear structures (e.g. halos and sub-halos) 
are very similarly reconstructed at all resolutions, 
while structures at about the critical density  (e.g. cosmic filament) start being resolved only at
sufficiently high DM mass resolution. The opposite trend
appear in grid codes, where large scale patterns are soon reconstructed at all resolutions,
while a clear modeling of the smaller halos and cluster satellites 
is achieved only approaching the highest available resolutions.

\begin{figure*}
\begin{center}
\includegraphics[width=0.33\textwidth]{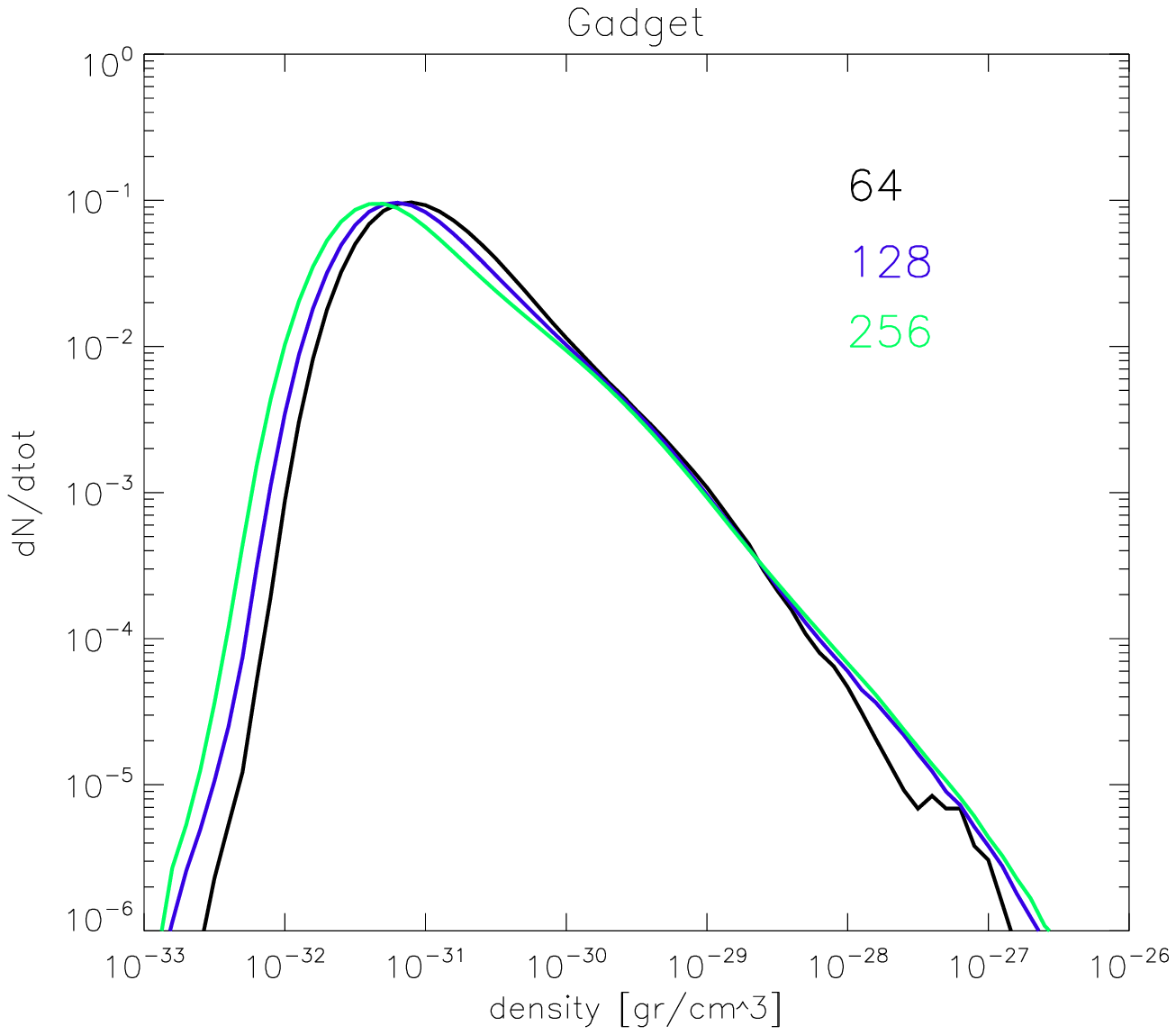}
\includegraphics[width=0.33\textwidth]{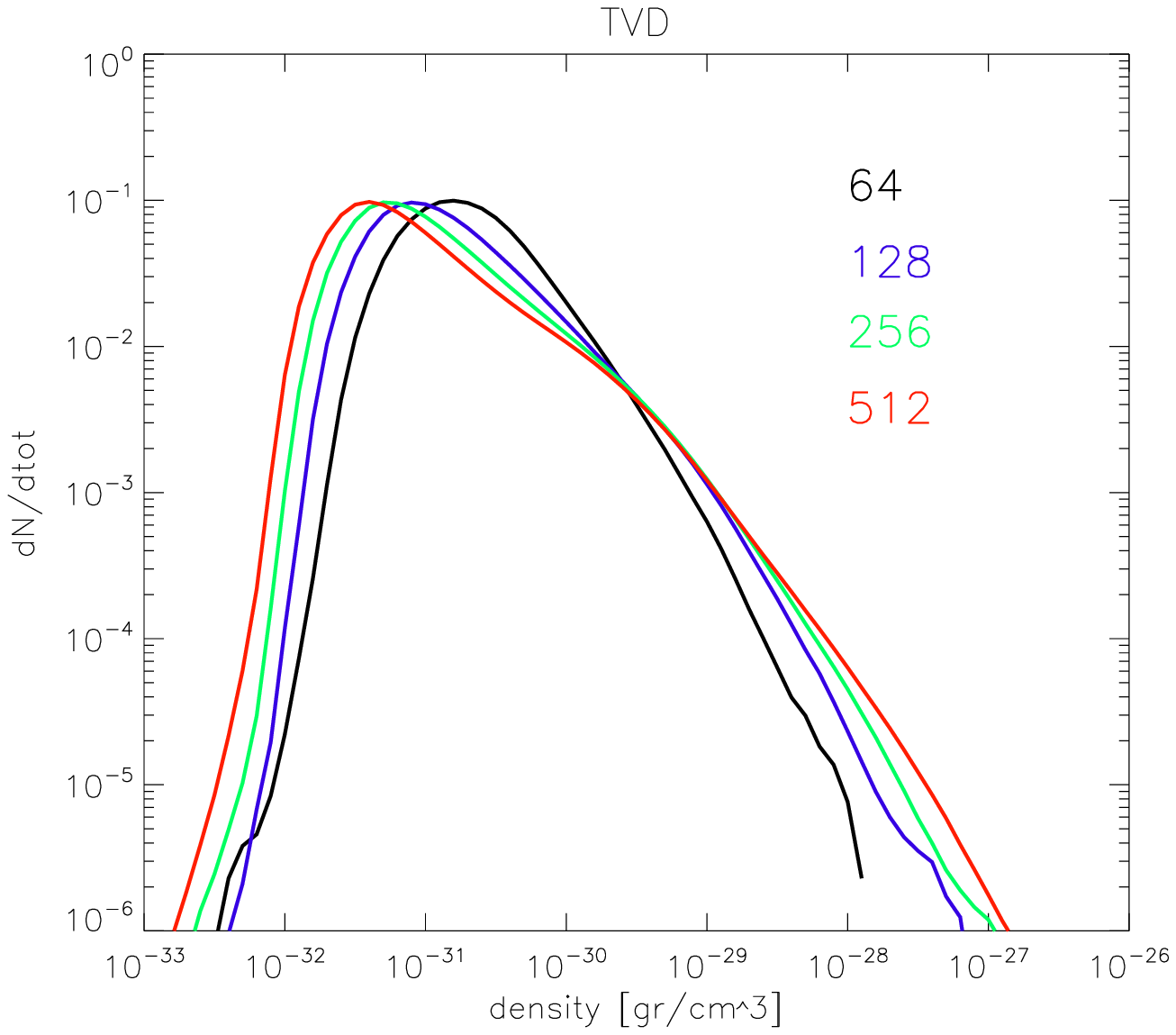}
\includegraphics[width=0.33\textwidth]{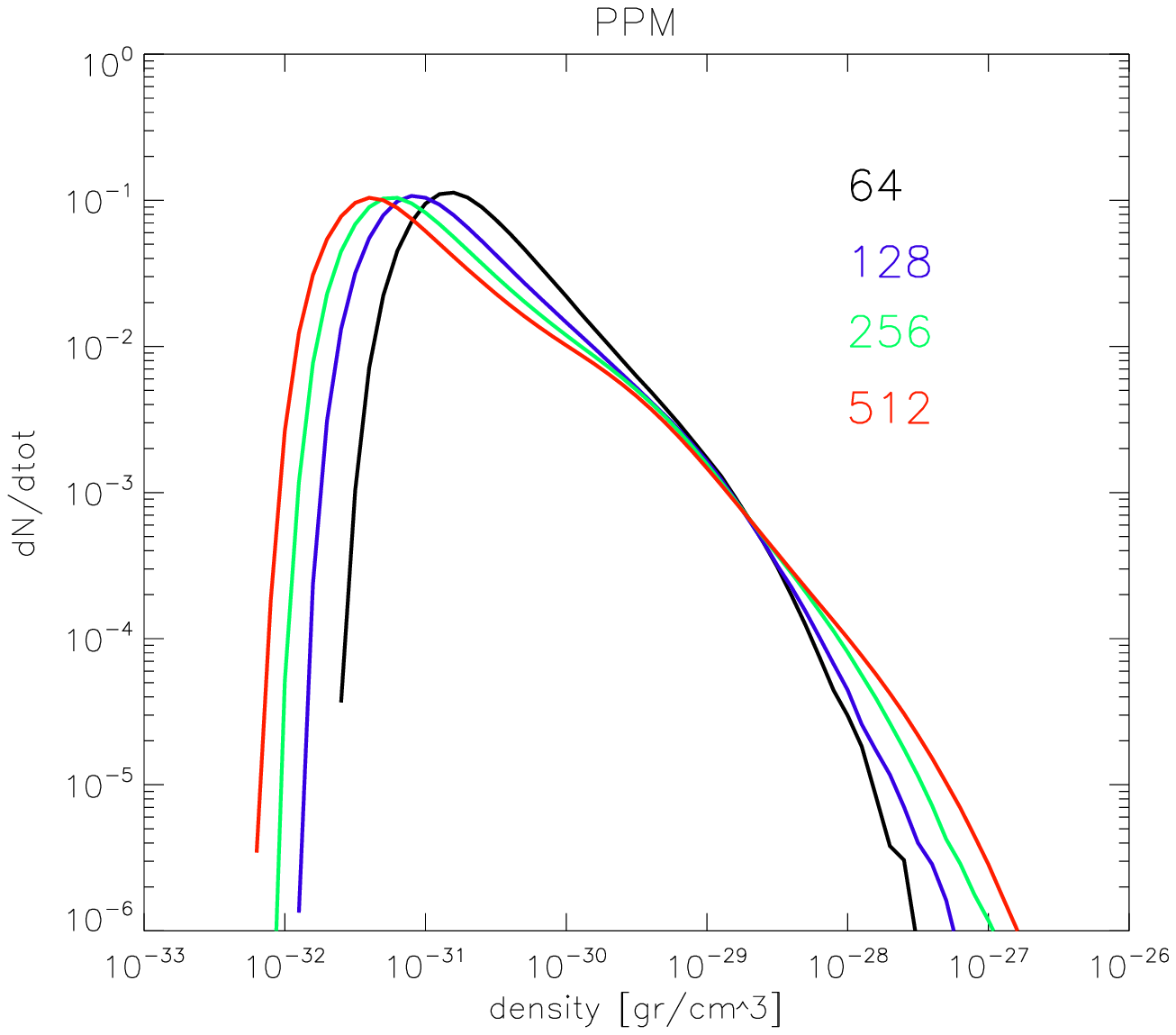}
\includegraphics[width=0.33\textwidth]{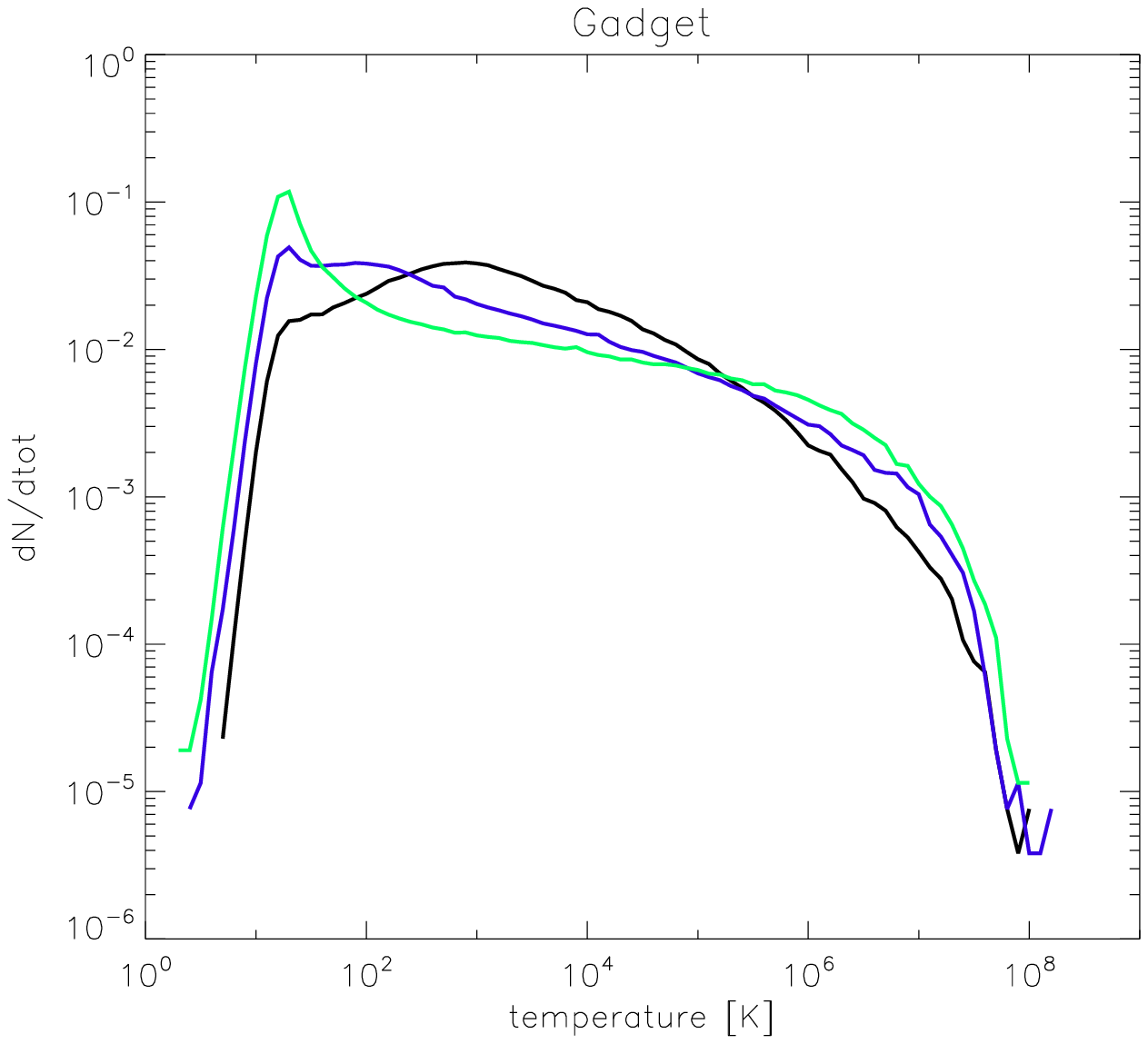}
\includegraphics[width=0.33\textwidth]{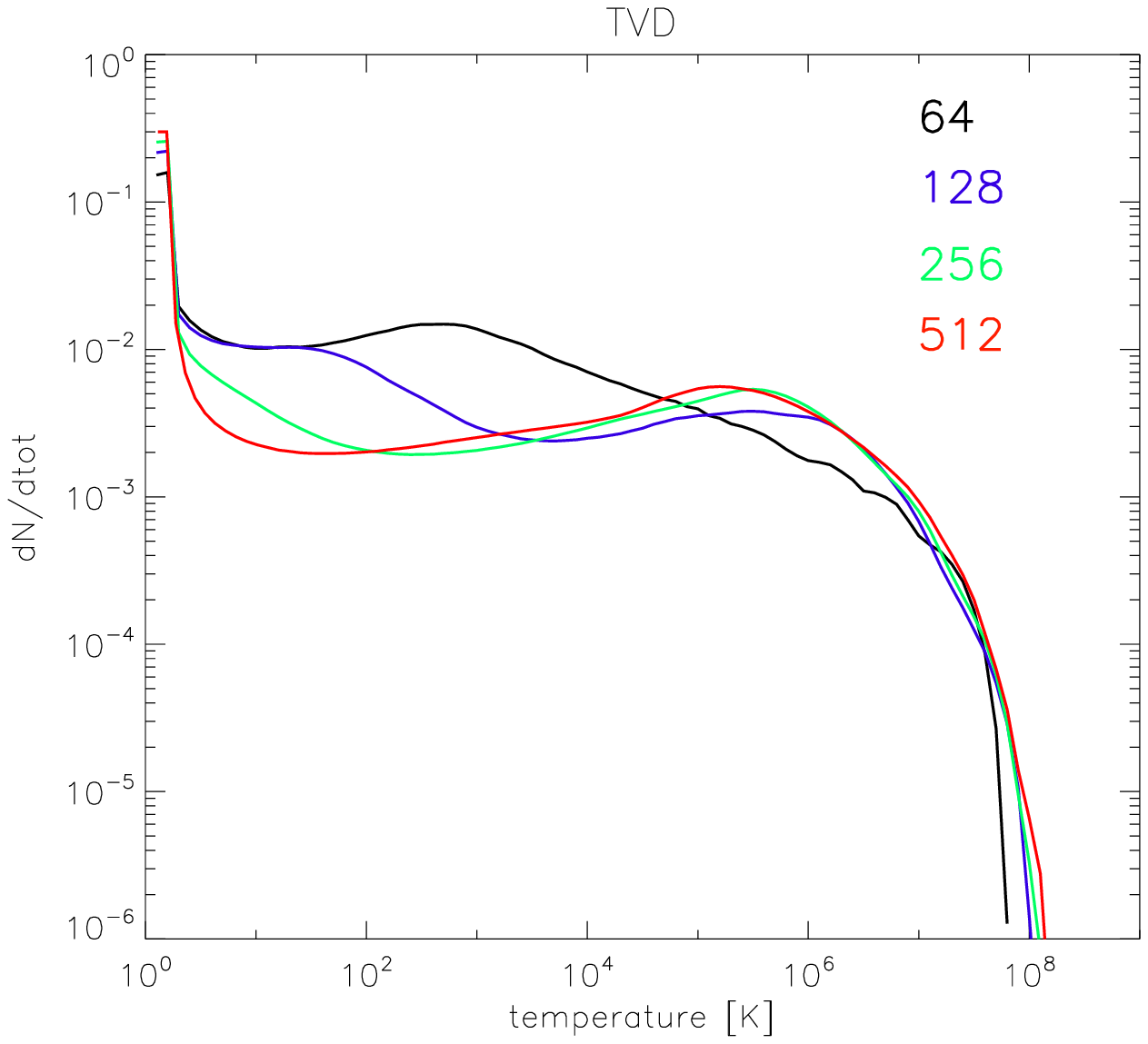}
\includegraphics[width=0.33\textwidth]{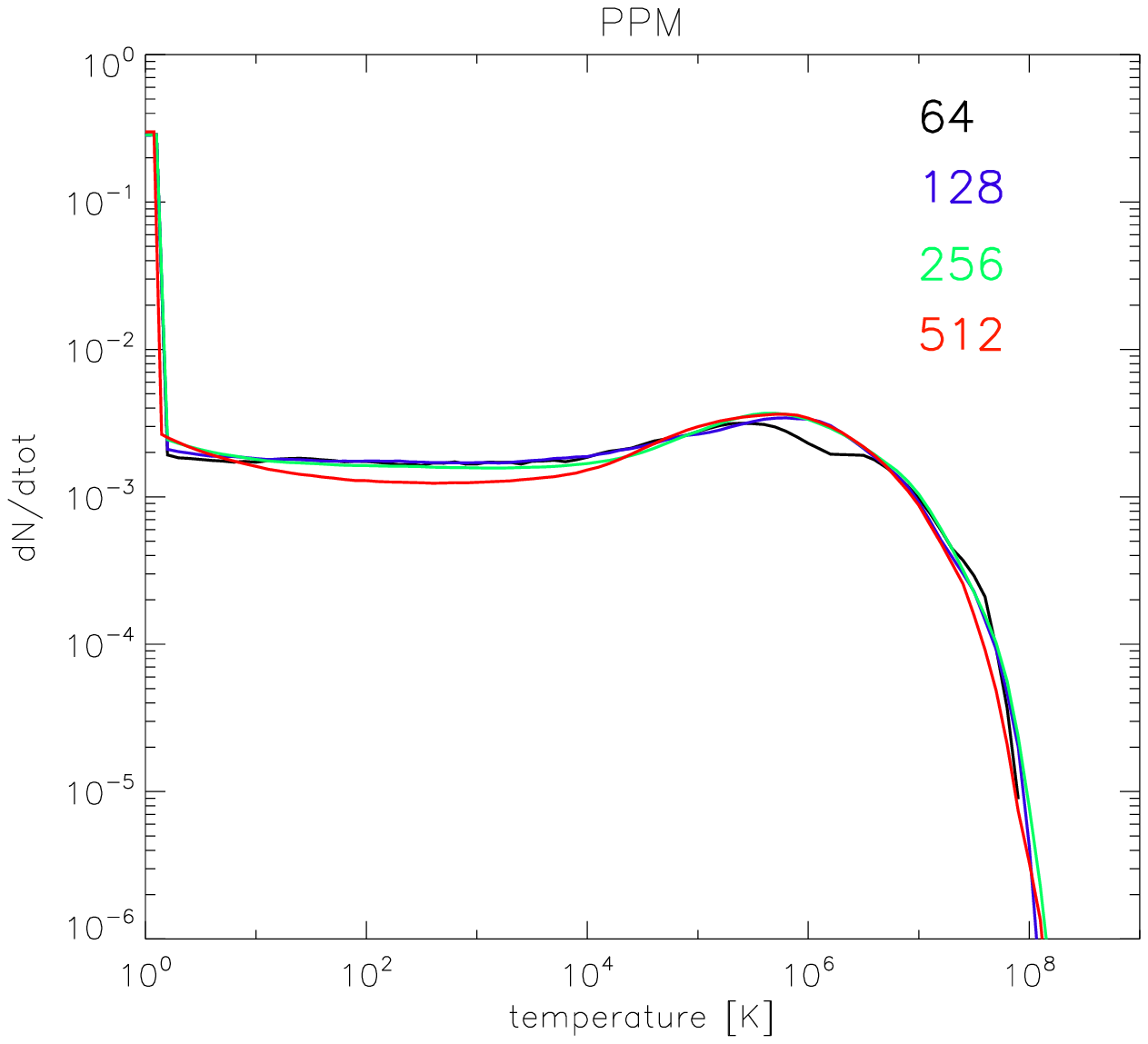}
\caption{Gas density and volume-weighted gas temperature distributions for all resolutions and all codes. The critical density of baryons is $\rho_{\rm cr,b} \approx 4.0 \cdot 10^{-31} g/cm^{3}$.}
\label{fig:res_pdf}
\end{center}
\end{figure*}

\begin{figure}
\begin{center}
\includegraphics[width=0.49\textwidth,height=0.35\textwidth]{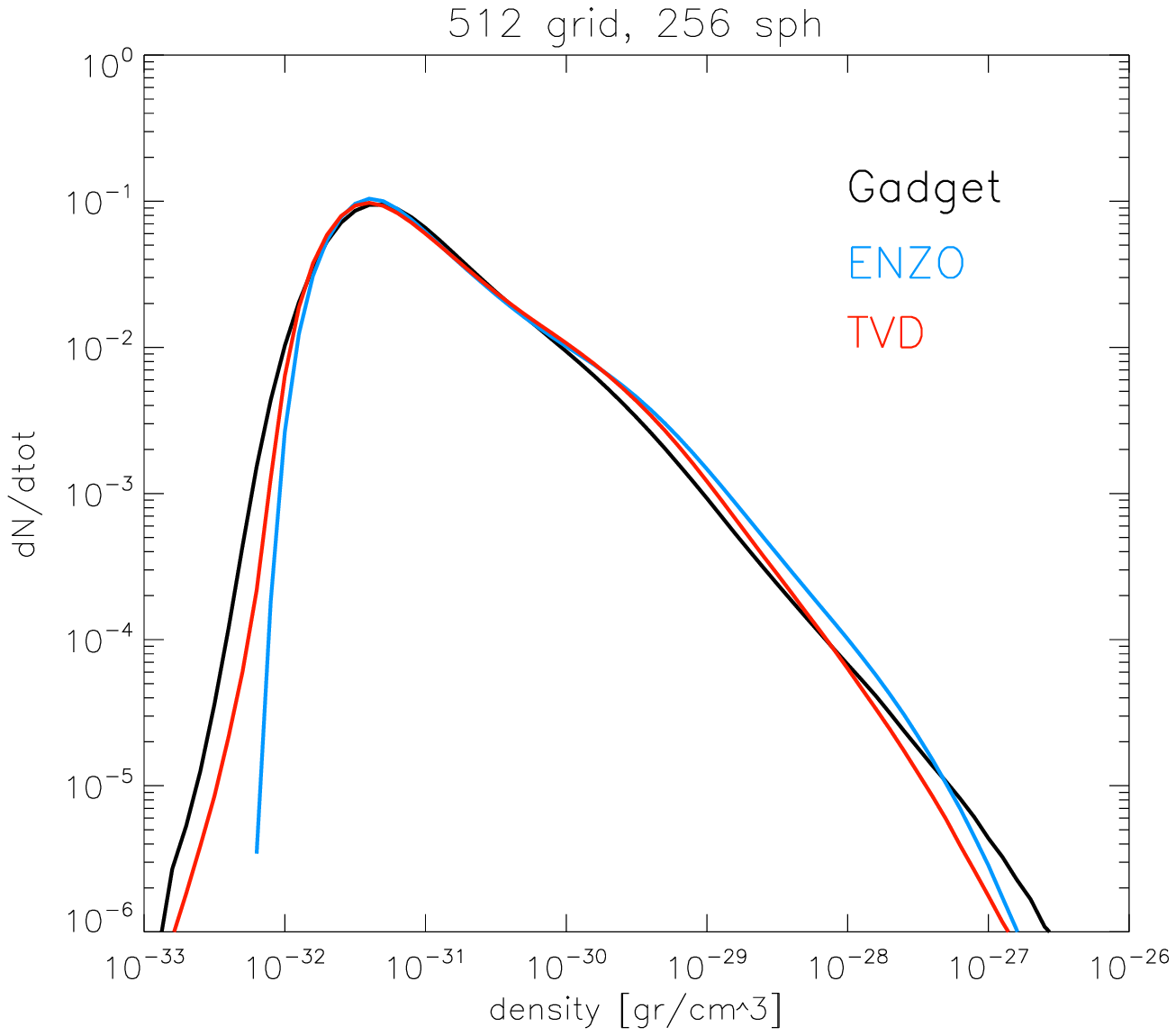}
\includegraphics[width=0.49\textwidth,height=0.35\textwidth]{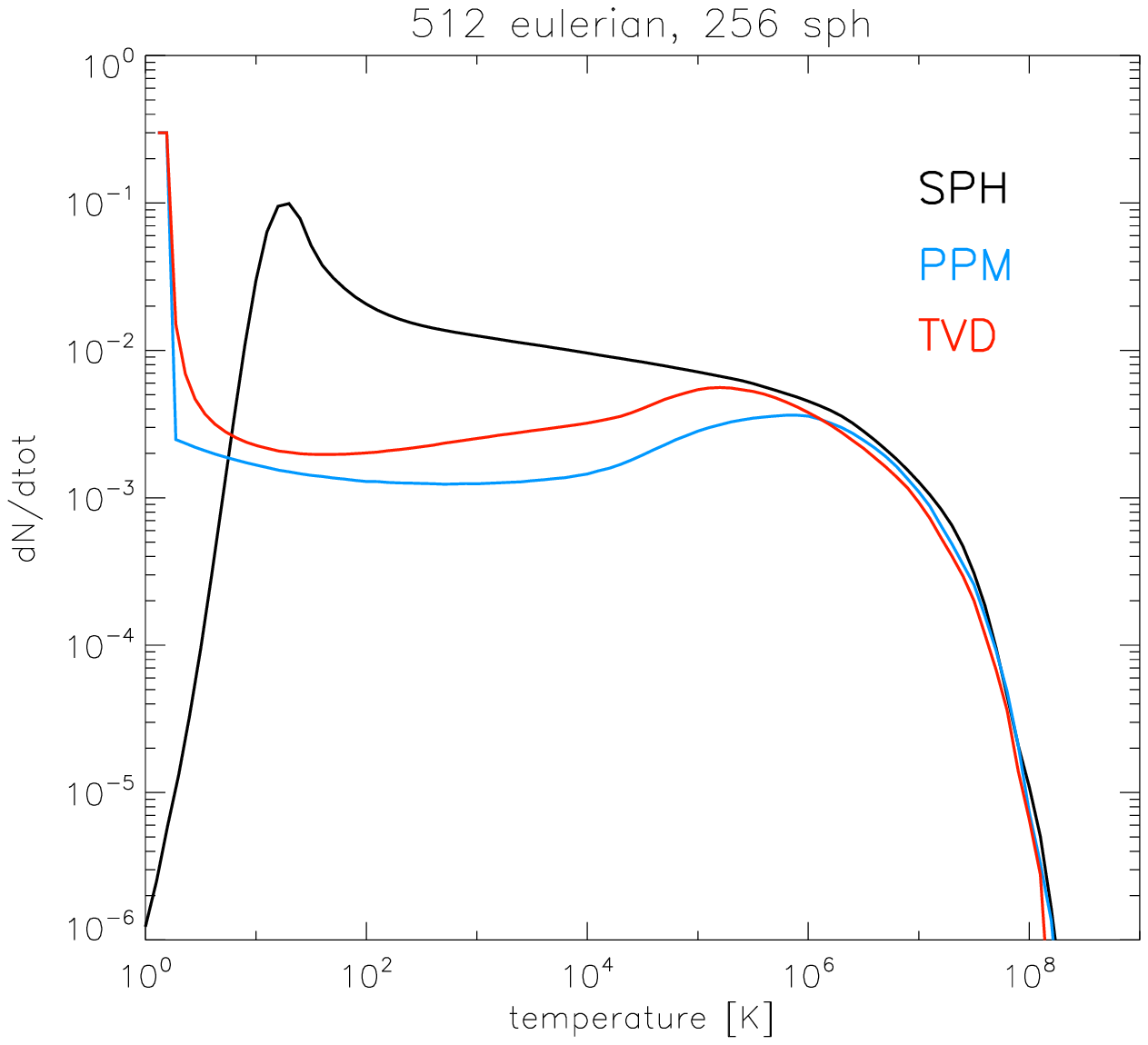}
\caption{Cross convergence of volume weighted gas density and gas temperature distributions for GADGET3 run with $256^{3}$ and grid runs with  $512^{3}$.}
\label{fig:all_pdf}
\end{center}
\end{figure}

\subsection{Distribution Functions}
\label{subsec:pdf}

A quantitative analysis of the differences between the codes
is performed by studying the volume-weighted distribution functions of gas density and gas
temperature at increasing resolution, as shown in Fig.~\ref{fig:res_pdf}.
Figure \ref{fig:all_pdf} further shows the cross-comparison between the highest resolution
runs available for each code. 

In this case, we adopt volume-weighted statistics for each
bin in gas density/temperature.
Despite the obvious fact that volume-weighted distributions cannot
be translated into observable quantities (since the
convolution of the two does not provide the total
gas energy within the simulated volume) we find this 
approach useful to focus on the properties
 of the low density, volume-filling baryon gas
 around large scale structures.
 Our purpose here is to highlight the differences in the 
 modeling of the lower density baryon gas at large
 scales (which encompasses filaments and clusters
 of galaxies) in the different numerical methods. This 
 can also be readily compared with the early comparison
 work of Kang et al.  (1994). 
 In addition, these volume filling regions are expected 
 to be an important site of acceleration of relativistic
particles, via direct shock acceleration at strong
 shocks (e.g. Miniati et al.  2001; Ryu et al.  2003;
 Pfrommer et al.  2006; Vazza, Brunetti \& Gheller 2009). 

In the following Section (Sect.~\ref{subsec:clusters_cp})
we will rather refer to mass-weighted profiles of 
gas density and gas temperature, since they are closely
related to the thermalization properties of internal merger shocks inside clusters. 
  
As expected, the cross--convergence between different codes is more satisfactory when resolution is increased:
the density distributions runs with $\geq 256^{3}$ DM particles
(i.e. with 
$m_{dm} \leq 4.5 \cdot 10^{9} M_{\odot}/h$), 
have the same average value in all codes within a $20-30 $ per cent scatter. The largest 
and the smallest gas densities are similar within a factor of $\sim 2$, 
and GADGET3 produces the most extreme values in both cases. 
GADGET3 runs are also the ones which provide the largest
degree of self-convergence, with very similar outputs at all
investigated resolutions.

In the case of temperature distributions, ENZO presents
the larger degree of self-convergence (within a factor of $\sim 10$ per cent)
at all resolutions, while the other codes show significant 
evolution with
resolution, especially at temperatures below $T < 10^{4-5} ~K$.

We note that different floors in the value of temperature were adopted in the three codes, to 
limit the lowest temperature available to cells/particles
of the simulated volume. For each code we used the temperature
floor usually adopted by each simulator: a minimum temperature of $T_{\rm o}=1~K$
in allowed in ENZO, $T_{\rm o}=2~K$ in TVD and $T_{\rm o}=24~K$ in GADGET3. This explains the 
different piling of cells/particles in the temperature distributions below $T<50~K$; we also made sure that the adoption of different 
floor in temperature does not affect in any way the temperature
distribution above the adopted $T_{\rm o}${\footnote{It should be stressed that all most recent simulations model the
action of the re-ionization background, hence increasing the minimum temperature in
the simulations to much larger values, $T_{\rm o} \sim 10^{3}-10^{4} ~K$ (e.g.
Vazza, Brunetti \& Gheller 2009). Therefore the analysis of the temperature distribution we present here is meant
to pinpoint the numerical problems of the various methods, while the differences
between runs employing re-ionization would be much smaller.} }.

On the other hand, the temperature distributions found in the simulations 
become quite similar for 
$T > 10^{6} ~K$, which would correspond to the typical virial 
temperatures of collapsed halos; this is in line with the
early findings reported by Kang et al. (1994), and later by O'Shea
et al.  (2005).

We conclude that even if the gas mass distribution within halos is rather convergent in all codes for (for a DM mass resolution
of $m_{dm} \leq 4.5 \cdot 10^{9} M_{\odot}/h$), the convergence in 
the gas temperature distribution is generally not yet reached,
and the
cross-convergence between codes is not achieved for all
regions where  $T<10^{5}-10^{6}$, for the resolutions investigated
in this project. 

In these regimes, some amount of spurious numerical heating can 
be expected due to the graininess of DM mass distributions, which makes two-body
heating a likely channel of (un-physical) energy transfer from the DM particles
to the baryon gas (Steinmetz \& White 1997). The effect of two-body heating is expected to decrease with the number of DM particles in the simulation, so the trend with resolution in all codes qualitatively suggests that at least part of the different temperature below $T<10^{4}~K$ is related to this effect. However, the evolution of 
gas temperature with resolution in ENZO runs is extremely small compared to all other codes.  

Interestingly, a similar trend was noticed by O'Shea et al.  (2005), by comparing 
the temperature distributions obtained with GADGET2 and ENZO (both using the PPM
version of the code, or its formulation with artificial viscosity, i.e. ENZO-ZEUS). 
The authors  suggested that the reported trend were consistent with an 
increasing action of the effective viscosity employed in the hydro solver of the
three codes, going from ENZO-PPM to ENZO-ZEUS to GADGET2. This explanation
is also likely in our case; we will come to this point again in Sect.~\ref{subsec:phas_comp}, in connection with the study of 
phase diagrams for the shocked cells/particles in the various
runs.

\begin{figure*}
\begin{center}
\includegraphics[width=0.32\textwidth,height=0.25\textwidth]{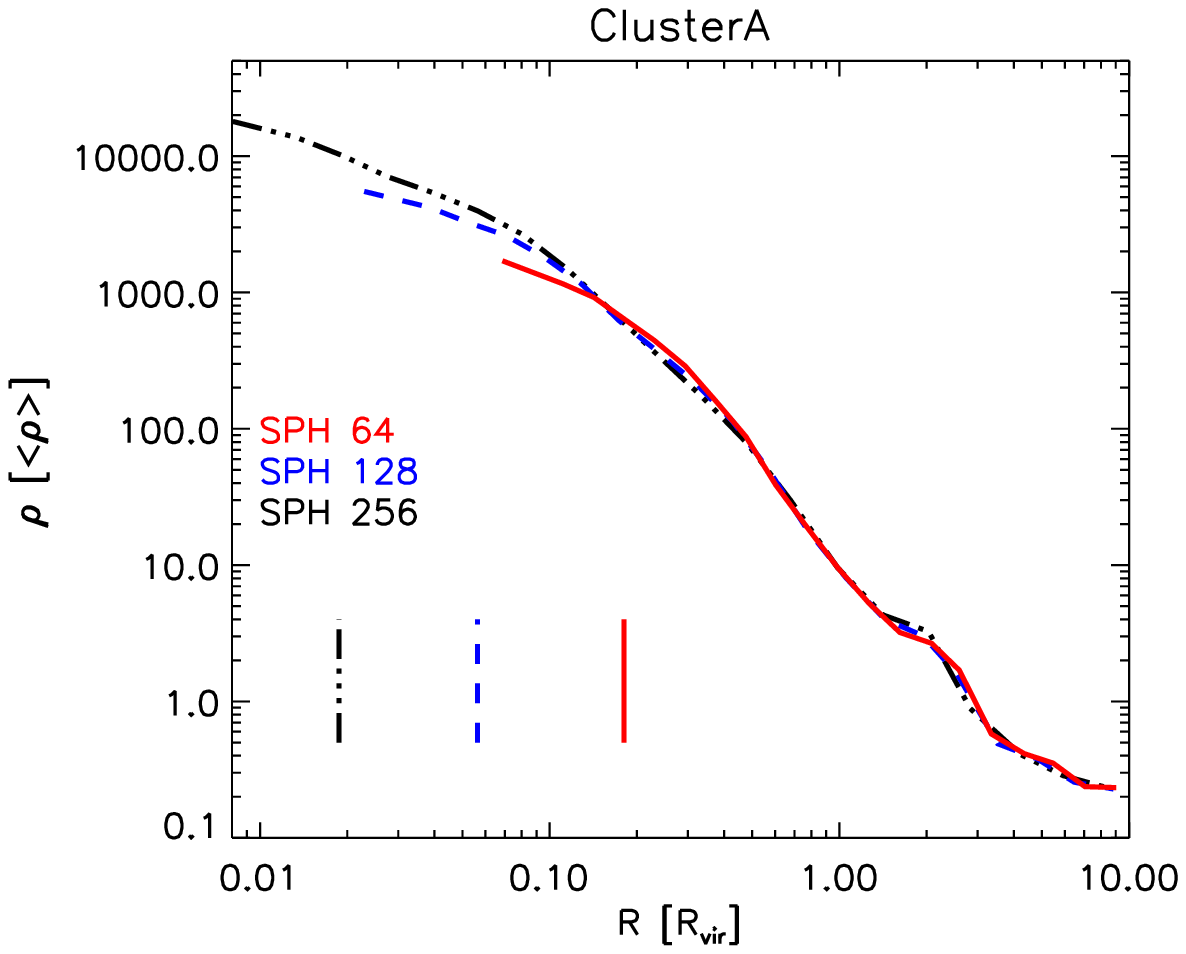}
\includegraphics[width=0.32\textwidth,height=0.25\textwidth]{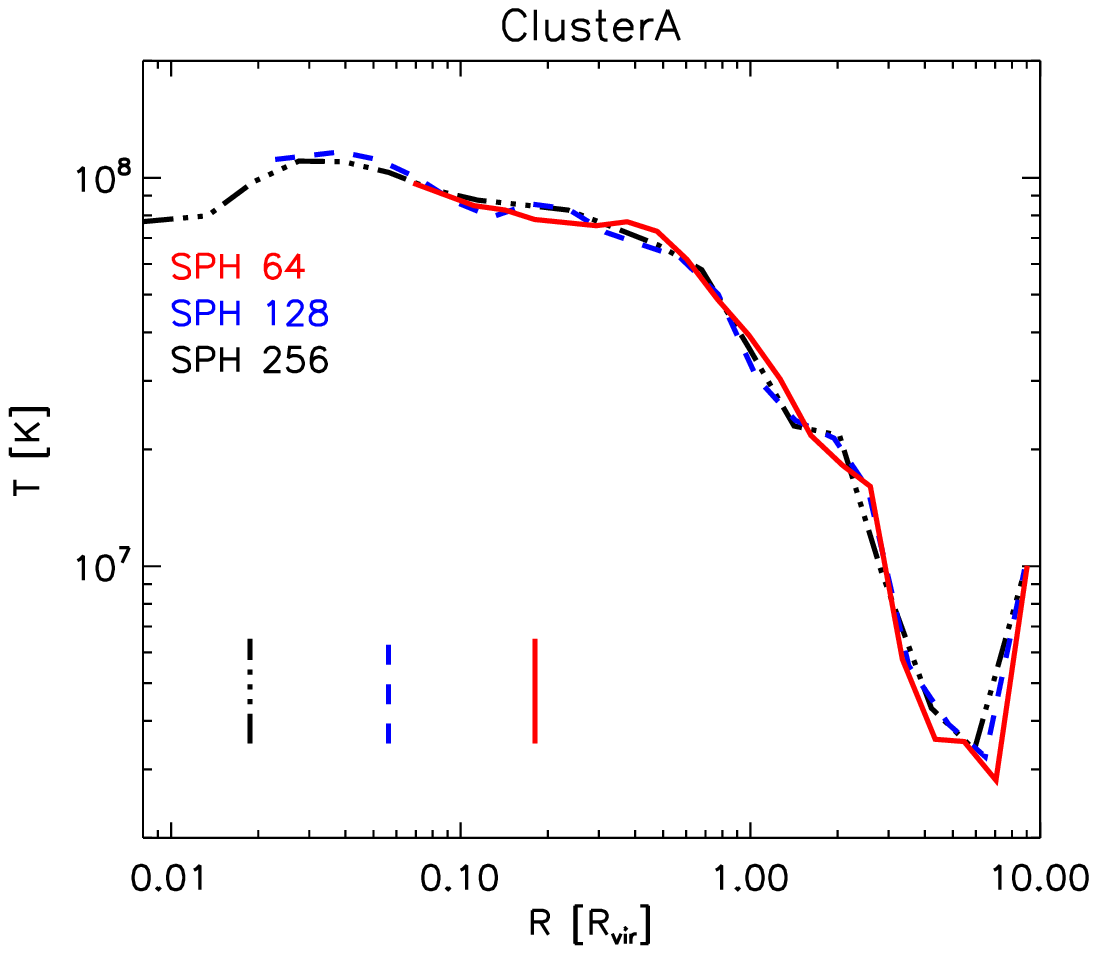}
\includegraphics[width=0.32\textwidth,height=0.25\textwidth]{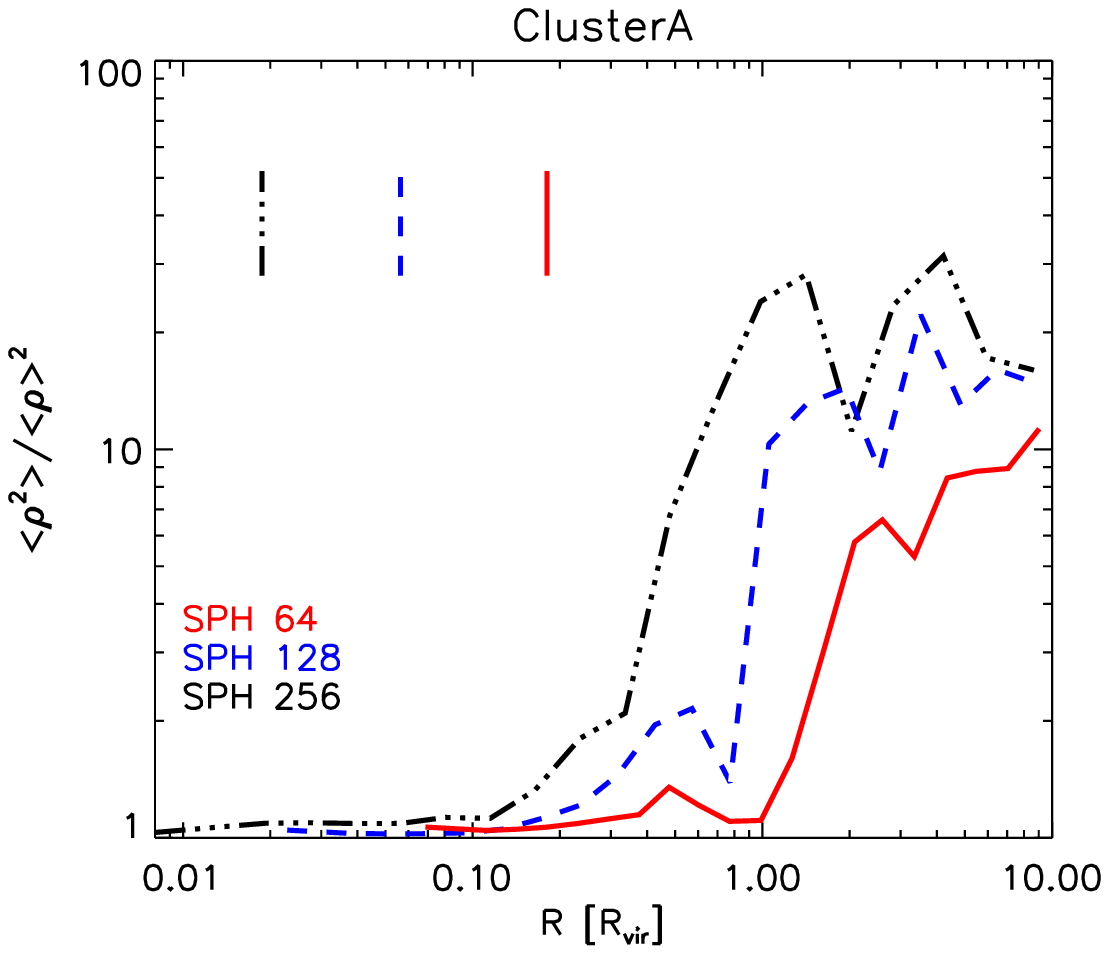}
\includegraphics[width=0.32\textwidth,height=0.25\textwidth]{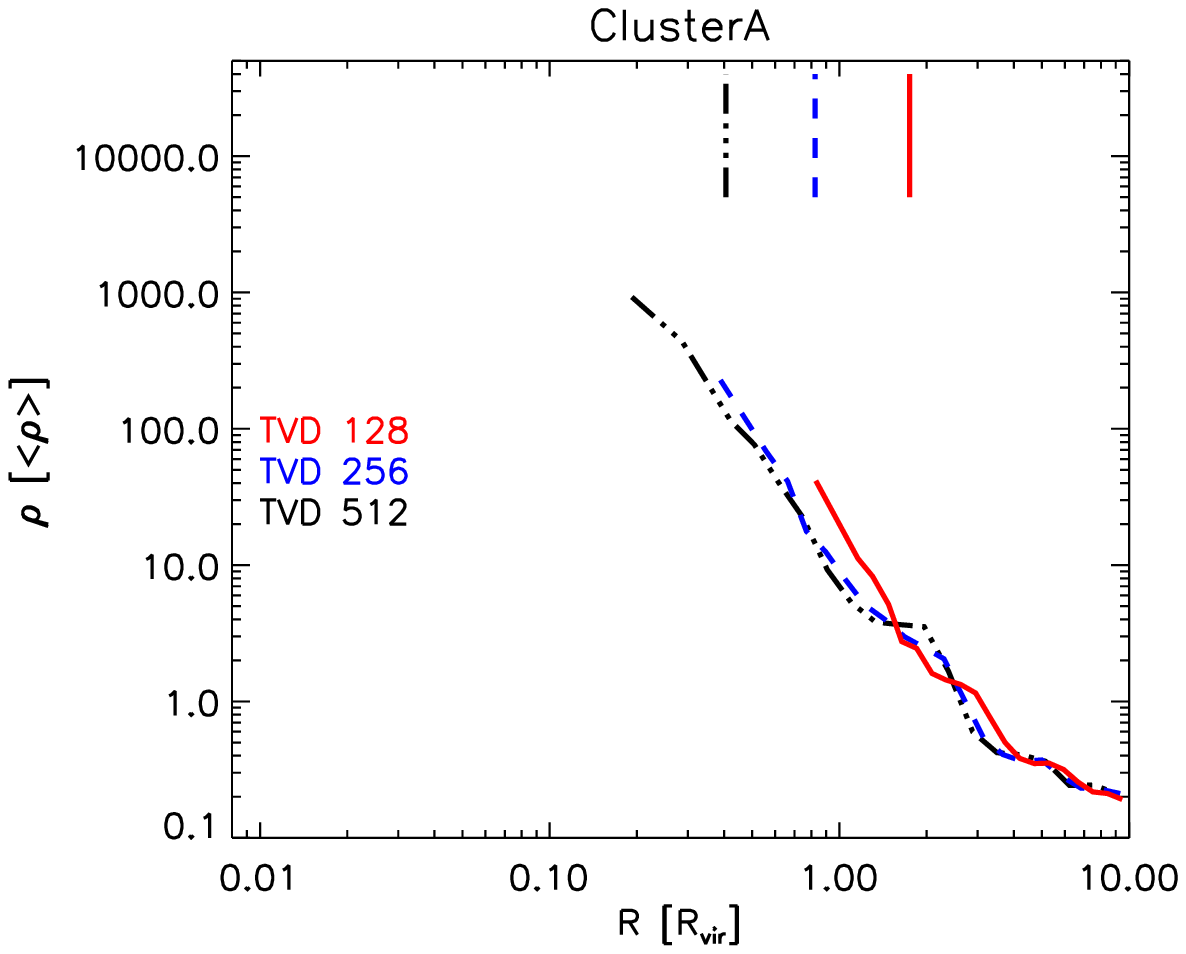}
\includegraphics[width=0.32\textwidth,height=0.25\textwidth]{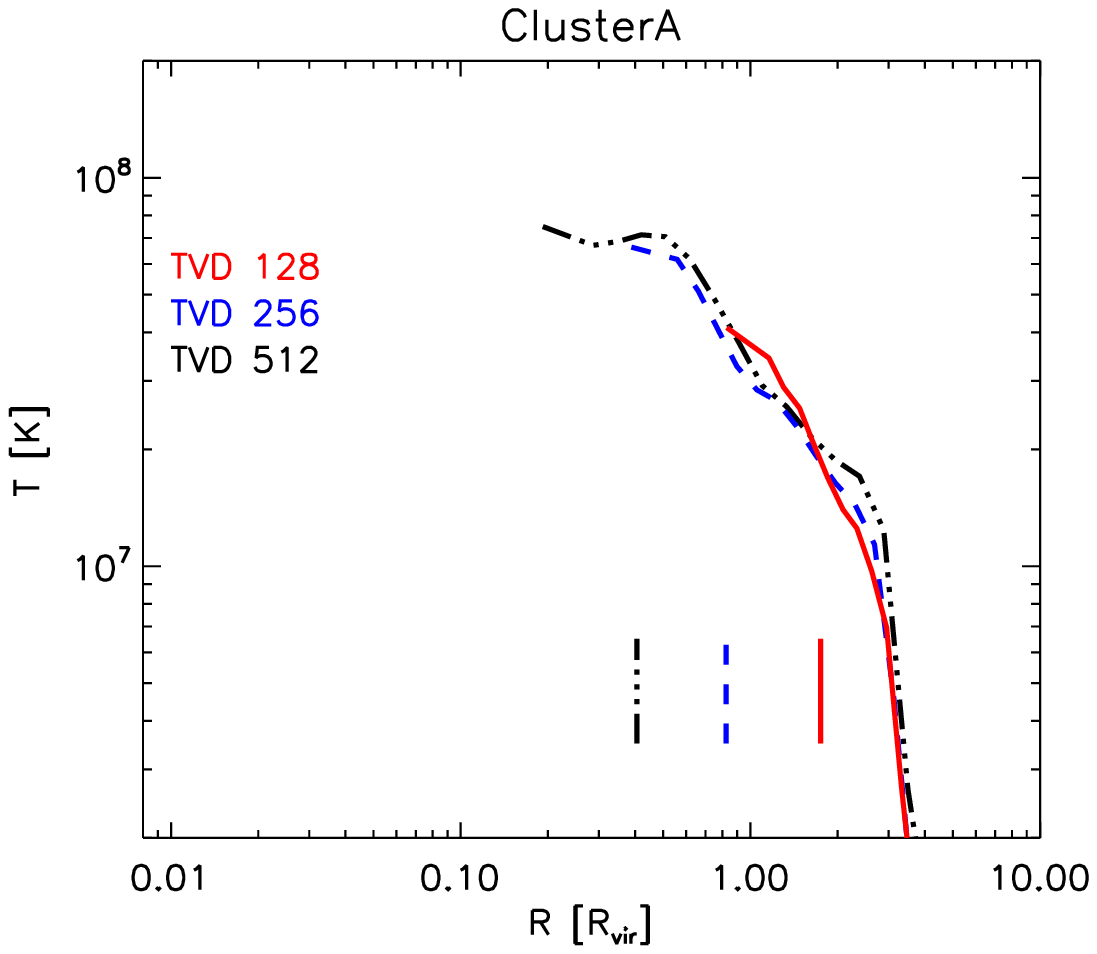}
\includegraphics[width=0.32\textwidth,height=0.25\textwidth]{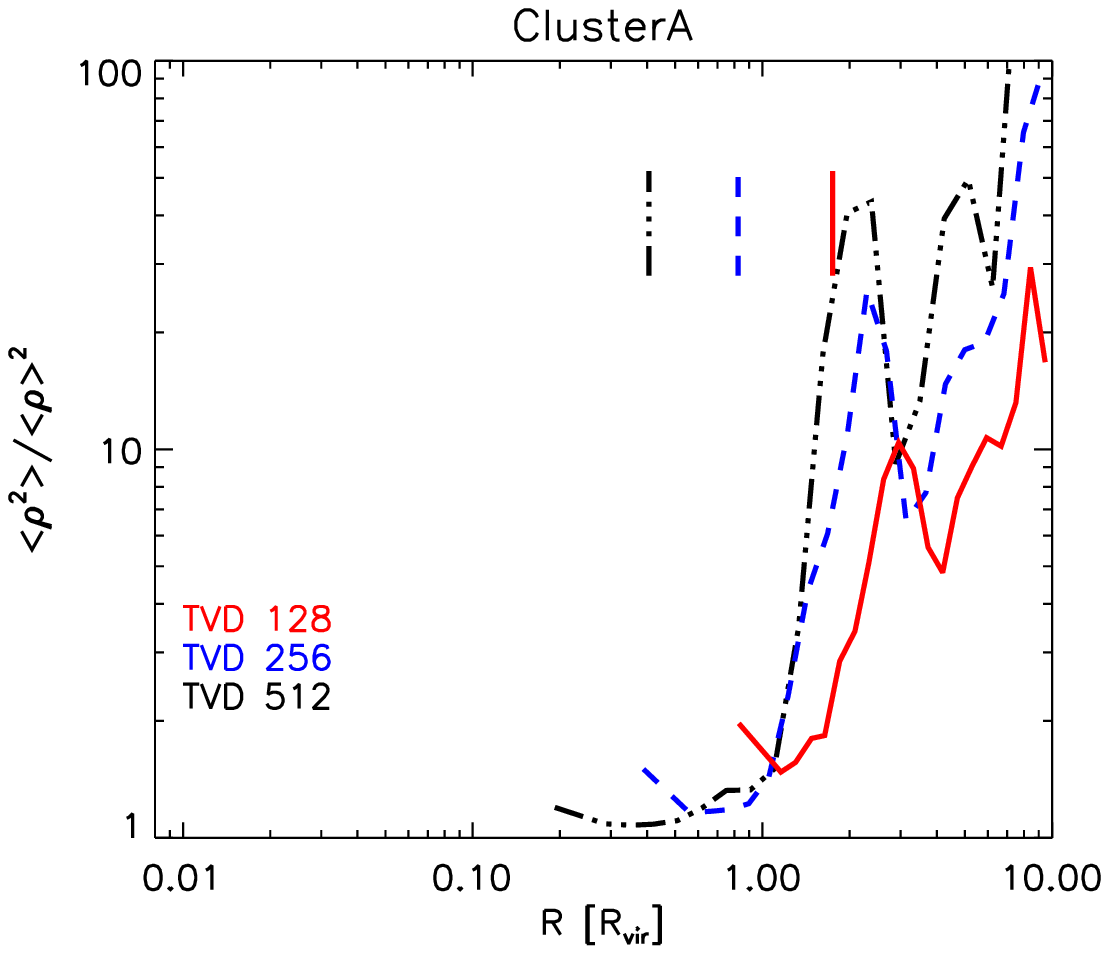}
\includegraphics[width=0.32\textwidth,height=0.25\textwidth]{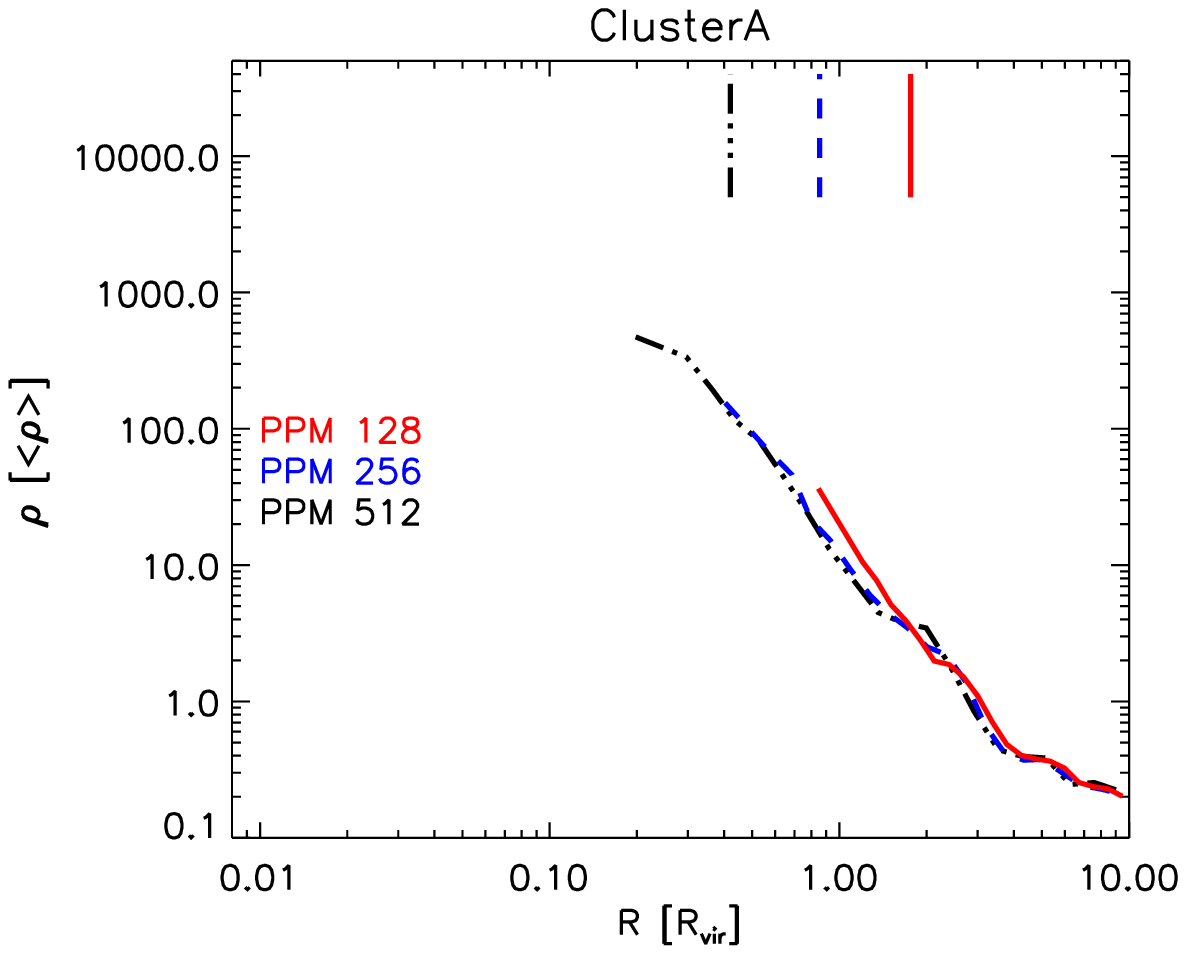}
\includegraphics[width=0.32\textwidth,height=0.25\textwidth]{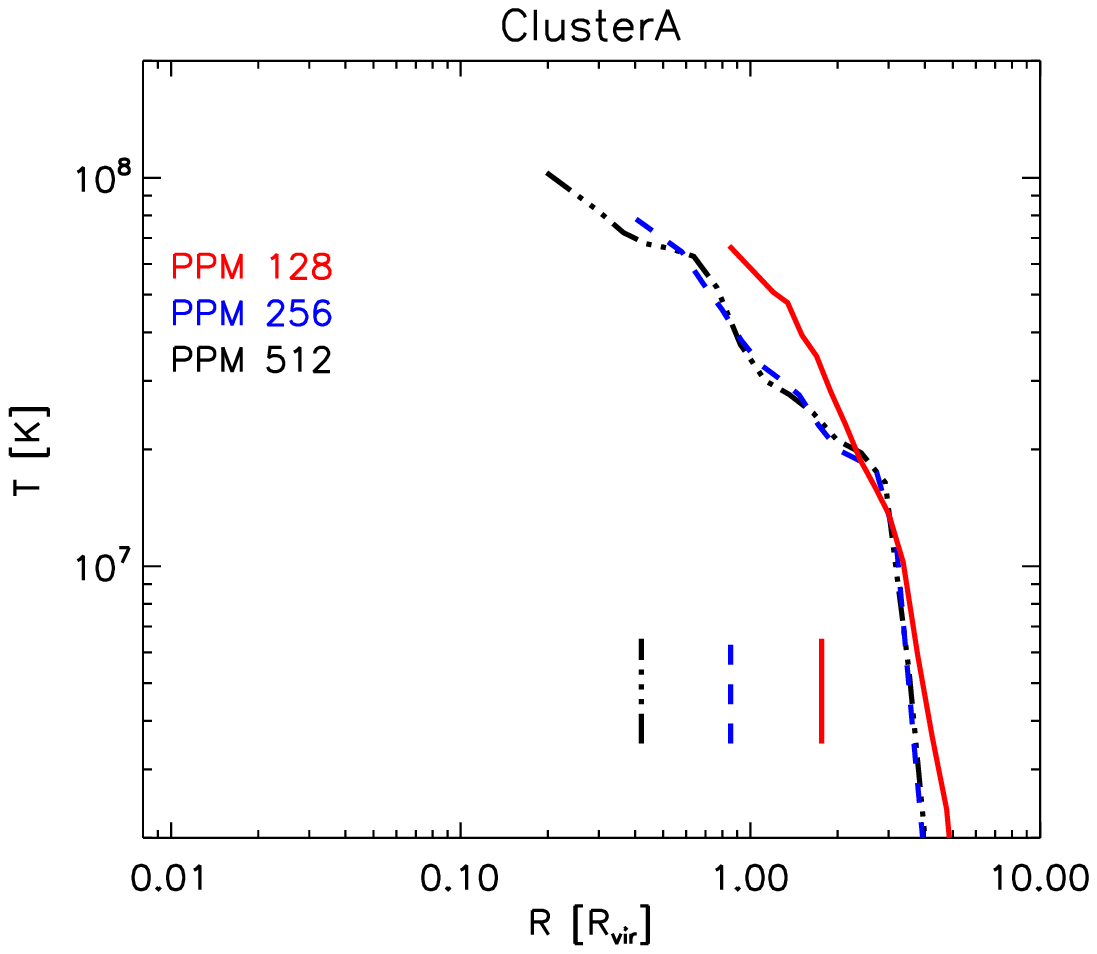}
\includegraphics[width=0.32\textwidth,height=0.25\textwidth]{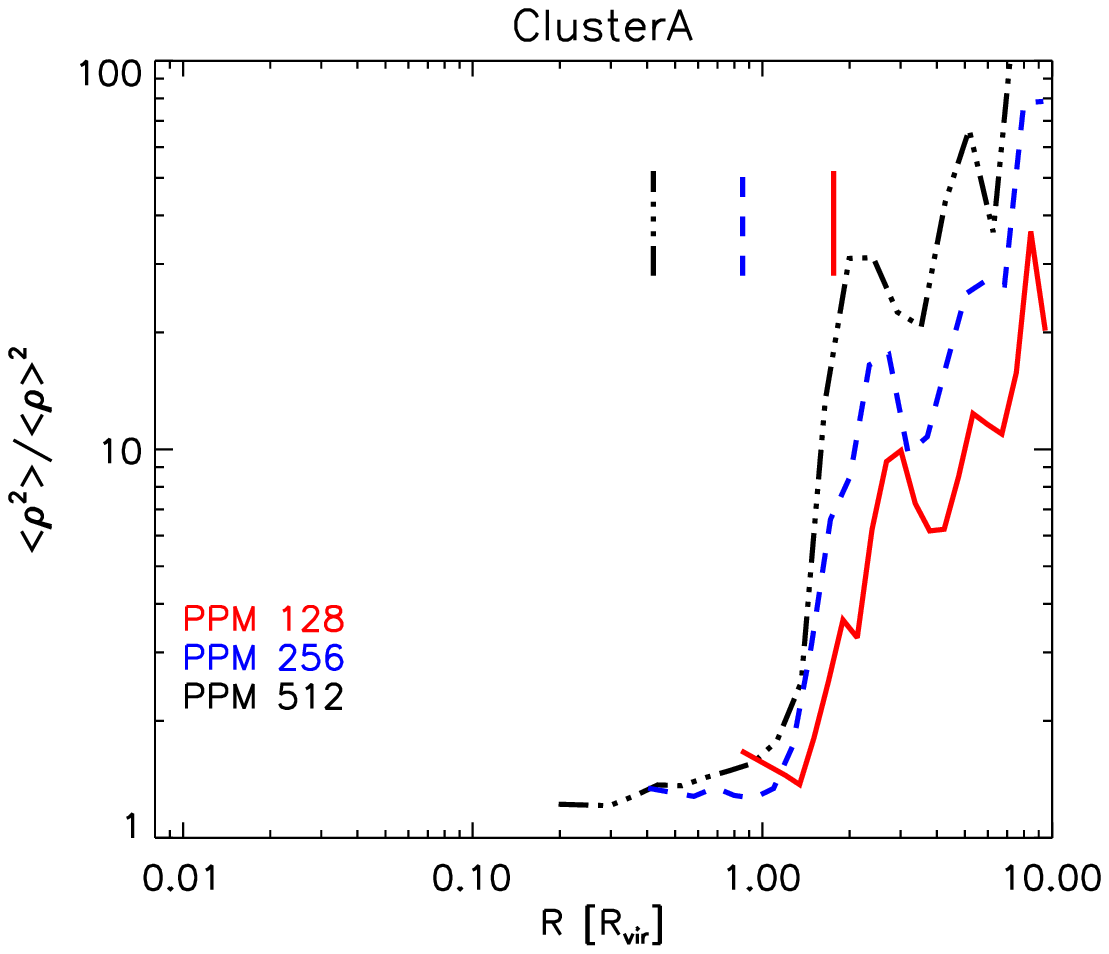}
\caption{Mass weighted profiles of gas density ({\it left column}), gas temperature ({\it center column}) and gas clumping factor ({\it right column}) for Cluster A at various
resolutions. GADGET runs are in the upper row, TVD runs are in the middle
and ENZO runs are in the bottom row. Vertical dashed lines show the minimum
radius enclosing the minimum mass suitable for convergence studies, as introduced in Sec.\ref{sec:dm}.}
\label{fig:prof_cpA}
\end{center}
\end{figure*}

\begin{figure*}
\begin{center}
\includegraphics[width=0.32\textwidth,height=0.25\textwidth]{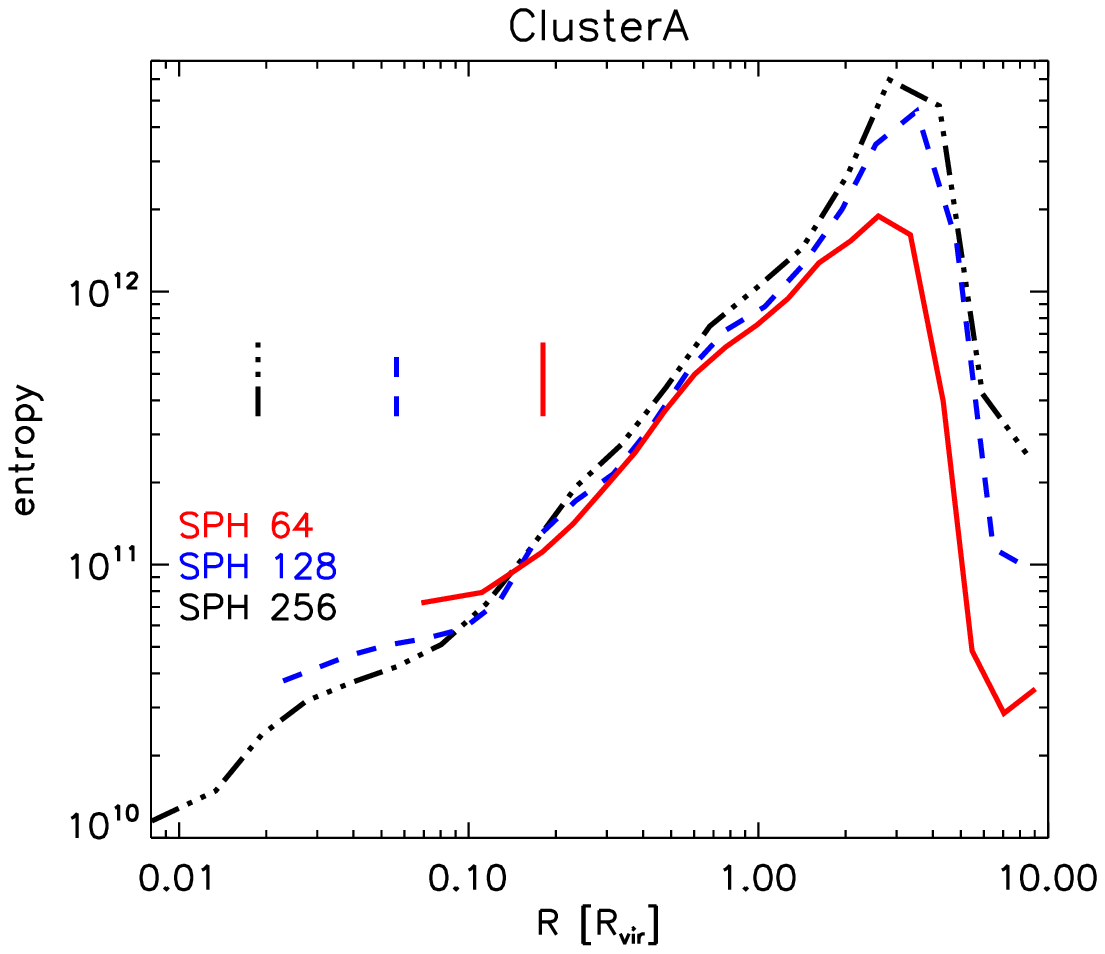}
\includegraphics[width=0.32\textwidth,height=0.25\textwidth]{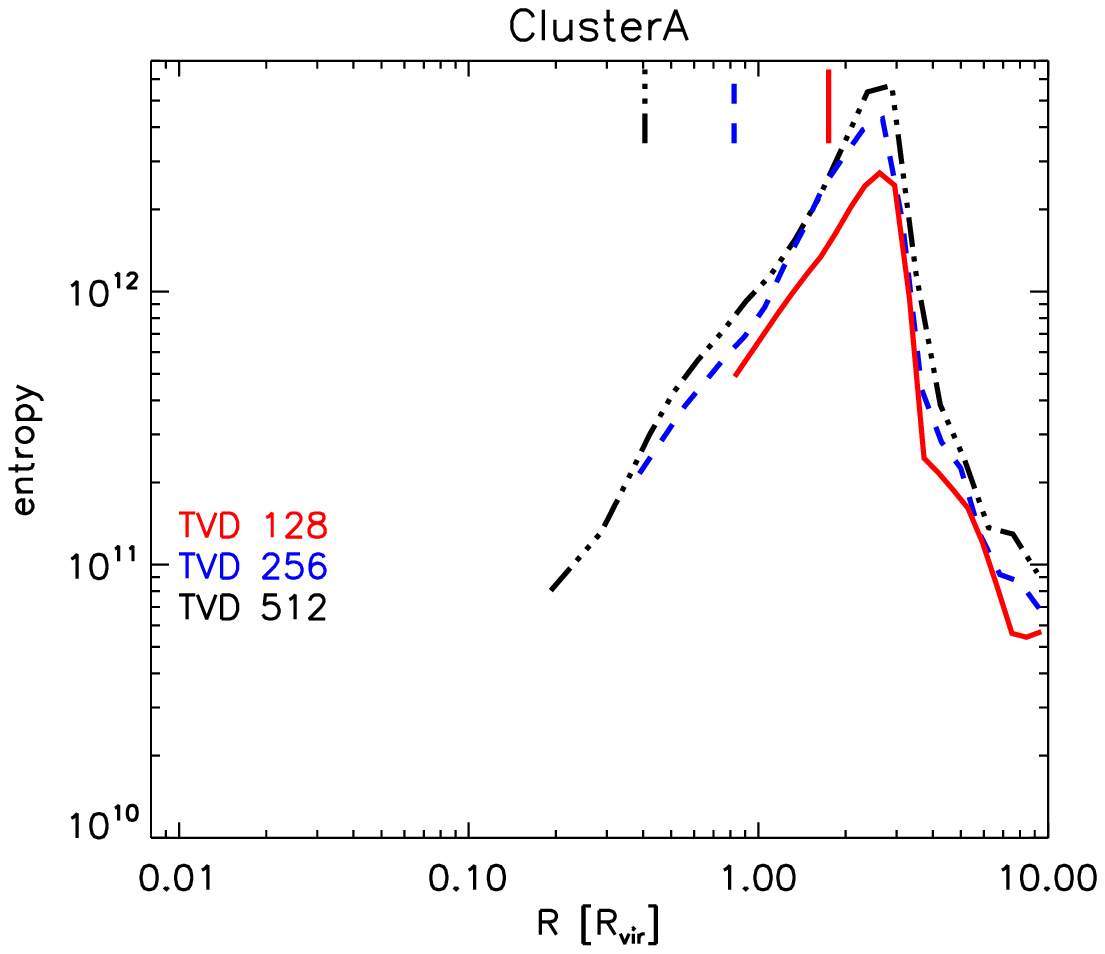}
\includegraphics[width=0.32\textwidth,height=0.25\textwidth]{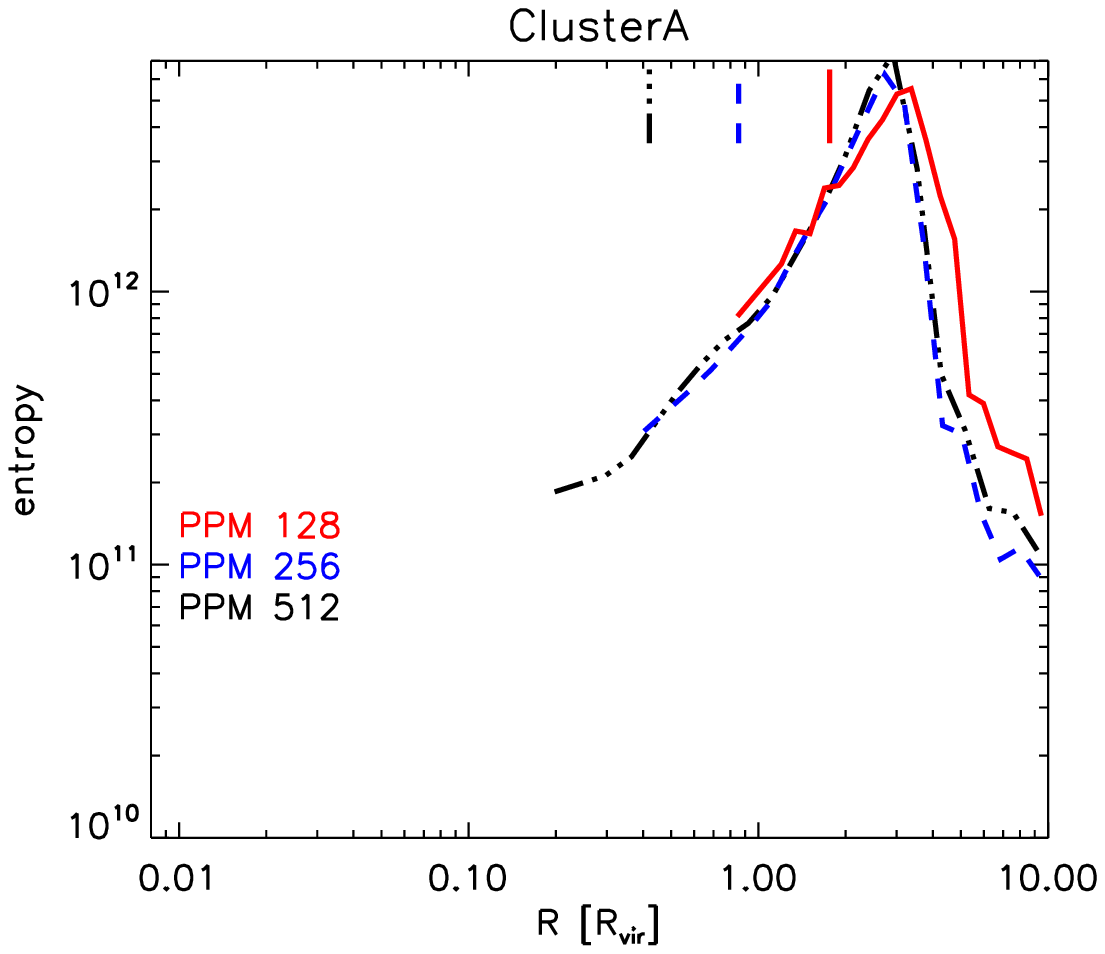}
\caption{Volume weighted profiles of gas entropy
(in arbitrary code units) for Cluster A at various
resolutions. The vertical dashed lines show the minimum
radius enclosing the minimum mass suitable for convergence studies, as introduced in Sec.\ref{sec:dm}.}
\label{fig:prof_entr}
\end{center}
\end{figure*}

\subsection{Properties of Galaxy Clusters}
\label{subsec:clusters_cp}

Differently from the case of gas density and gas temperature
distributions in the whole simulated volume, for which large
statistics is available, our setup does not allow us to study the convergence
with resolution of cluster statistics for a large number
of objects. 
Given the minimum requirement of mass and spatial resolutions outlined in the previous Sections, we must expect that only a few galaxy clusters in our $(100 \Mpc/h)^{3}$ box are sampled with 
enough particle/cells to allow the monitoring of thermodynamical distributions inside the virial radius, being free from resolution effects, namely the two most massive clusters within the 
sample:

\begin{itemize}

\item  cluster A: a system of total mass $M = 1.36 \cdot  10^{15} ~M_{\odot}/h$ and $R_{\rm vir} = 2.32 \Mpc/h$, in a fairly relaxed dynamical stage;

\item cluster B: a system of total mass $M = 1.64 \cdot 10^{15} ~M_{\odot}/h$ and $R_{\rm vir} = 2.47 \Mpc/h$, in an ongoing merger phase.

\end{itemize}

We preliminary checked that the total masses at all resolutions and in all codes are in agreement 
within a $\sim 6 $ per cent level within $R_{\rm vir}$, so that
the general parameters defining the systems are nearly identical
in all investigated resolutions.

However, we still have a minor source of scatter in the detailed
comparison of data, given by the fact that the different codes adopt different time stepping criteria, and even if the
cosmic time of the outputting of data is formally the same, 
tiny differences of the order of a few $\sim 10 ~Myr$ can be
expected in the data. This is expected to be problem only for the comparison of small scales in the cluster profiles, for which a perfect synchronization is impossible. 

This is issue is particularly relevant for the cluster merger B: at $z=0$ the exact positions of
the thermal features linked to the merger event are 
spread at different distances from the cluster
center, as an effect of tiny differences in the internal
timings of the codes. In the case of the relaxed cluster A, the spatial
locations of sub-clumps is much more similar in 
all codes. 

In Fig.~\ref{fig:prof_cpA} (upper panels) we show the mass-weighted profiles of
mass density, temperature and of the gas mass clumping factor, $\delta_{\rm \rho} \equiv <\rho^{2}>/<\rho>^{2}$ , for cluster A at $z=0$.

We define here the mass density profile as $\sum_{\rm i} m_{\rm i}/V_{\rm shell}$, where $m_{\rm i}$ is the mass associated to each particle/cell 
in the simulation, and $V_{\rm shell}$ is the volume of each radial
shell along the radius. The profile thus defined is
independent of the differences in the properties of clumping
within each shell, and allows us to investigate how the matter 
is distributed in the the different simulations. The computation
of the clumping factor then provides the complementary
information about the distribution function of gas matter within
each radial shell{\footnote {We notice that constructing the radial profiles
at large distance from the center of clusters can be affected by tessellation problems in the case of SPH runs, if the smoothing length of
the particles is large compared to the width
of the shell used to compute the
profile. The discreteness of grid cells (whose
edges may intersect more than a single radial shell) may be
regarded as a small source of uncertainty for the computation
of the radial profiles in the lower resolution grid runs. 
Correcting for these effect is non trivial, and complex tessellation 
techniques may be adopted in order to minimize the above effect. 
We notice however than the trends
reported in our work are generally much larger
than the uncertainties associated with these issues.}}.

In this case, the weighting by gas mass ensures that the profiles are closely related to the thermal energy
of the gas inside clusters, which in turn depends on the statistics of energetic and low 
Mach number internal shocks (see also Sect.~\ref{sec:shocks}).

The profiles of density and temperature converge with resolution
rather steadily, with an agreement better than a $\sim 20$
per cent between the profiles at all radii, when different
resolution are compared. This is reassuring, since the combination of the above profiles
gives the profiles of the thermal energy distribution within the clusters, and this is
a rather well converged finding in all codes.
On the other hand the profiles of the gas clumping factor shows a much slower convergence even within each code, 
with sizable evolution at all radii from
the cluster center. In all runs the clumping factor increases with radius,
and reaches $<\delta_{\rm \rho}> \sim 10$ outside of $R_{\rm vir}$. At the best available resolutions
the self-convergence between for each code is yet to be reached, despite the
fact that the profile of gas matter density is much better behaved. 

In Figure~ \ref{fig:prof_entr} we report the volume weighted profiles of the entropic function
($S=T/\rho^{2/3}$) for each particle/cell, for all codes and resolution.
The weighting by the volume here is chosen to focus more on the
trend of the entropy associated with the smooth, volume filling accretion
around clusters.
Compared to the more standard "entropy" profile, based on the ratio between temperature and 
$\rho^{2/3}$ profiles, we consider the profile of the entropic function more useful to
characterize the tiny differences of entropy which could be very locally associated with different dynamical accretion pattern in the different codes.

\begin{figure}
\begin{center}
\includegraphics[width=0.235\textwidth,height=0.23\textwidth]{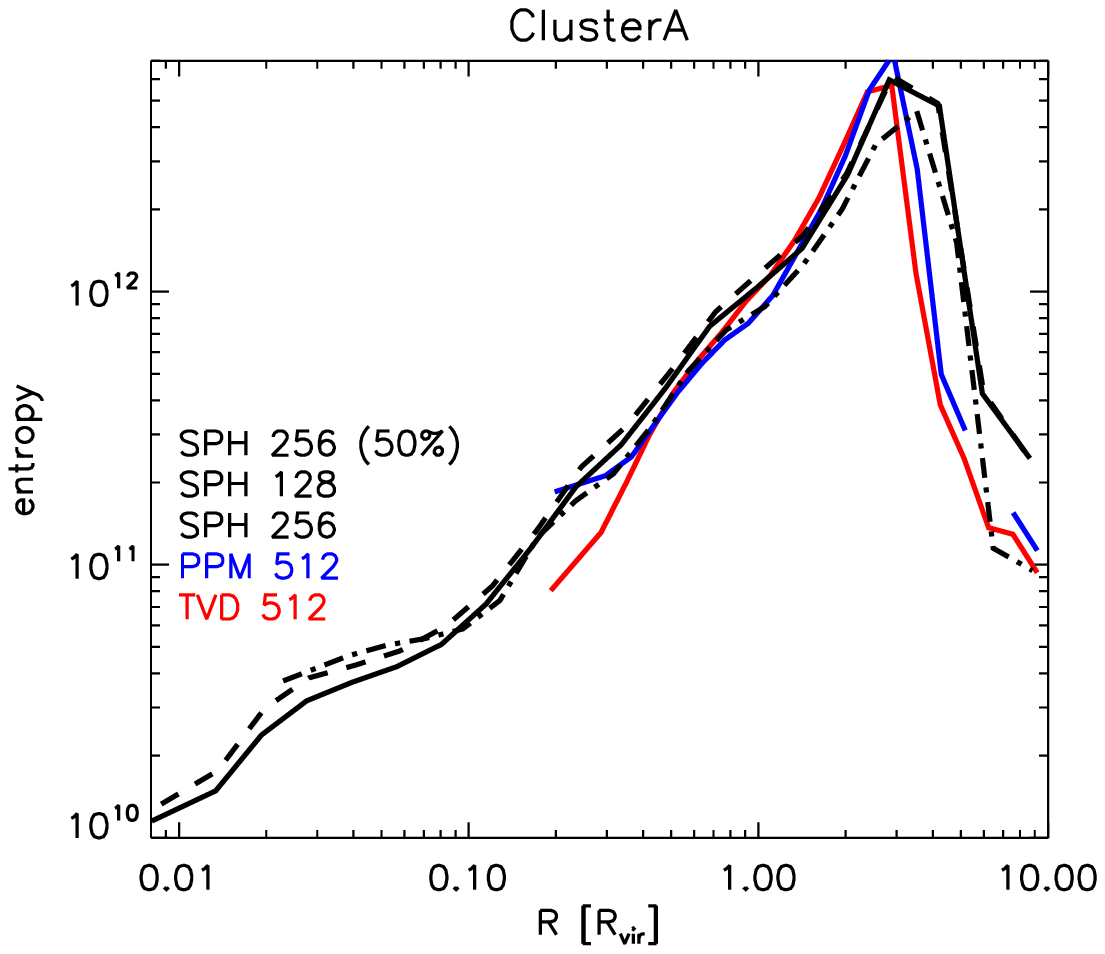}
\includegraphics[width=0.235\textwidth,height=0.23\textwidth]{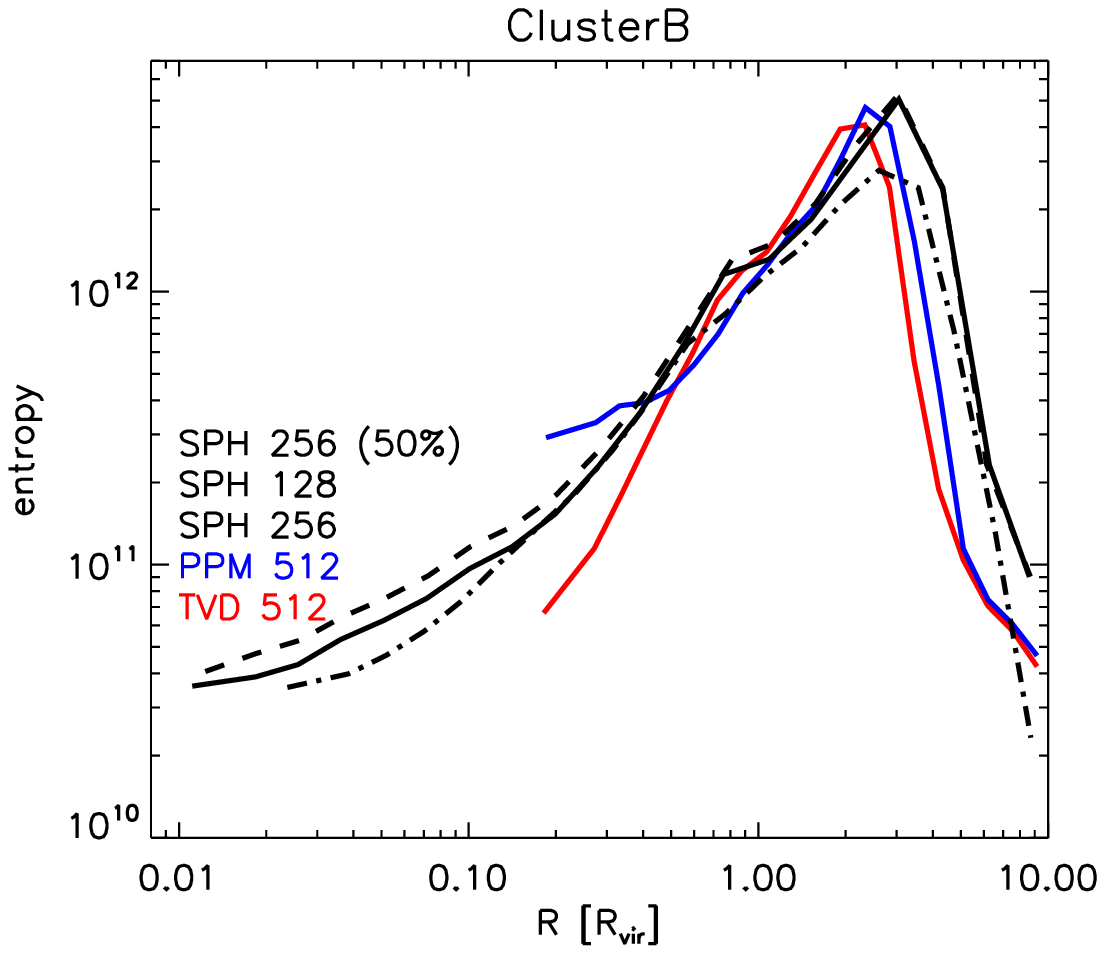}
\caption{Cross comparison of the volume-weighted gas entropy profiles (in arbitrary code units) for cluster A ({\it left column}) and for cluster B ({\it right column}). GADGET3 runs at 128 are reported in dot-dash, while
the 256 runs are in solid; the long dashed lines report the profiles for GADGET3 runs at 256, but
considering only the 50 per cent less dense particles.}
\label{fig:prof_cpA_cfr}
\end{center}
\end{figure}

The study of mass-weighted entropy distributions in a smaller re-simulation of this
project will be discussed in Sect.~\ref{subsec:tracers}, where we investigated the entropy generation associated with the clumps
of matter in clusters.
In this case, GADGET3 runs are those characterized by the slowest
resolution compared to grid methods, and they also present a peak of the entropy gradient
at a significant larger distance compared to TVD and ENZO runs. We also report the interesting trend
that, compared to grid methods, the increase in spatial resolution causes
a significant smoothing of the entropy jump in GADGET3 clusters (see Fig.~\ref{fig:prof_cpA_cfr}).
At the best available resolution, the full-width-half-maximum of the entropy "jump" in 
grid methods is significantly smaller than in GADGET3 ($\sim 2 R_{\rm vir}$ in TVD and ENZO versus
$\sim 3 R_{\rm vir}$ in GADGET3).  
To check if differences in the clumping of gas matter is responsible for the
above differences, 
we also computed the profiles for the 256 GADGET3 run by considering
only the 50 per cent less dense particles (Fig.~\ref{fig:prof_cpA_cfr}), but no significant differences can be found.
This difference in GADGET3 can only be partially explained by SPH smoothing effects, since the observed broadening is considerably larger than the smoothing length at these over densities. 
The dynamics of shock waves on large scale accretion pattern around clusters are however expected to play the major role here: we will further explore this issue in 
Sect.~\ref{sec:shocks}. 

\bigskip

A second interesting feature of entropy profiles is the hint of a flattening of the entropy profile at $\approx 0.3 R_{\rm vir}$ in clusters
simulated with ENZO compared to GADGET3 runs. This is
in line with a number of existing results in the literature 
(e.g. Frenk et al. 1999; Wadsley et al.  2008; Tasker et al. 2008; Mitchell et al. 2009),
even if the grid resolution here is too coarse to show conclusive evidence.
However, tests employing efficient adaptive mesh refinement with ENZO have recently shown that that the extreme flatness of the entropy profile in these cluster runs inside $0.1 R_{\rm vir}$ is 
a very robust feature against numerical and mass/spatial resolution
effects (Vazza 2011).

Based on the literature, it seems likely that the differences in the inner
entropy profiles are produced by the different integrated mixing role played by artificial viscosity which is enhanced in grid codes compared to SPH (e.g. Wadsley et al.  2008; Mitchell et al.  2009). 
With our setup we tested in detail the way in
which entropy is advected inside clusters in ENZO and GADGET with a further re-simulation, discussed in the Section below (Sec.\ref{subsec:tracers}.

\bigskip

On the other hand, it is very likely that the leading mechanism which sets the shape of the entropy distribution beyond $R_{\rm vir}$
is the action of shock waves. Gravitationally 
induced motions of gas matter are the leading drivers of shock waves in these simulations,
and therefore a detailed analysis of the
distribution of gas matter in the outer shells
of simulated clusters is helpful to
understand the reported differences.
While we defer to Sect.~\ref{sec:shocks} a detailed
study of the morphologies and statistics
of accretion shock around clusters, while here
we study in detail the simple 
gas matter distribution in the outer cluster regions, comparing different codes and resolutions. 

\begin{figure*}
\begin{center}
\includegraphics[width=0.32\textwidth,height=0.25\textwidth]{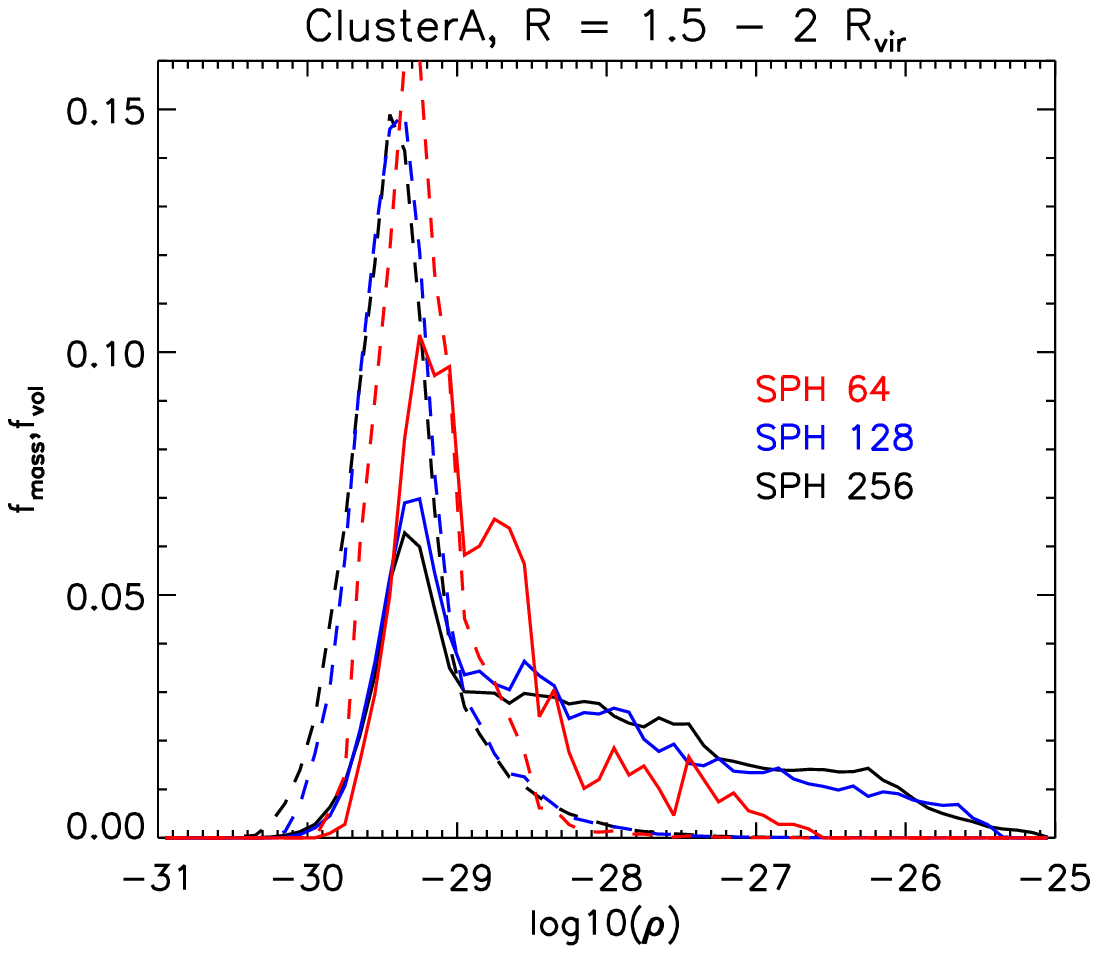}
\includegraphics[width=0.32\textwidth,height=0.25\textwidth]{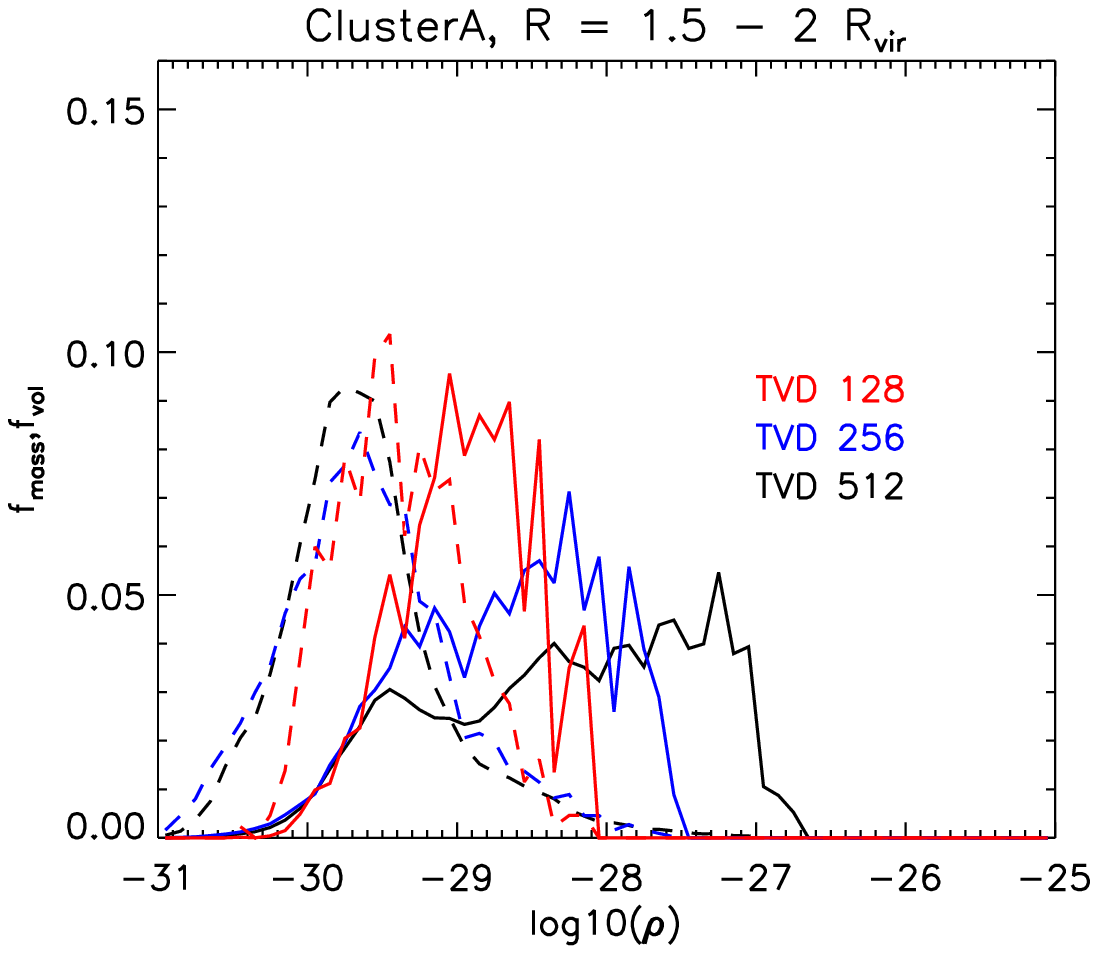}
\includegraphics[width=0.32\textwidth,height=0.25\textwidth]{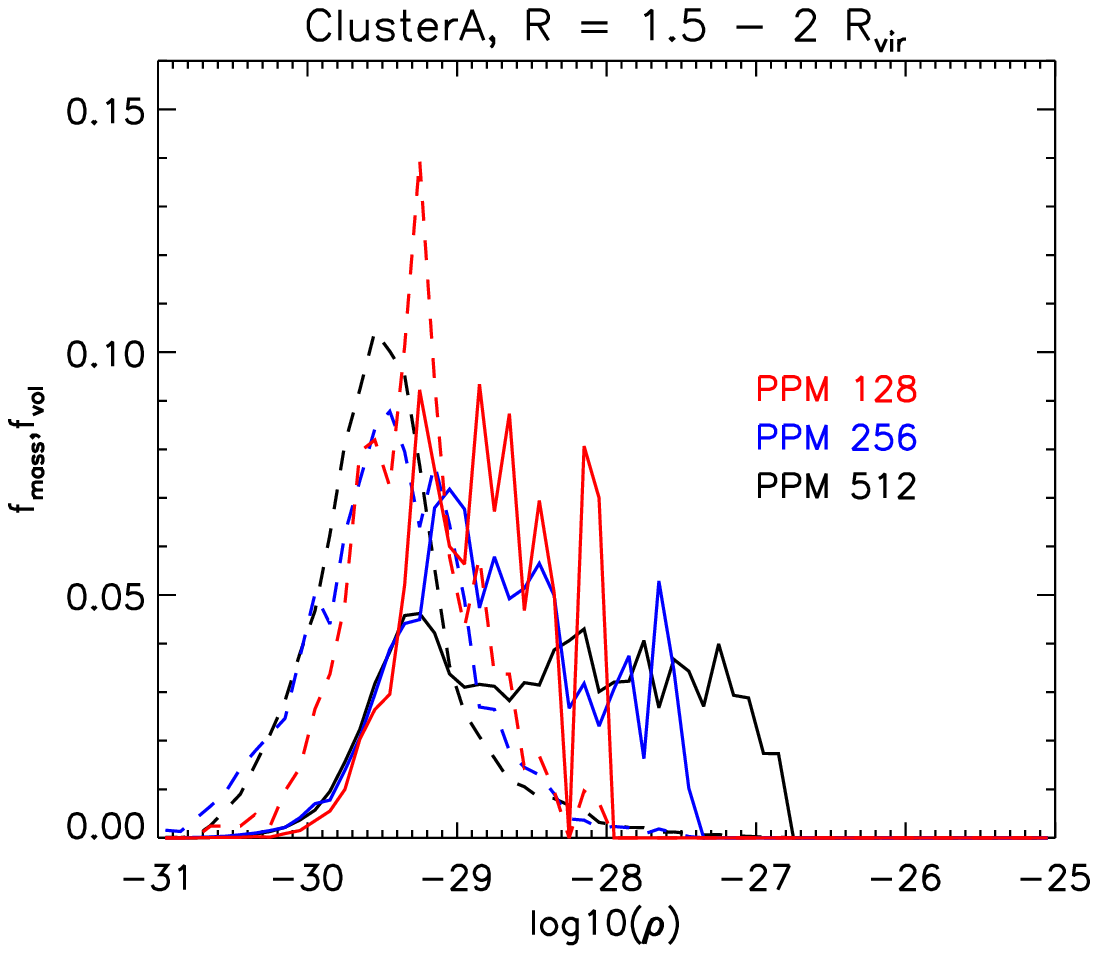}
\caption{Mass-weighted (solid lines) and volume-weighted 
(dashed) distributions of gas density within the shell
$1.5 \leq r/R_{\rm vir} \leq 2$ around cluster A, for
all simulated runs. }
\label{fig:distr_clump1}
\end{center}
\end{figure*}

The panels in Fig.~\ref{fig:distr_clump1} 
present the mass-weighted and volume-weighted
distribution of gas matter within the radial
shell $1.5 R_{\rm vir} \leq r \leq 2 R_{\rm vir}$ outside of cluster A, for the same 
runs of previous Figures. 

As expected, the volume weighted distributions show that the grid codes are able to
resolve more structures (e.g. smooth filaments of gas) in the low density regions; on the other hand it can be seen in 
the mass  weighted distributions that GADGET3 resolves much more collapsed objects, which are
absent in the grid codes. This corresponds
to the larger number of gas clumps that
can be visually seen in the projected
maps of Fig.~\ref{fig:maps_zoom}.
However, such material
in grid codes produces also an excess of baryon gas in the range $10^{-29} ~g/cm^{3} \leq \rho \leq 
10^{-27} ~g/cm^{3}$, compared to GADGET3. These two excesses in grid codes and in SPH produce signals of a similar order, which explains why 
the average clumping factors reported in 
Fig.~\ref{fig:prof_cpA} are quite similar, despite the fact that
the differential distribution have rather different shapes. 
The differential distribution of gas matter in the outer
shells provide a preliminary suggestion that the shock waves
associated with these accretions can be 
significantly different in the two methods.
Indeed, a larger contribution from stronger shocks (driven by "smooth", rather than "clumpy" accretions) should be expected
on average in grid codes, at the same radius. The
larger entropy jumps associated with these strong shocks around
smooth accretions may then well explain the
differences of shape in the outer entropy profiles of cluster A and cluster B. 
For recent works employing higher spatial and mass resolution to characterize in detail the clumping and azimuthal scatter properties of gas matter in the outer region of galaxy clusters, we address the reader to Roncarelli et al.  (2006), Burns  et al.  (2010), Vazza et al.  (2011) and Nagai et al.  (2011).  The issue of matter clumping in the 
outskirts of galaxy clusters has also recently become a topic available
to X-ray observations (e.g. Simionescu et al.  2011; Urban et al.  2011), 
and therefore the predictions of different numerical methods, even
at $\sim R_{\rm vir}$, are going 
to be likely tested with observations in the next future.

\bigskip

%\begin{figure}
%\begin{center}
%\includegraphics[width=0.45\textwidth,height=0.25\textwidth]%{imag_comp/clump_dist_comp.ps}
%\caption{Cross-comparison of the mass-weighted %(solid lines) and volume-weighted 
%(dashed) distributions of gas density within %the shell
%$1.5 \leq r/R_{\rm vir} \leq 2$ around cluster %A.} 
%\label{fig:distr_clump2}
%\end{center}
%\end{figure}

\begin{figure*}
\begin{center}
\includegraphics[width=0.95\textwidth]{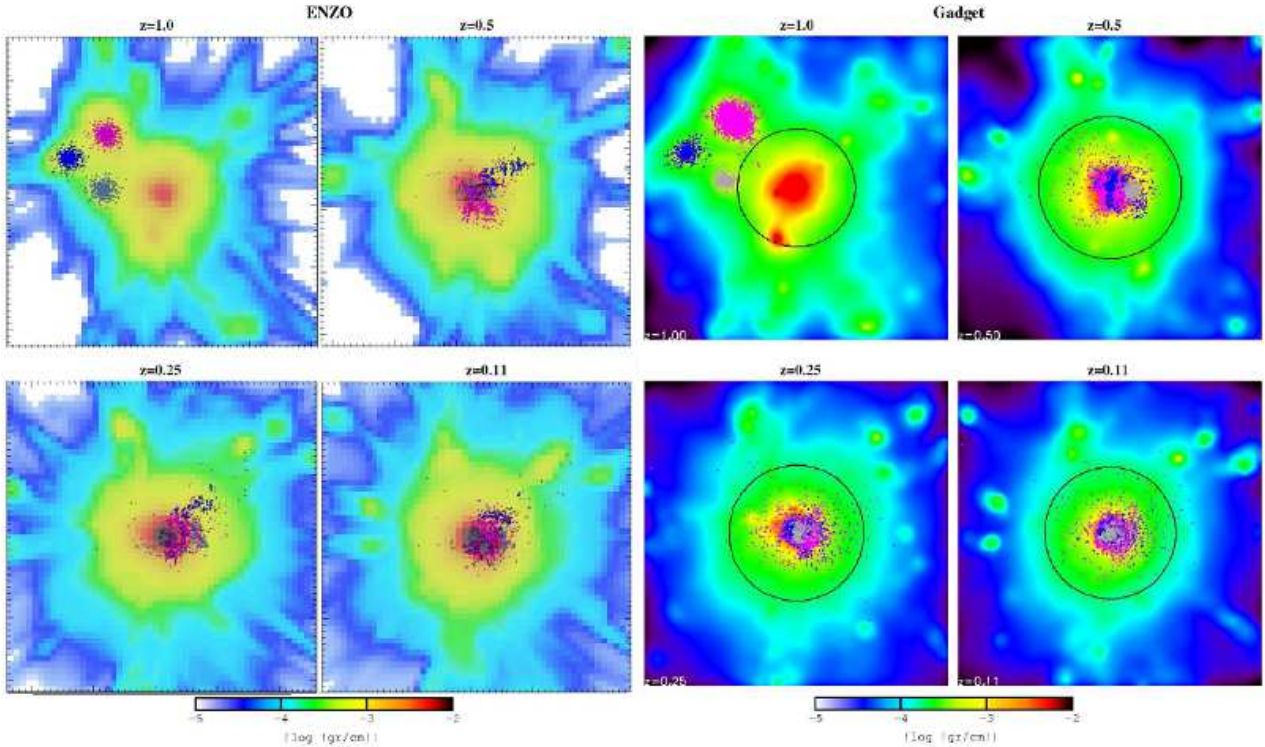}
\caption{{\it First two columns:} projected map of gas density (colors) and tracers positions for 4 time steps in ENZO 256 run, for the test simulation described in Sec.\ref{subsec:tracers}. {\it Last two columns:} same as in the first 4 panels, but for GADGET3 256 run. The side of the images and the line of sight are comoving $12\Mpc/h$ in all cases.\label{fig:trac_maps}}
\end{center}
\end{figure*}

\begin{figure*}
\begin{center}
\includegraphics[width=0.45\textwidth]{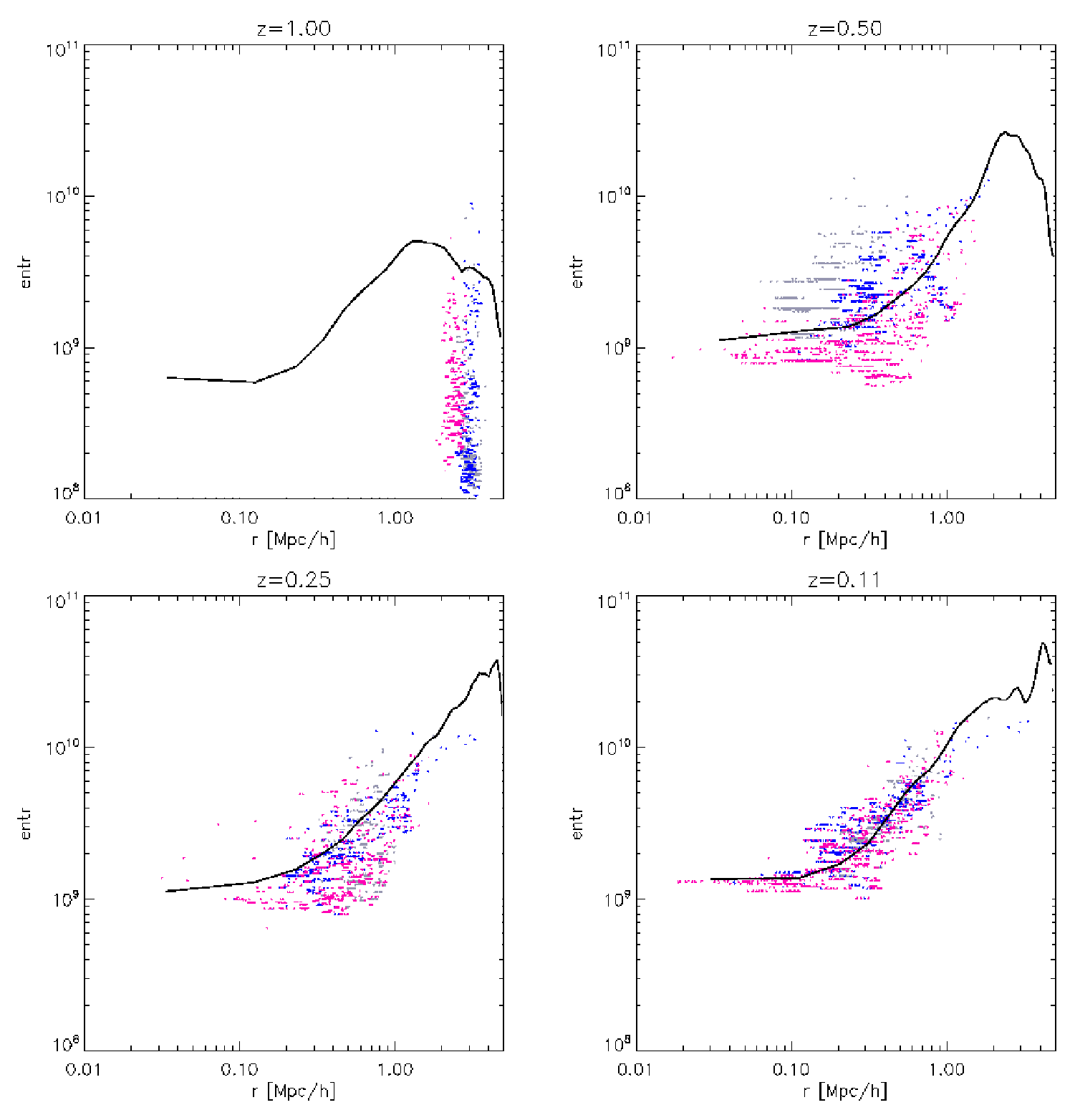}
\includegraphics[width=0.45\textwidth]{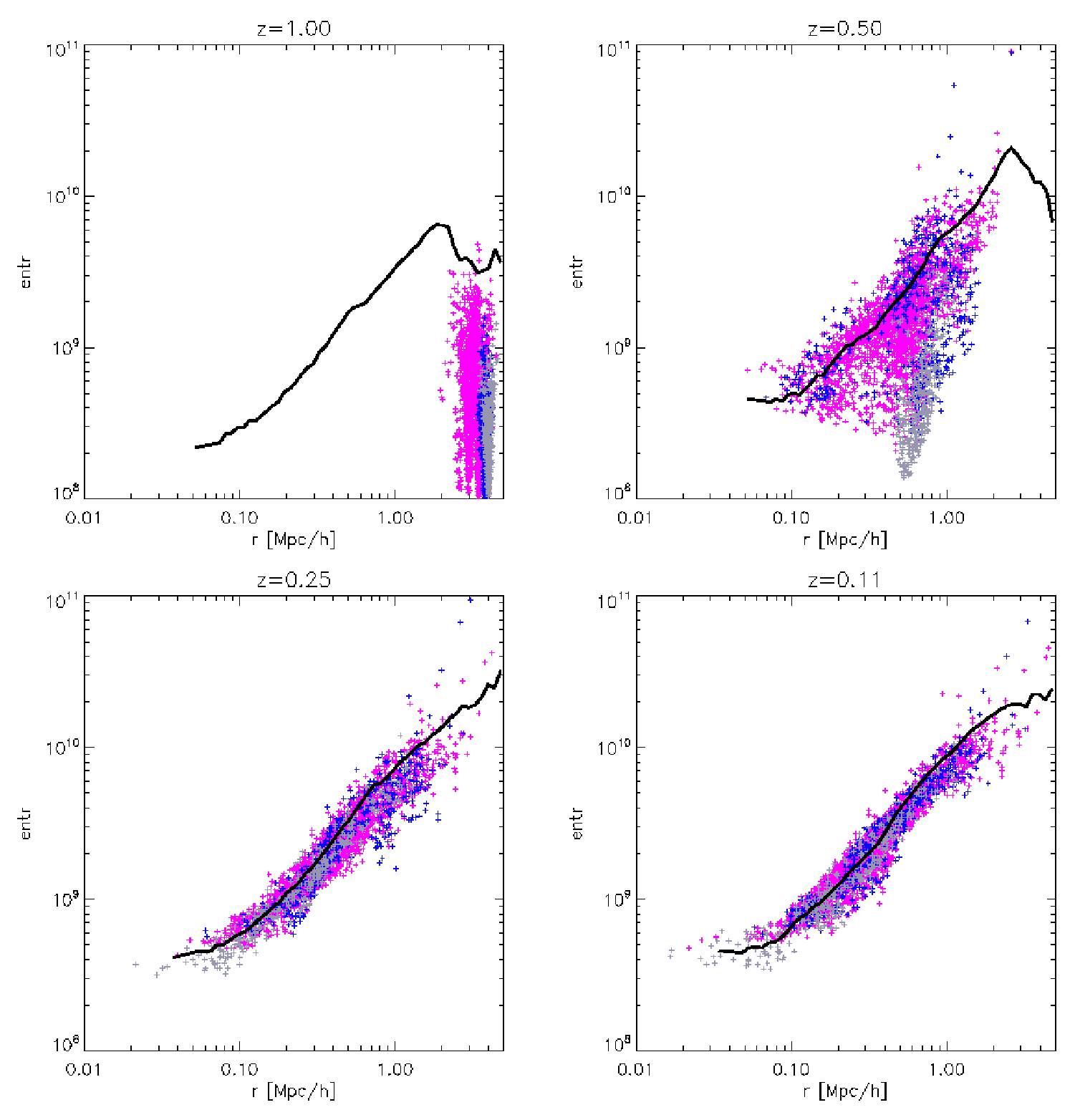}
\caption{Radial profiles of the physical gas entropy (in arbitrary code units) corresponding to all panels in Fig.~\ref{fig:trac_maps}. The solid lines show the average mass weighted entropy profiles for the complete GADGET3/particles (left 4 panels) and ENZO/cells distributions (right 4 panels), while the overlaid colours show the contribute from particles/tracers initially located within the 3 selected sub-halos.\label{fig:trac_entr}}
\end{center}
\end{figure*}

\subsection{A Test with Tracers}
\label{subsec:tracers}

In order to analyze in a more conclusive way the
differences in the entropy profiles of cluster, we performed 
a re-simulation study which followed in detail
how the entropy of gas is build inside one massive
clusters during its evolution. 

To this goal we simulated
a smaller volume of side $40\Mpc/h$, 
whose initial conditions
were produced in a similar way as in Sec.\ref{sec:sim}; 
in this case an even larger normalization for the 
matter power spectrum parameter were used, $\sigma_{8}=1.6$,
in order to form a $M \sim 10^{15}M_{\odot}$ cluster inside this small volume{\footnote{The initial conditions for the $40\Mpc/h$ box, at
different resolutions, can be found at this URL:  http://canopus.cnu.ac.kr/shocks/case1/.}}.

Since the entropy profiles of grid codes were found to
be very similar, for simplicity we tested here only the 
and ENZO run with $256^{3}$ cells (corresponding to 
a spatial resolution of $156 \kpc/h$ with a GADGET3 run with $256^{3}$ 
gas particles.

We are interested in the evolution of gas entropy linked to
the matter accretion history of the cluster, and
 we identified
all gas sub-halos in place at $z=1$ outside of the
main cluster in the volume, and  
we followed their evolution in time.
The location of their 
centers (based on a spherical over-density halo finder) is
in agreement in both simulations within a $200 \kpc/h$ accuracy. 
We selected all particles
belonging to the 3 sub-halos in GADGET3 runs, while (mass-less) tracer particles were placed inside the corresponding
cells in ENZO run. 
The distribution of tracers was generated using an number density profile corresponding
to a King profile, using a sampling of $\sim 0.1$ of the
cell size. We checked that the final tracers distributions are statistical independent of the particular profile adopted for the initial generation (see also Vazza, Gheller
\& Brunetti 2010).

The gas tracers in ENZO were then evolved
by updating their positions according to the underlying
Eulerian velocity field, with the same procedure of Vazza, Gheller \& Brunetti (2010). In summary, the three-dimensional velocity field was interpolated at the location of tracers
using a Cloud In Cell kernel, and the positions were 
updated every 2 time steps of the simulation with a first-order
integration. The entropy assigned to the tracers at 
each time step corresponds to the entropy of 
the cells where each tracer sits at the time of observation.

The visual inspection of projected tracers/SPH particles positions 
as a function of redshifts (Fig.~\ref{fig:trac_maps}) clearly shows that the accretion of gas clumps is a different process in the two runs.

Even if the initial positions of the clumps centres are equal down to the cell
resolution, soon after their accretion through $R_{\rm vir}$ their trajectories
differ considerably: the particles from sub-halos in GADGET3 soon mix with
the main cluster atmosphere after accretion, and most of the particles
from sub-halos end up in the dense and low-entropy cluster core.  In ENZO the
tracers mix more slowly at the beginning, and most of accreted gas component is
bound to the infalling clumps even after the crossing $R_{\rm vir}$.  In
particular, most of tracers initially located in two clumps (colored in blue and
in gray) never penetrate inside the core of the main cluster, but find their selves settling at
larger cluster radii, $\sim 0.2-0.3 R_{\rm vir}$.

The analysis of the entropy profiles of the main cluster 
and of SPH particles/tracers is presented
in Fig.~\ref{fig:trac_entr}, and confirm the difference
in the accretion history of the two methods. In this case, since we are interested in the
evolution of gas clumps, the weighting by gas density of the entropic function is adopted here. 

In GADGET3, only a fraction of the matter from clumps is 
shock heated to higher entropy, and the un-shocked low entropy material can be delivered to the low entropy center of 
the main cluster, where it remains until the end
of the simulation. Already at $z=0.25$ ($\sim 5$ Gyr after 
their accretion inside $R_{\rm vir}$) the entropy of SPH particles from sub-halos is
nearly identical to the entropy of the main cluster.
On the other hand in ENZO run the gas from clumps
is soon shock heated to higher entropy values (compared to
particles in sub-halos in GADGET3), and it
retains its entropy for a larger
time, placing on average on radii external to the 
cluster core. In the ENZO run, there is still a relevant
scatter in the entropy of tracers at $z=0.1$, compared to the
main profile of the cluster, which is very different
from GADGET3 results.

Our results suggest that the following different
mechanisms are at work in the two methods: 
a) in SPH, accreted clumps soon loose their gas because of the interaction with the ICM of the main
cluster, the entropy of their gas gets quickly in 
an equilibrium with the atmosphere of the host cluster and
many particle from the sub-halos can end up within the low
entropy core of the main cluster; 
b) in PPM, accreted clumps are efficiently shock heated while entering the atmosphere of the main 
cluster, they reach more slowly an equilibrium with the average
entropy of the main cluster atmosphere and most of the
accreted material sets to an higher
adiabat in the cluster profile (compared to the SPH run), avoid
to concentrate within the cluster core.

In both cases, we observe that the shock heating and mixing 
motions following the matter accretions 
from small satellites (i.e. minor mergers) are not
efficient processes in changing 
the overall shape of the entropy profile within the
main cluster, which is already in place at $z \sim 1$. 

On the other hand, we can speculate that the different trajectories and thermodynamical evolution of the gas matter accreted by
sub-clumps in the two methods highlights the sizable differences
of transport phenomena in the two schemes, which are relevant
to many astrophysical topics in galaxy clusters (e.g. metal
enrichment, cosmic ray transport, non-thermal emissions).

Since we do not make use of adaptive mesh refinement in ENZO simulations here, 
the spatial resolution is too poor to study fluid instabilities
and cluster turbulence (for studies of tracers in high resolution ENZO runs with adaptive
mesh refinement, see Vazza, Gheller \& Brunetti 2010 and Vazza 2011). 
However we notice that at this point it is clear that the flatness 
of inner cluster entropy profile generally found in
PPM codes is not a product of employing AMR itself, but it is a more
fundamental feature linked to shocks and mixing
inside clusters.

In their seminal work Mitchell et al. (2009)
investigated the production of cluster entropy in
a binary clusters merger with GADGET and the PPM
code FLASH (Fryxell et al.  1998), and found that
the most important factor which produces the differences
seen in the two numerical methods is the early
mixing of entropy during the collision of cluster
cores, driven by fluid instabilities, which is 
much more pronounced in PPM than in SPH. 
Our test here shows that the way in which fluid instabilities
 and shocks follow the accretion
of smaller subunits of cluster also differ in the
two approaches, and lead to dissimilar entropy tracks
for the accreted gas.

\begin{figure*}
\begin{center}
\includegraphics[width=0.99\textwidth]{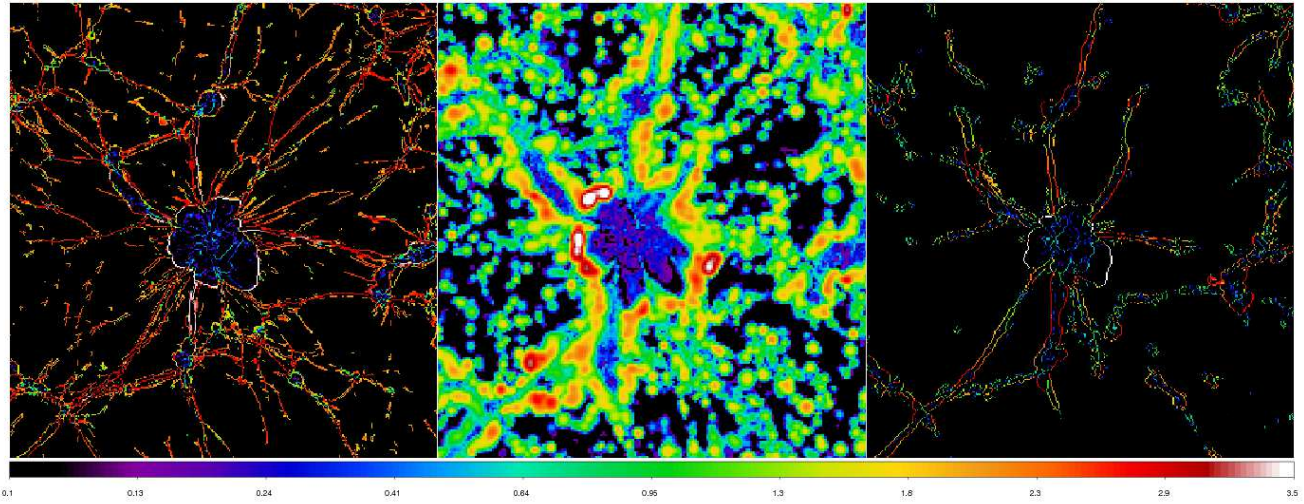}
\caption{Projected maps of shocks (in $log M$) for a slice of $75 \Mpc/h$ in the simulated volume
for the most resolved runs of the sample (left:ENZO at $512^{3}$; center: GADGET at $256^{3}$; right: TVD
at $512^{3}$). We adopt a weighting by volume for each particle/cells, and a fixed width of $\approx 550 \kpc$ along the line of sight in all maps.}
\label{fig:maps_shocks}
\end{center}
\end{figure*}

\begin{figure*}
\begin{center}
\includegraphics[width=0.95\textwidth]{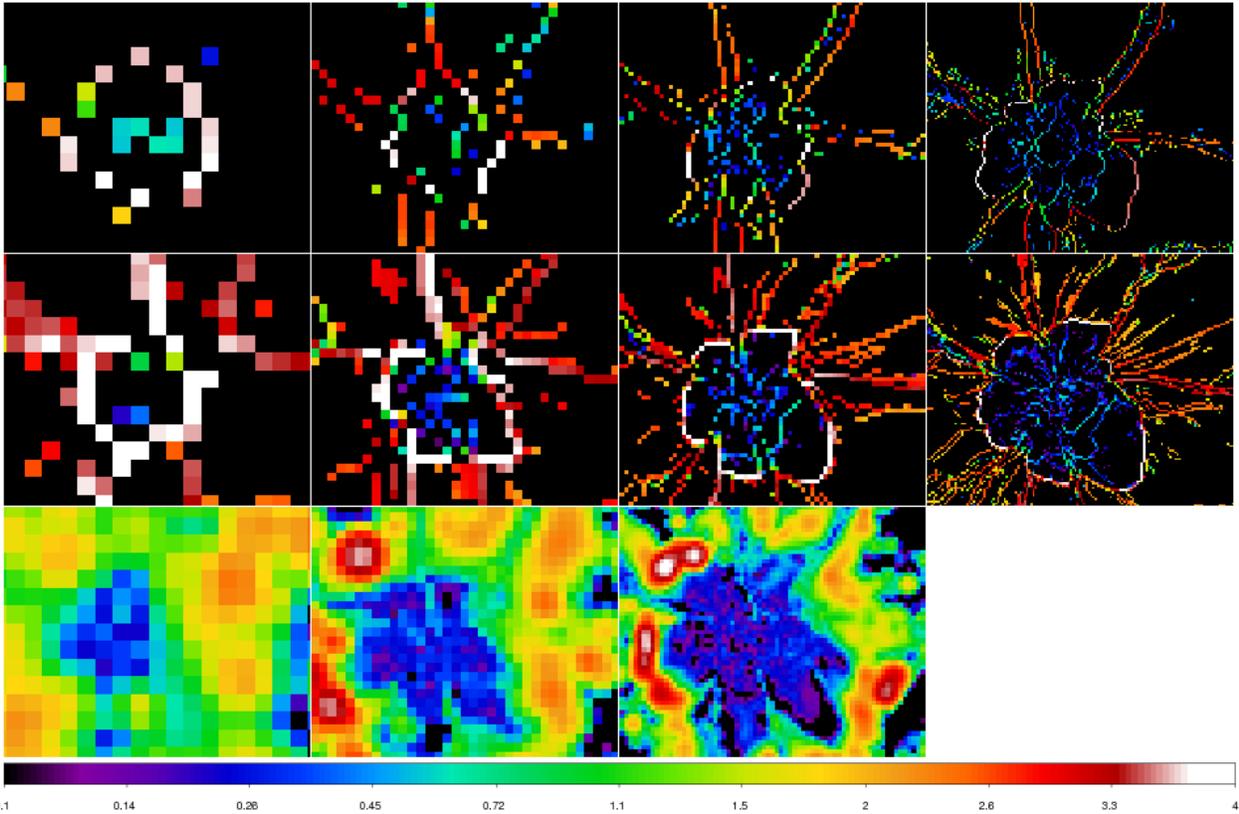}
\caption{Map of shocks (in $log M$) for a slice with the side
of 25 \Mpc/h through the center of cluster B.
The first row reports the results of the TVD runs and the Temperature Jump shock finder as a function of resolution, the second row reports the results for the PPM runs and the Velocity Jumps shock finder, the third row reports the results fro the SPH runs and the Entropy Jump shock finder. From the left to 
right column, the width along the 
line of sight is $2200$, $1100$, $550$ and $275\kpc$ respectively.}
\label{fig:maps_shocks_zoom}
\end{center}
\end{figure*}

\begin{figure*}
%\begin{center}
\includegraphics[width=0.9\textwidth]{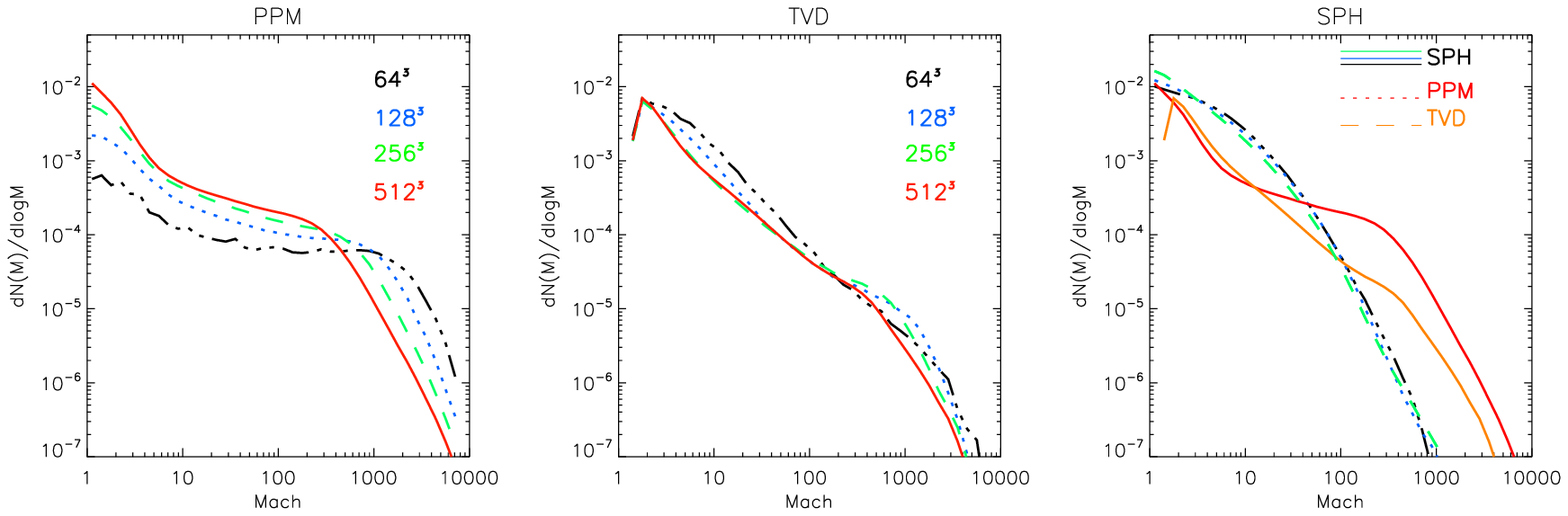}
\caption{Volume weighted number distributions of shocks at all resolutions and 
for all simulations. In the right panel also the results from the grid 
codes (at the maximum available resolution) are shown for comparison }.
\label{fig:histo_shocks}
%\end{center}
\end{figure*}

\section{Shock Waves in Cosmological Simulations}
\label{sec:shocks}
 
Many of the
differences previously found between the codes, such as the temperature structures in low density
environments and entropy distributions in the innermost and in the outer
regions of clusters, are likely connected to the dynamics of matter
accretion processes in the accretion regions of large scale
structures. 
In these regions, the activity of strong shock waves is the 
leading driver of thermalization, entropy generation and
possibly of cosmic ray acceleration in large scale structures,
(e.g. Ryu 
et al.  2003), via the diffusive shock acceleration mechanism 
(e.g. Blandford \& Ostriker 1978).

The numerical modeling of shock waves is among the most
important tasks that cosmological codes must correctly
perform in run-time; several different numerical techniques,
involving the use of ad-hoc numerical viscosity (as in SPH) or the solution of the Riemann problem through explicit methods (as in
PPM or TVD), have been adopted for this task 
(e.g. Dolag et al.  2008 for a review). 

All these methods generally perform well in the case
of rather simple shock problem (e.g. Tasker et al.  2008), while
their performances in
the very complex environment of large scale structure
simulations are more uncertain.
To date, no detailed comparison of the statistics of shocks
developed in the various numerical method have ever been published;
our sample of runs thus offers the optimal framework to test the 
outcomes of the different methods in the cosmic volume. 

In order to readily compare the statistics of shocks in each
simulations, a shock finding method is needed to detect and
measure the strength of shocks in the simulations. To this end we start by
 presenting the shock detecting method
explicitly developed to work on each specific code in our
project.

\subsection{Shocks Capturing Algorithm}

\label{subsec:basic}

The Rankine--Hugoniot jump conditions allows one to evaluate the shock Mach
number, $M$, from the thermodynamical state of the pre-shock and post-shock
regions (under the assumption of a pre-shock medium at rest and in thermal and
pressure equilibrium). If the adiabatic index is set to $\gamma = 5/3$ one has
the well known relations (e.g. Landau \& Lifshitz 1966):

\begin{equation}
\frac{\rho_{2}}{\rho_{1}}=\frac{4M^{2}}{M^{2}+3},  
\label{eq:dens}
\end{equation}

\begin{equation}
\frac{T_{2}}{T_{1}}=\frac{(5M^{2}-1)(M^{2}+3)}{16M^2}
\label{eq:temp}
\end{equation}

and

\begin{equation}
\frac{S_{2}}{S_{1}}=\frac{(5M^{2}-1)(M^{2}+3)}{16M^{2}}
\left(\frac{M^{2}+3}{4M^{2}}\right)^{2/3},  
\label{eq:entropy}
\end{equation}
with indices $1,2$ referring to pre and post--shock quantities,
respectively, and where the entropy $S$ is $S=T/\rho^{2/3}$.  

In practice measuring $M$ of shocks in cosmological simulations 
is more problematic than in this ideal case:
matter falling in the potential wells
drives chaotic motions and the temperature distribution 
around shocks is usually patchy due to the continuous accretion of cold clumps and filaments into hot halos. 
These complex behaviors establish complex pattern of pre-shocks velocity, temperature
and density fluctuations which makes problematic to measure Ranking-Hugoniot
jumps in a clean way.
To overcome this problem, 
detailed analysis strategies have been conceived over the last years,
with the goal of recovering the measure of $M$ in fully cosmological
simulations in the most
accurate way. 

\subsubsection{The Temperature Jumps Method - TJ}
\label{subsec:Ryu0}

The analysis of jumps in temperature is a powerful way
of measuring
the strength of shocks in Eulerian cosmological simulations, and its application was first discussed in Miniati et al.  (2001), with a more sophisticated formulation in Ryu et al.  (2003).
The
cells hosting a possible shock pattern are preliminarily tagged by two conditions: 

\begin{itemize}
\item $\nabla T \cdot \nabla S > 0$;
\item $\nabla \cdot {\bf v} < 0$.
\end{itemize}

The additional condition on the strength of the temperature 
gradient across cells is also customary requested:

\begin{itemize}
\item $\mid \triangle log T \mid \geq 0.11$;
\end{itemize}

\noindent
(specifically $\mid \triangle log T \mid \geq 0.11$ 
filters out shocks with a Mach number $M < 1.3$, Ryu et al. 2003).

It is customary to simplify the process of identification of shocked cells by using a one--dimensional procedure
applied successively in three orthogonal directions.
In the case of multiple shocked cells in close contact, 
the center of shocks, which can be spread across 2--3 zones, 
is placed where $\nabla \cdot {\bf v}$ is minimum.
Then the 
Mach number is calculated based on Eq~.\ref{eq:temp}, where 
$T_2$ and $T_1$ are the
post and pre--shock temperature across the shock region{\footnote {We note that Skillman et al. (2008) pointed out that the application of a split
coordinate approach to the TJ method may lead to an overestimate in the 
number of shocks, compared to an unsplit TJ method, in ENZO AMR simulations. The bulk of the thermalized
energy at shocks, however, is only marginally affected by the above differences}}.
In the following Sections, we will refer to this method as to the {\it TJ} method.

In this work, we applied the TJ method following the original formulation of Ryu
et al.  (2003), with the exception that we do not employ the temperature floor of
$T_{\rm o}=10^{4}$ customarily used to mimic the effect of re-ionization, in order to
readily compare with the outcomes of the other simulations of the project.

\subsubsection{The Velocity Jumps Method - VJ}
\label{subsec:vj}

A similar approach, based on the post-processing analysis
of velocity jumps across cells in grid simulations was proposed in  
Vazza, Brunetti \& Gheller (2009) for the analysis
of ENZO simulations.
Conservation of momentum in the reference frame of 
the shock yields: 

\begin{equation}
\rho_{1}v_{1}=\rho_{2}v_{2},
\label{eq:momentum}
\end{equation}

with the same notation used in Eqs.\ref{eq:dens}--\ref{eq:entropy}.
In the ideal case in which the pre-shocked medium is 
at rest and in thermal and pressure equilibrium, 
the passage of a shock with velocity $v_s$ leaves a $\Delta v$ in-print 
as a velocity difference between the shocked and pre--shocked cells.
In the lab frame a relation holds between $\Delta v$ and $M$,
which can be obtained by combining
Eqn.~\ref{eq:momentum} with Eqn.~\ref{eq:dens}:

\begin{equation}
\Delta v =\frac{3}{4}v_{s}\frac{1-M^{2}}{M^{2}}.
\label{eq:mach_v}
\end{equation}

where $v_{s} = M c_{s}$ and $c_{s}$ is the sound velocity computed
in the pre--shocked cell.

The procedure to identify shocks in 3--D with the VJ method follows these
steps:

\begin{itemize}
\item candidate shocked cells are selected as 
$\nabla \cdot {\bf v} < 0$ (calculated as 3--dimensional velocity 
divergence);

\item if more candidate shocked cells are found together, the one with the
minimum $\nabla \cdot {\bf v}$ is considered as the shock center;
 
\item the three Cartesian axes are scanned with 1--D sweeps and  
$\Delta {\bf v_{x,y,z}}$ jumps along the axis of scan are measured, between cells located at a  $\Delta l$ 
distance on opposite side of the shock center. In ENZO PPM we can safely use $\Delta l=1$, therefore $M$ is measured across 3 cells (e.g. Vazza, Brunetti \& Gheller 2009 for a detailed discussion).

\item the sound speed is taken from the cell in the tagged patch which shows the lower temperature,
and based on this the Mach number along each direction is computed from 
Eqn.~\ref{eq:mach_v};

\item we finally reconstruct the 3-D  Mach number in the shocked cell with 
$M = (M_{x}^{2}+M_{y}^{2}+M_{z}^{2})^{1/2}$.

\end{itemize}

\noindent

In the following we refer to this procedure as 
the velocity jump (VJ) method.

Vazza, Brunetti \& Gheller (2009) reported 
overall consistency between VJ and TJ method in ENZO simulations with fixed grid resolution, with minor differences in the most rarefied environments. In Vazza et al. (2009) and Vazza et al.  (2010) the application of the VJ method is extended to ENZO runs with Adaptive Mesh Refinement. 

The application of a qualitatively similar method, working on the velocity field of SPH particles
in GADGET3 simulations, has also been presented by
Hoeft et al. (2008)

\subsubsection{The Entropy Jumps Method - EJ}
\label{subsubsec:EJ}

A method to measure the Mach number of gas flows in GADGET
runs was presented in Pfrommer et al. (2006). In this 
method, a run-time algorithm monitors in run-time the
evolution of entropy for each particles, and from the 
entropy jump (in time) the Mach number of the shock can be 
inferred. 

The instantaneous
injection rate of the entropic function due to shocks for each SPH particle is
$dA(S) /dt$, where $A$ is the entropic function, defined
by $P = A(S)\rho^\gamma$ (where $P$ is the gas pressure).  
If the shock is
broadened over a scale of order the SPH smoothing length $f_h h$ ($f_h \sim
2$ is a factor which has to be calibrated with shock-tube tests), one can
roughly estimate the time it takes the particle to pass through the broadened
shock front as $\Delta t = f_h h/v$, where $v$ can be approximated with
the pre-shock velocity $v_{1}$. Assuming that the present particle temperature
is a good approximation for the pre-shock temperature, it is possible 
to replace $v_{1}$ with $M_{1} c_{1}$.

Based on these assumptions and using $\Delta A_{1}\simeq\Delta t d A_1/dt$, 
the jump of the entropic function of the particle crossing a shock will
be:

\begin{eqnarray}
\label{eq:A2/A1_1}
\frac{A_2}{A_1}  &=& \frac{A_1 + \Delta A_1}{A_1} = 1 + \frac{f_h h}{M_1 c_1 A_1}
\frac{dA_1}{d t}, \\
\label{eq:A2/A1_2}
\frac{A_2}{A_1}  &=& 
\frac{P_2}{P_1} \left(\frac{\rho_1}{\rho_2}\right)^{\gamma}
= f_A(M_1),
\end{eqnarray}
where, using Equation \ref{eq:dens} and \ref{eq:temp} one has: 

\begin{equation}
f_A(M_1) \equiv \frac{2\gamma M_1^2 - (\gamma-1)}{\gamma+1}
\left[\frac{(\gamma-1) M_1^2 + 2}{(\gamma+1) M_1^2}\right]^\gamma,
\end{equation}
that combined with Equations~\ref{eq:A2/A1_1} and \ref{eq:A2/A1_2}:
\begin{equation}
\left[f_A(M_1) - 1 \right] M_1 = 
\frac{f_h h}{c_1 A_1} \frac{d A_1}{d t}.
\label{eq:f4}
\end{equation}

The right-hand side of Eqn.~\ref{eq:f4} can be estimated individually for each
particle, and Eqn.~\ref{eq:f4} allows to estimate their Mach number (see Pfrommer
et al. 2006 for details).  In the following we will refer to this method as EJ
method.  

The EJ method has been applied in a series of papers to characterize shocks on the fly,
inject CRs with a Mach number-dependent acceleration efficiency, account for the
non-linear back reaction of the CR pressure on the hydrodynamics and following
the transport of CRs during GADGET3 simulations of cosmological structure
formation, galaxy and galaxy cluster formation (Pfrommer et al.  2006, 2007,
2008; Pfrommer (2008); Jubelgas et al. (2008); Pinzke \& Pfrommer 2010; Pinzke et
al. 2011).  In our work here, the original EJ scheme has been applied in
run-time to GADGET3 runs, and the measured distributions of Mach numbers for the
gas particles have been analyzed in post-processing.

\begin{figure*}
\begin{center}
\includegraphics[width=0.95\textwidth,height=0.3\textwidth]{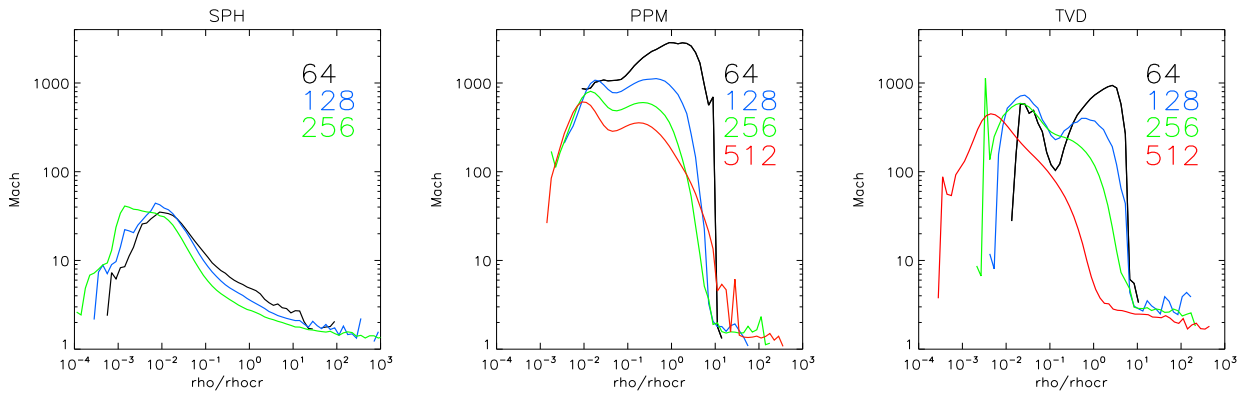}
\caption{Volume weighted mean Mach number as a function of gas density, for 
all runs of the project.}
\label{fig:rhog_mach}
\end{center}
\end{figure*}

\subsection{Shocks Maps and Morphologies}
\label{subsec:shock_1_cp}

We measured the strength of shocks in our simulations, by applying
the TJ method in post-processing to TVD runs, the VJ method in 
post-processing  to ENZO run and the EJ method in run time
for GADGET3 runs. 

The panels in Figure~ \ref{fig:maps_shocks} show the large scale pattern of shock
waves for a thin slice (of 550 kpc) in the simulated box at z=0,
for the best available resolutions in all codes. 
Only for display purposes, 
the Mach numbers measured in GADGET3
have been interpolated onto regular grids with resolution corresponding
to a $256^{3}$ mesh. 

Even at the best available resolution, the morphological distributions of shocks
in the various run looks less similar than what is generally found for the
density-weighted maps of temperature (Fig.~\ref{fig:maps}-\ref{fig:maps_zoom}). 
In all runs innermost region of clusters and filaments hosts only
weak shocks, $M \sim 2 - 5$, while the strongest shocks are located outside
cosmic structures. However, the strong external shocks are very sharp and 
regular in grid codes, while they seem to be grouped in clumps in GADGET3. 
While in GADGET3 runs the shocked structures are rather volume filling (due to the
smoothing kernel in less dense regions), in both grid methods the shocks
outside clusters are regular surfaces with radius of curvature $\sim 3 - 10 \Mpc$, with a very small volume filling factor.

We notice that this difference between SPH and grid methods depends on the different resolutions outside $R_{\rm vir}$,
however the general trend is that when the spatial and mass resolution of DM particles is {\it increased}, the differences
between grid codes and SPH are even more sizable. 

This is shown in the  panels of Fig.~\ref{fig:maps_shocks_zoom}, which zoom into the cluster region
at the center of the cosmological box.
Looking at the strong external shocks in the upper left sector of the
cluster, one can see that these features become increasingly 
sharper and more regular in grid methods, while they become stronger and more
clumpy in GADGET3 runs. On the other hand, the trend with resolution inside of the cluster
is quite similar in all codes, with increasingly thinner and weaker shocks as the resolution is increased.

\subsection{Mach Number Distributions}
\label{subsec:dn_comp}

The volume distribution of Mach numbers in the cosmological volume
is a simple statistical proxy that allows us to readily compare the different shock finder and underlying simulations. However, they cannot be directly translated into observational quantities, and therefore
 their study is just intended to be a useful to cross-check of numerical implementations, rather than
 a physical test.

Figure~ \ref{fig:histo_shocks} shows the volume-weighted distribution of 
shocks Mach number from all runs using our projects.

At the best available resolution, the distributions from the different methods are 
quite similar, showing a peak of shocks at $M \sim 1.5$ and a steep decrease at
stronger shocks. Compared to the peak, the average frequency of $M>1000$ shocks is 
$\sim 10^{-5}$ in GADGET, and $\sim 10^{-3}$ in ENZO and TVD.

GADGET runs present the best degree of self-convergence, with very little
evolution between runs $64^{3}$ and $256^{3}$. The VJ methods applied to
ENZO runs on the other hand shows the slowest degree of evolution, with
a particularly poor performance at the $64^{3}$ run; this is due 
to the difficulty of removing baryon bulk flows from velocity jumps
associate with shocks at very low grid resolutions. 
The TJ method present a noticeable self-convergence at all resolutions, 
although a the $64^{3}$-$128^{3}$ run present a different convexity
in the range $10 \leq M \leq 100$ (where the contribution from internal
and external shocks takes place), similar to the converged findings of
the EJ method applied to GADGET3.

In both grid codes, the increase of resolution always cause a progressive
weakening of the strongest shocks in the most rarefied environments; also
the bump of external shocks is progressively shifted towards lower $M$.

We notice that at the best available resolution here, the convergence all simulations (and most significantly 
in grid codes) is not yet reached, even if it looks approaching; the same is true also for the distribution
of thermal energy flux across shocks (Sect.\ref{subsec:en_comp}). 
Based on the tests in the literature, run with these same codes (e.g. Ryu et al. 2003; Skillman et al. 2008; Vazza, Brunetti \& Gheller 2009;
Vazza et al. 2009) one can see that a very good convergence (i.e. better than a $\sim 10$ per cent level)  in the most important shock statistics is expected 
to for a spatial resolutions of $\sim 50-100 kpc$, which are below our best resolution here. However the trend with resolution is usually very
regular, and the differences reported here are significant, despite the fact that a small evolution with resolution may still be present.
We also remark that an additional and unavoidable source of difference with resolution is due to the ways in which the shock-finder
methods work, because that the dependence on resolution of the different thermodynamical jumps used for the computation can be different,
especially for very coarse resolution. 

We also  notice here that
the modeling of a re-heating UV radiation from 
from massive stars and AGNs is crucial for a realistic estimate
of the baryon gas temperature outside of cosmic structures (e.g. Haardt \& Madau 1999). In order to measure realistic Mach number in the rarefied universe outside clusters, groups and filaments a re-ionization
temperature background is usually accounted in simulations, either
in post-processing (Ryu et al.  2003; Skillman et al. 2008; Vazza, Brunetti \& Gheller 2009) or in
run-time (Pfrommer et al.  2006; Vazza et al.  2010).
In this case the minimum temperature in all simulations is set by the low temperature
floor (see Sect.~\ref{subsec:pdf}); however the differences in the values adopted
in the different codes (from 1 K in the case of ENZO, to 24 K in the case of GADGET3) cannot account for the sizable differences in the distribution of Mach numbers.

The differences between the methods are highlighted 
when we plot the volume weighted 
average Mach number of shocks,  $\hat{M}$, as
a function of gas density (Fig.~\ref{fig:rhog_mach}). 
The results of the
different codes are consistent only for 
$\rho/\rho_{\rm cr} \geq 10$ regions (typical
of the outskirts of galaxy clusters and filaments), with $\hat{M} \sim 2$.
At lower densities we report the following trend:
in SPH $\hat{M}$ is smoothly increasing moving towards lower density regions,
while in grid codes the transition of $\hat{M}$ moving to lower densities is 
very sharp, and causes a net increase of $\hat{M}$ by 2 orders of magnitude
in both grid methods.
These large differences in the range
$\rho/\rho_{\rm cr} < 10$ mirror the different thermal structures 
of baryons in the outermost regions of LSS in grid codes
and in SPH (Sec.\ref{sec:thermal}). In these environments, the 
self convergence in grid codes is not yet reached even at the 
best available resolution $195 \kpc/h$.

\begin{figure*}
\begin{center}
\includegraphics[width=0.95\textwidth]{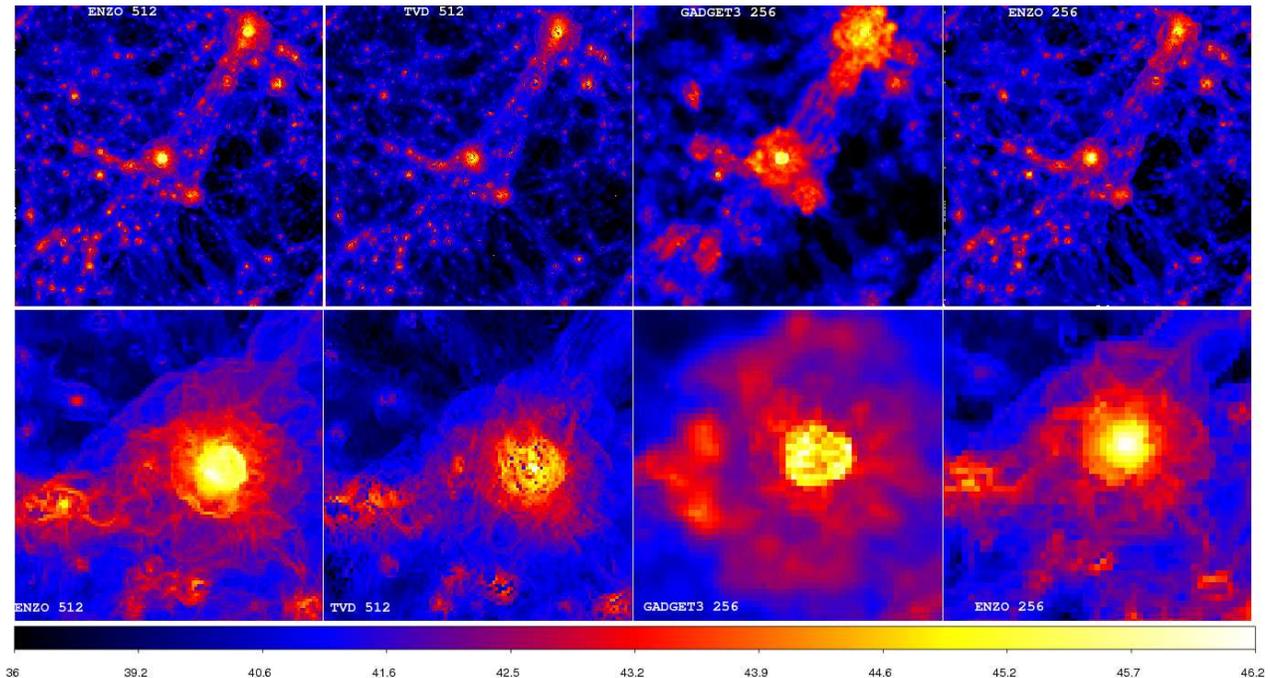}
\caption{Maps of projected of the thermalized energy flux at shock waves (in units of $log[erg/s]$) across
the whole simulated volume (top panels) and for a sub-region of side 25 \Mpc/h centered on the most massive galaxy
cluster of the sample (bottom panels).}
\label{fig:fth_shocks_map}
\end{center}
\end{figure*}

\begin{figure*}
\begin{center}
\includegraphics[width=0.9\textwidth]{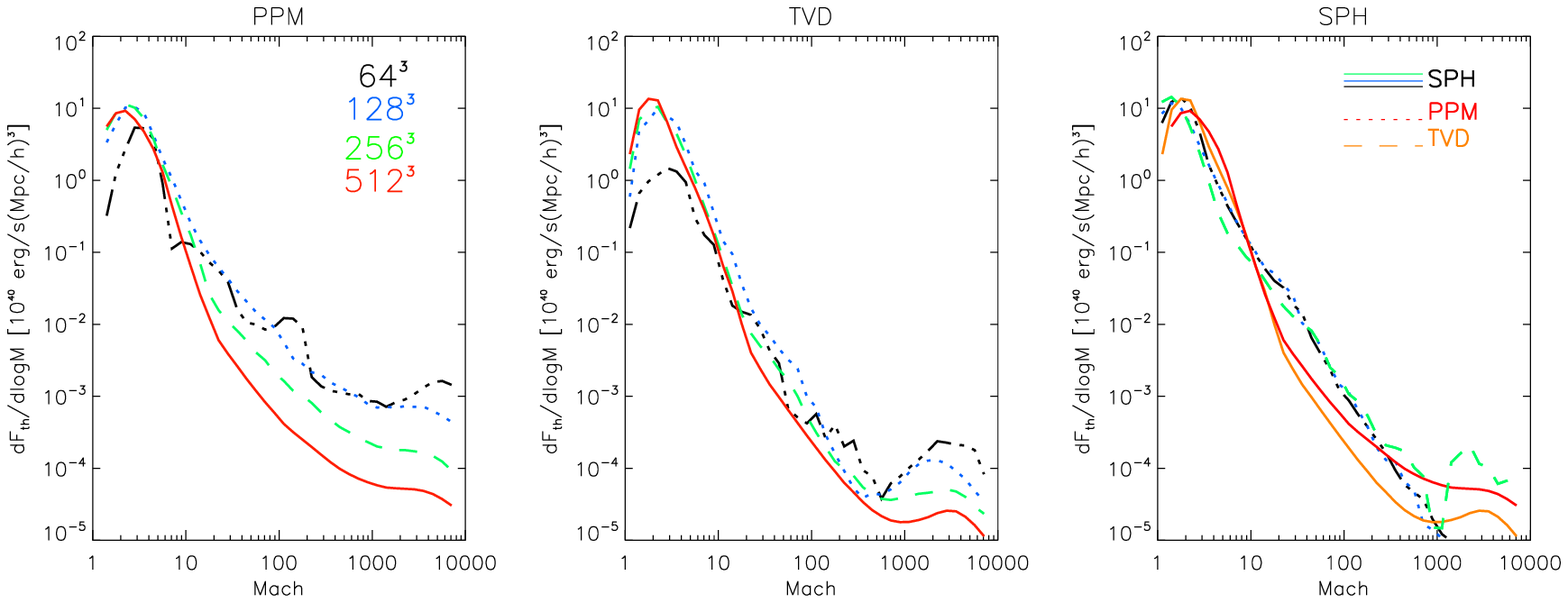}
\includegraphics[width=0.9\textwidth]{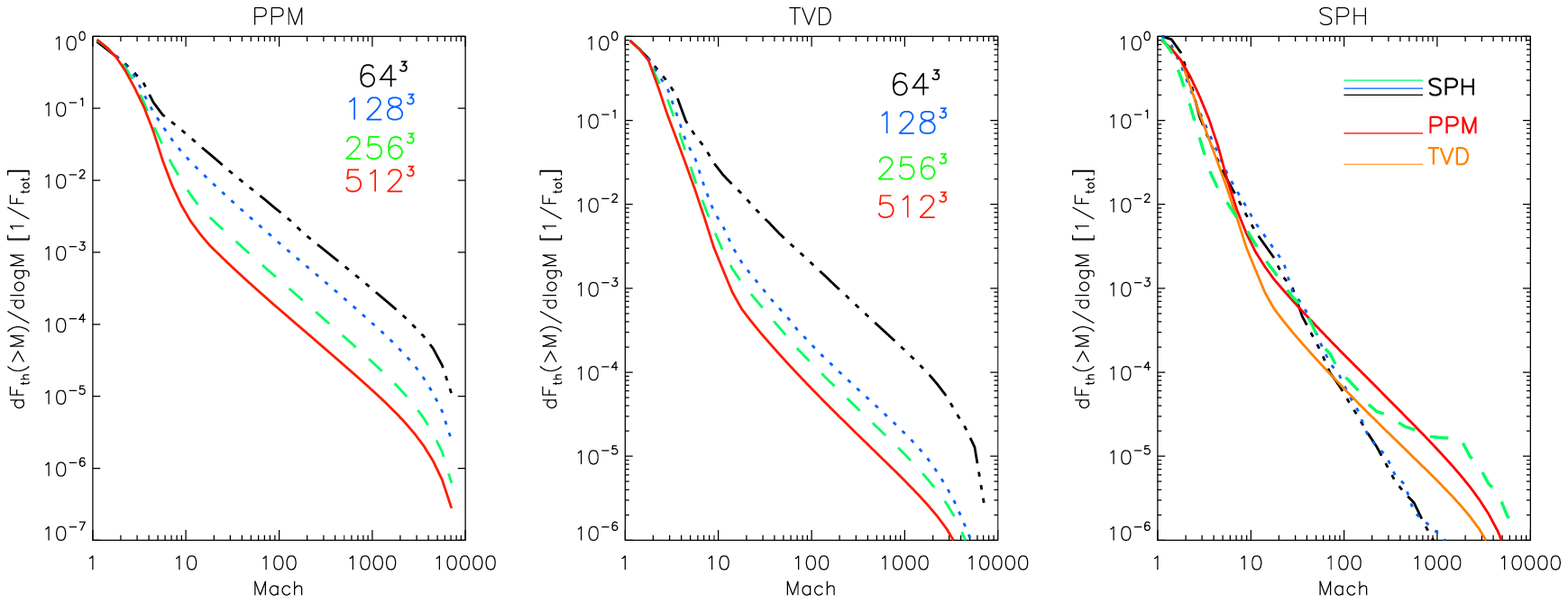}
\caption{Top panels: differential distributions of the thermalized energy flux through shocks at all resolution and for all codes. Bottom panels:
cumulative distributions for the same runs.}
\label{fig:fth_shocks}
\end{center}
\end{figure*}

\begin{figure*}
\begin{center}
\includegraphics[width=0.3\textwidth]{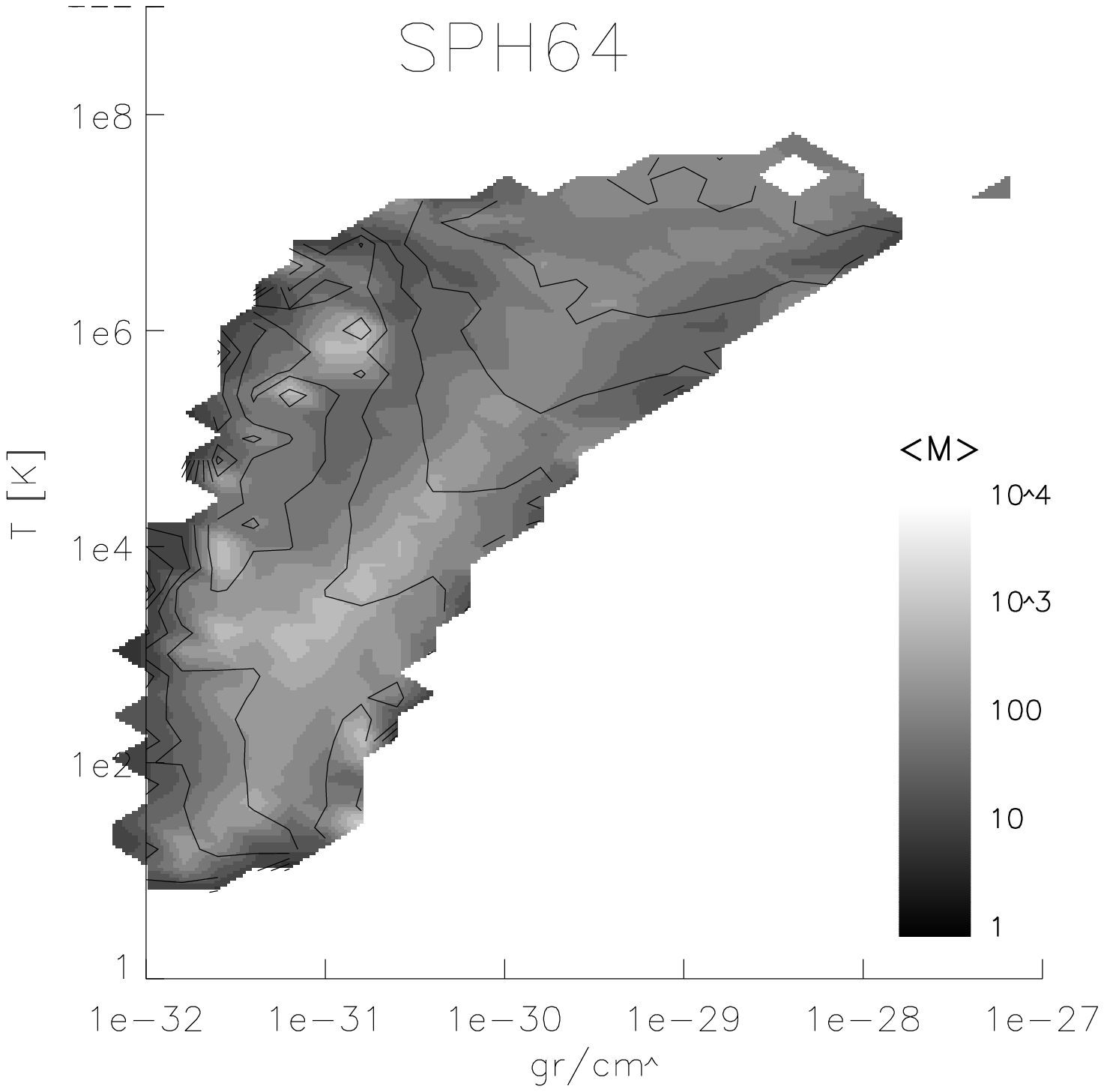}
\includegraphics[width=0.3\textwidth]{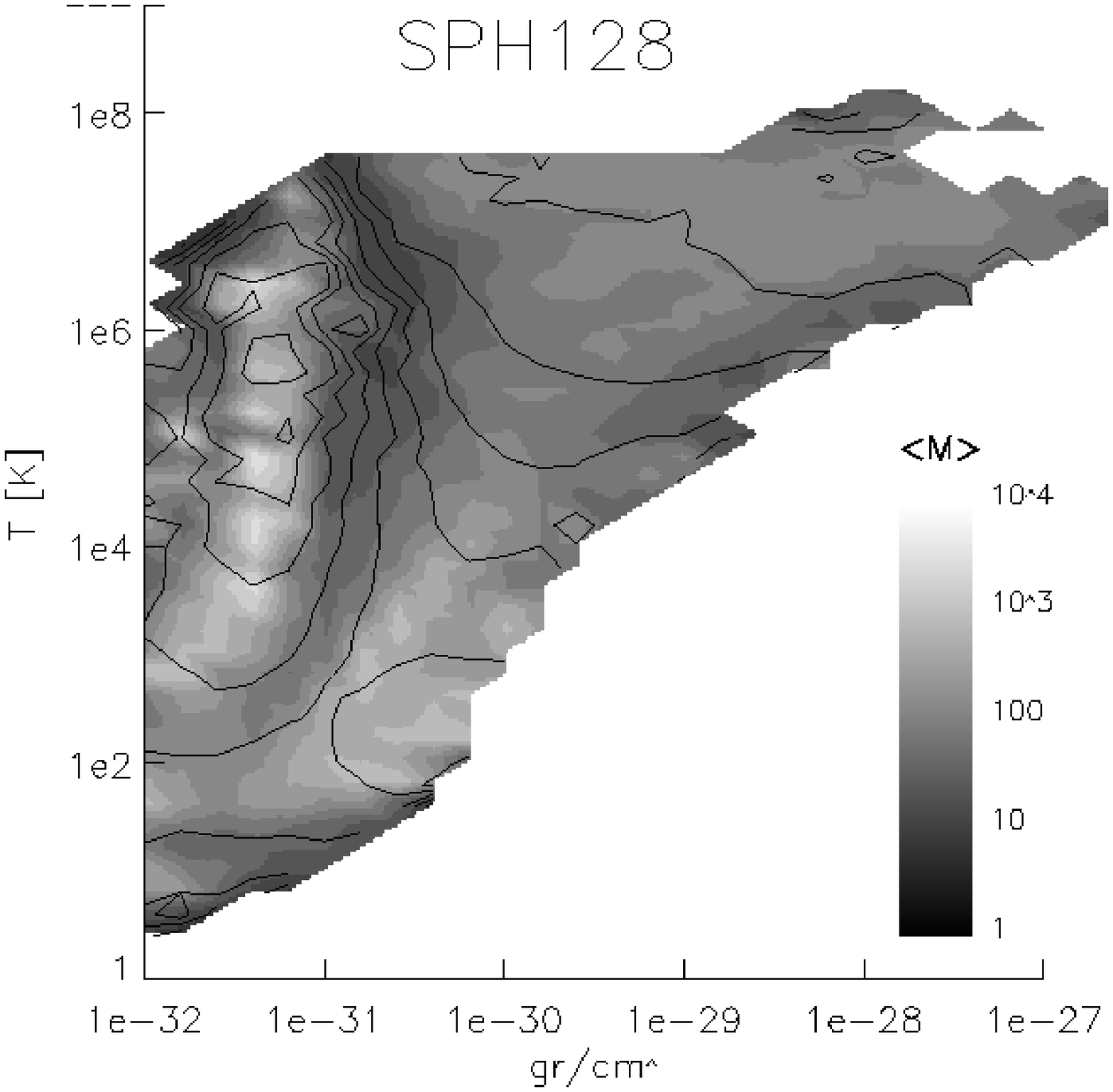}
\includegraphics[width=0.3\textwidth]{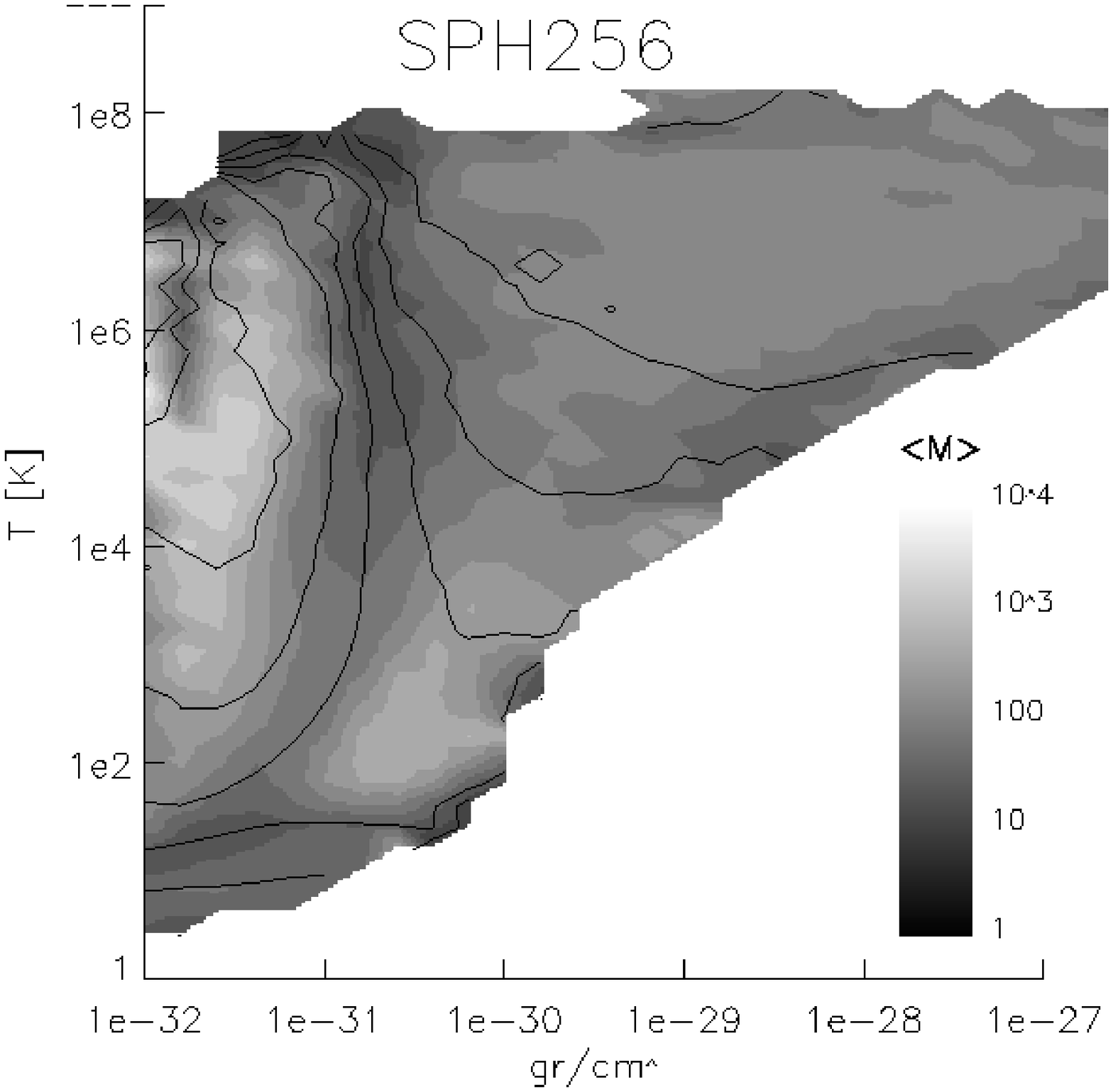}
\includegraphics[width=0.3\textwidth]{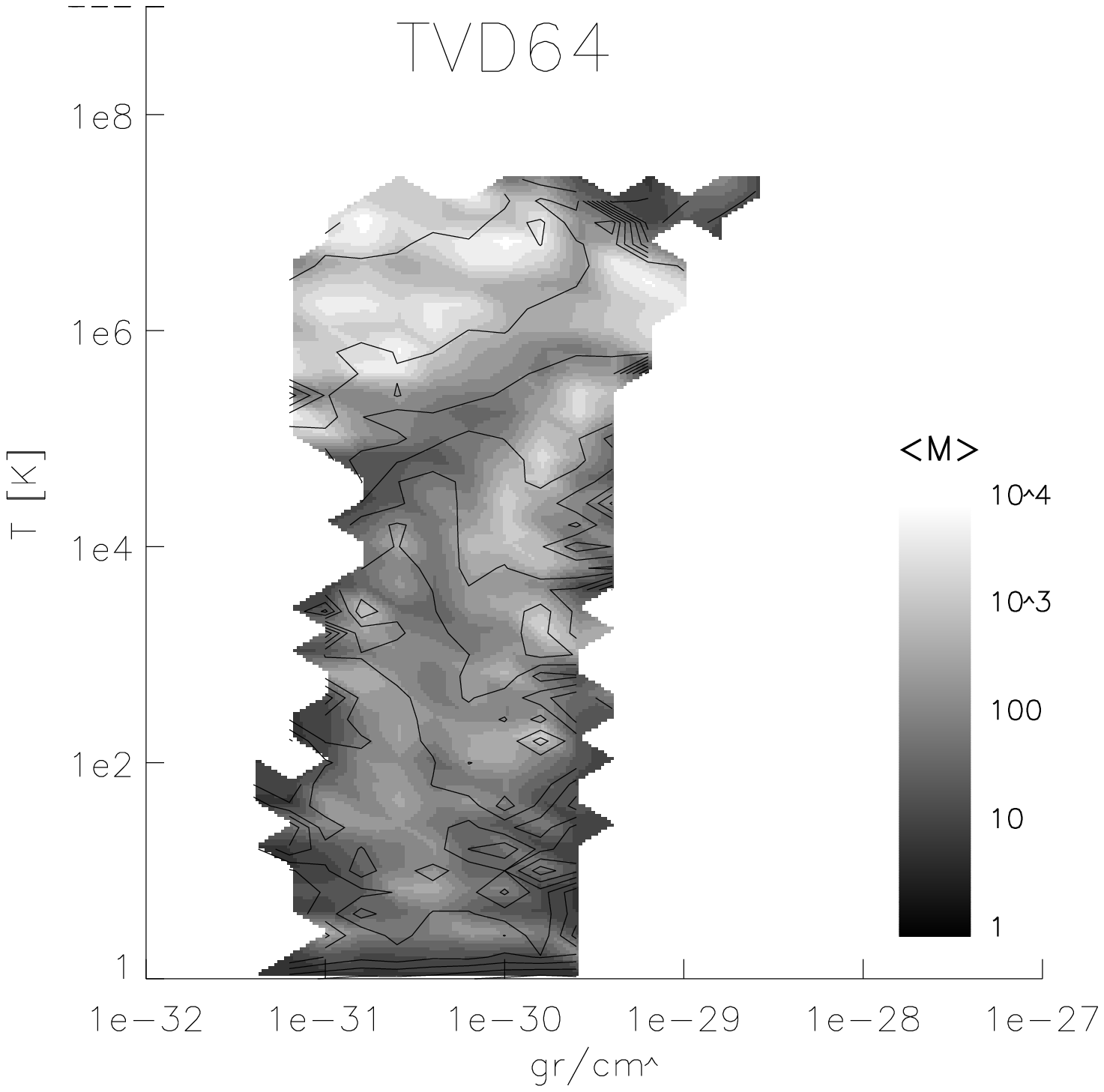}
\includegraphics[width=0.3\textwidth]{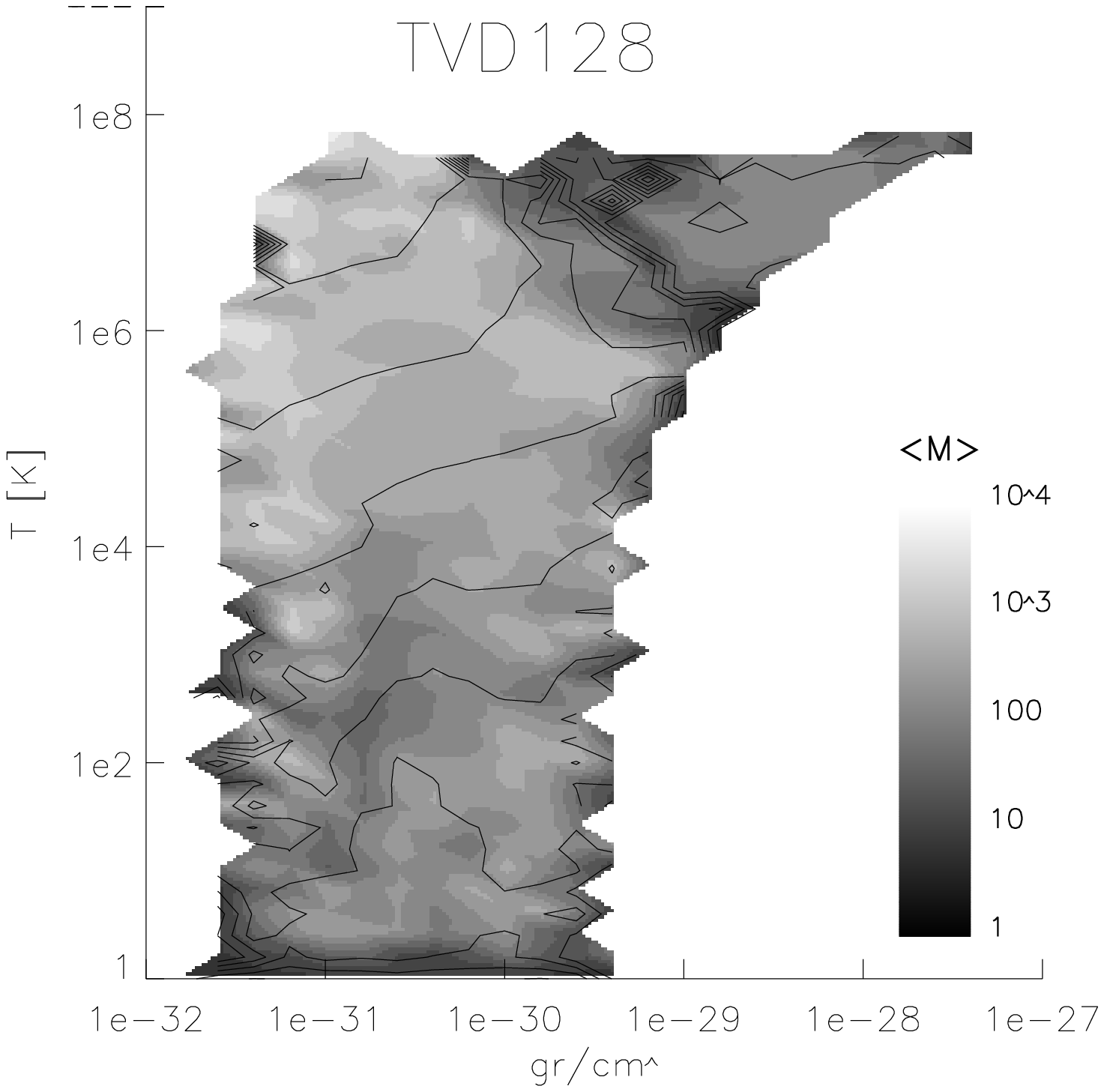}
\includegraphics[width=0.3\textwidth]{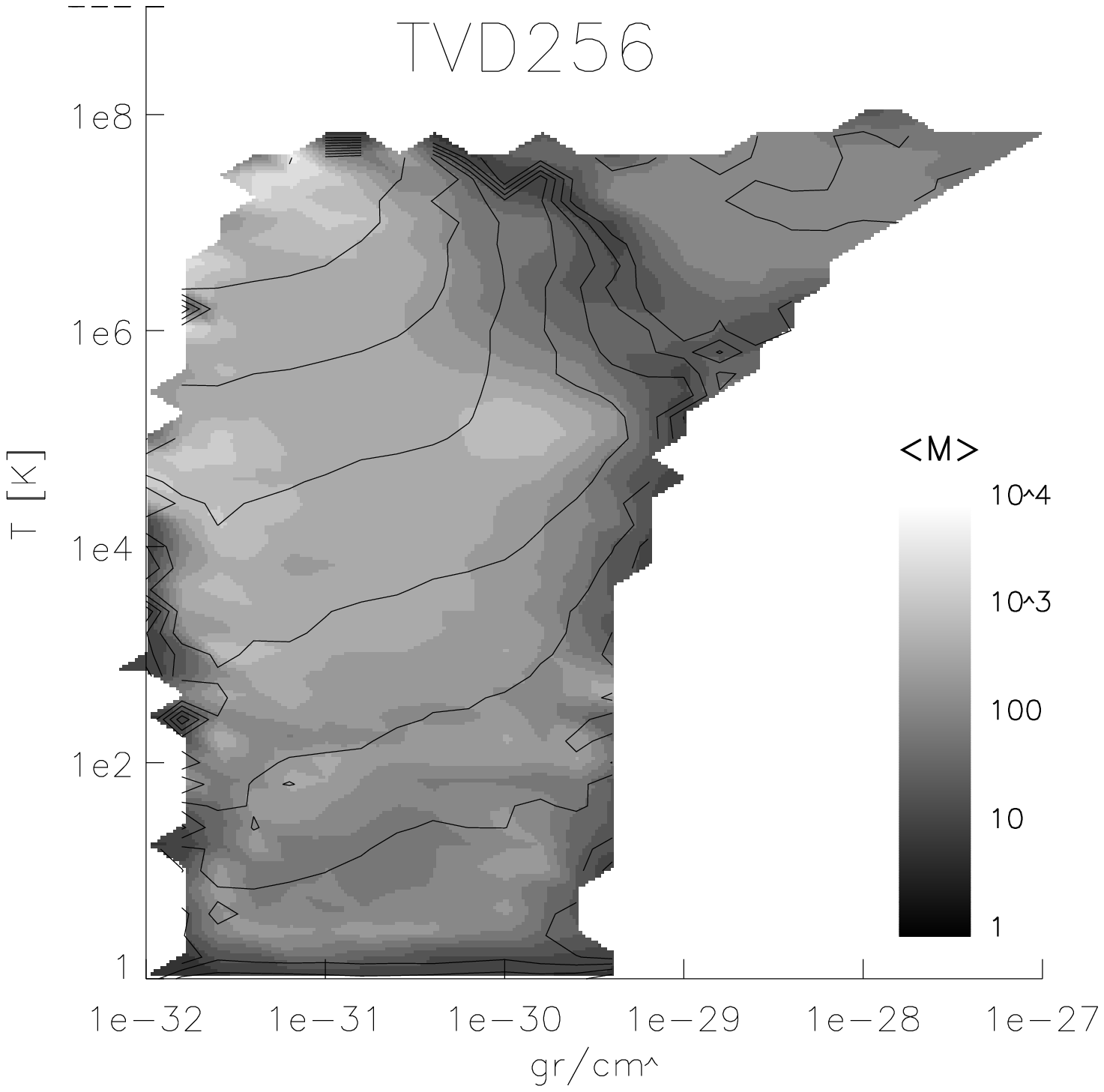}
\includegraphics[width=0.3\textwidth]{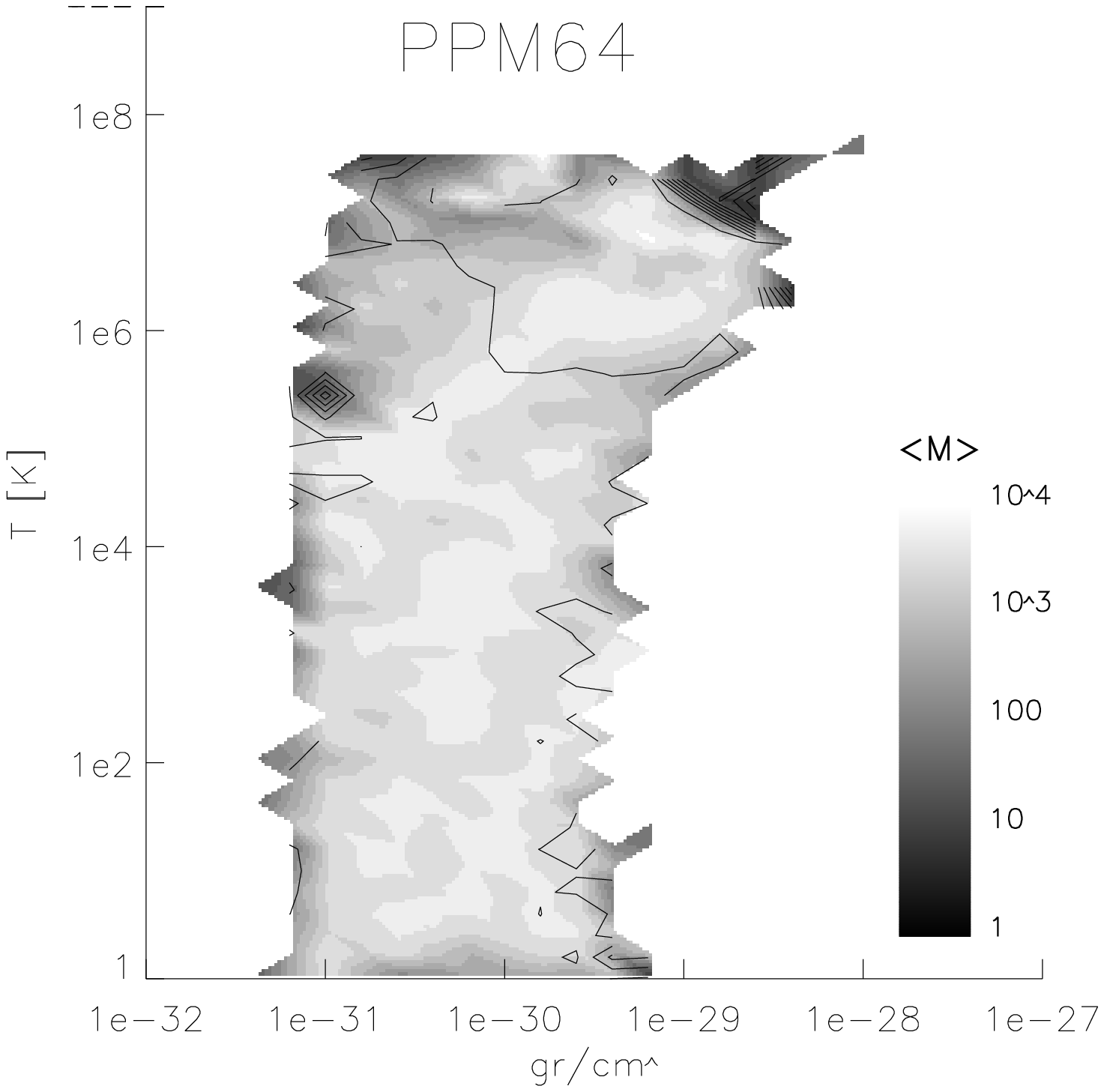}
\includegraphics[width=0.3\textwidth]{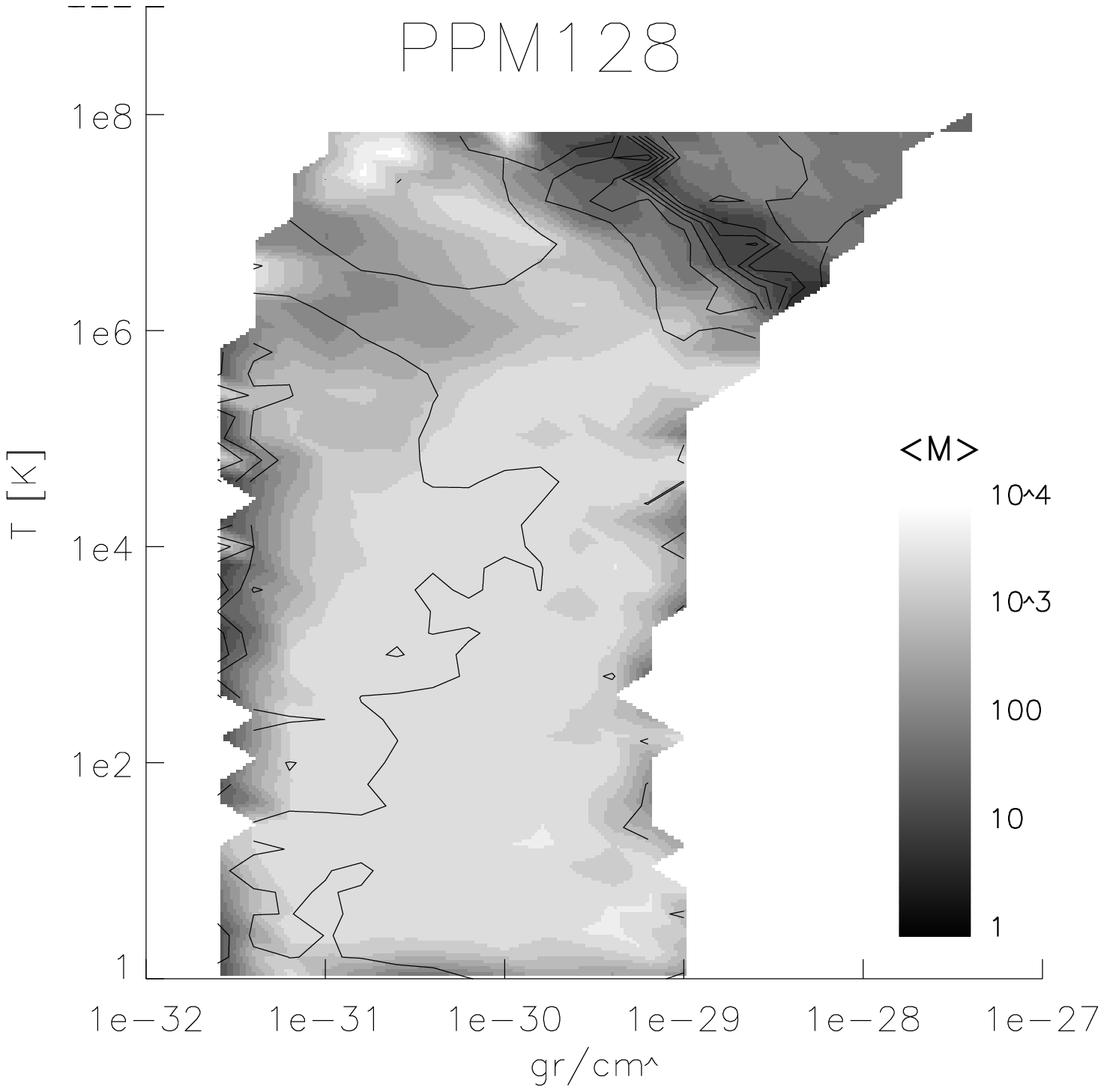}
\includegraphics[width=0.3\textwidth]{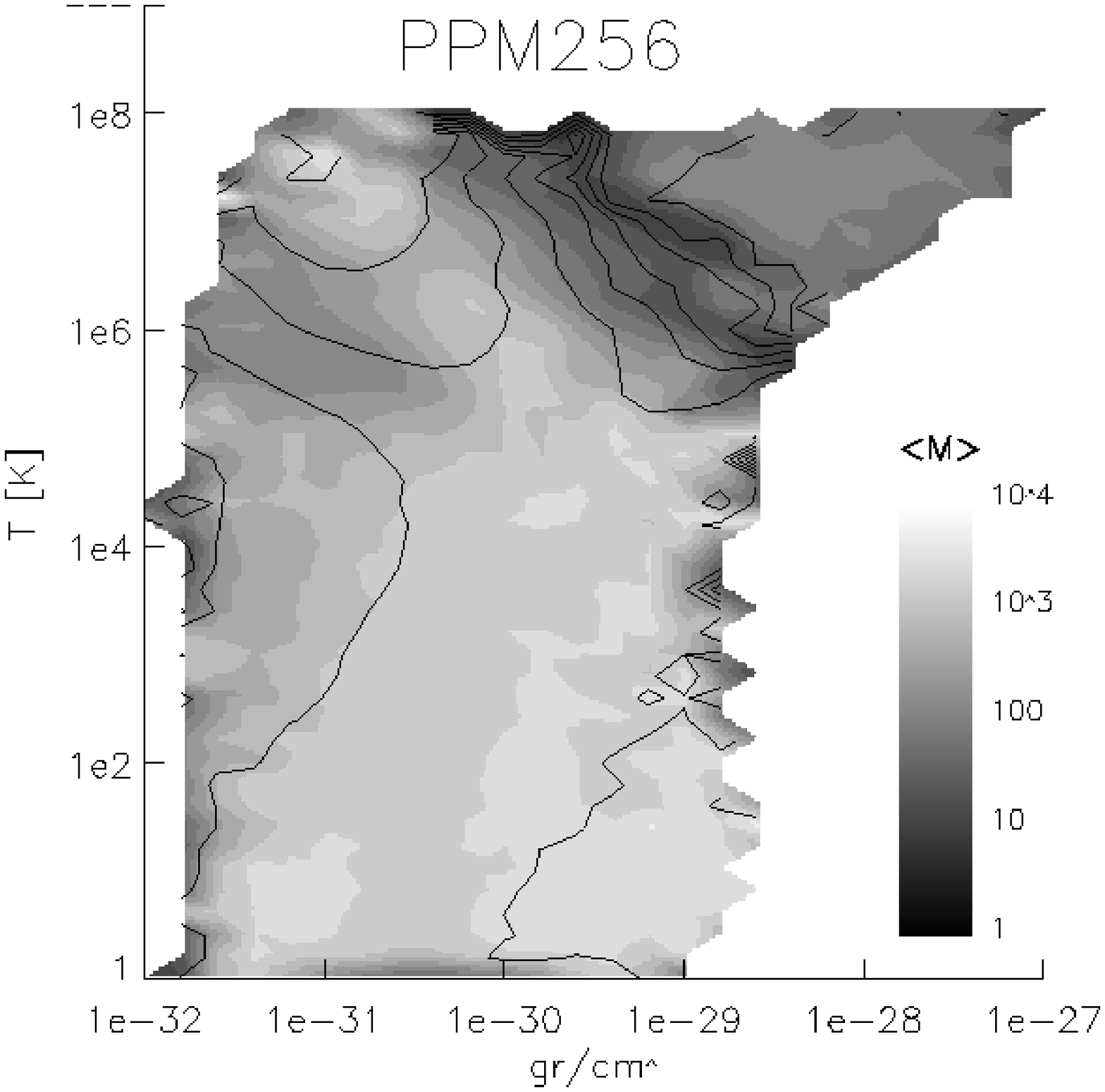}
\caption{Phase diagrams for shocked cells in the simulations, color 
coding shows the flux-weighted average Mach number. Additional 
isocontours with a coarse binning in  $\hat{M}$ space are shown for clarity.}
\label{fig:phase_shocks}
\end{center}
\end{figure*}

\subsection{Energy Distributions.}
\label{subsec:en_comp}

The thermal energy flux across each shock in the simulations is measured
as:

\begin{equation}
f_{\rm th} = \delta(M) \cdot \rho_{\rm pre} M^{3} c^{3}_{s}/2,
\label{eq:fth}
\end{equation}

where $\rho_{\rm pre}$ is the pre-shock gas density and $\delta(M)$ is a 
monotonically increasing function of $M$ which follows from
Rankine-Hugoniot jump conditions, whose formula can
be found for instance in Kang et al. (2007).

In the TJ and in the VJ methods this quantity is computed in post-processing
based on the shock direction, while in the EJ method $f_{\rm th}$ it is 
measured in run-time.

 We remark that in all 3 methods, the numerical recipes to compute the effective thermalization at the post-shock
are tuned to remove the effect of adiabatic compression of the gas in the post-shock region, which can provide sizable additional thermalization in the regime of weak shocks (see Ryu 
et al.  2003; Pfrommer et al.  2006).

We report in Fig.~\ref{fig:fth_shocks_map} the projected map of $f_{\rm th}$ across the
simulated volume in the best resolved runs (top panels), and for a zoomed
region of $25 \Mpc/h$ (bottom panels). We also 
report for comparison with the SPH run the corresponding ENZO $256^{3}$ run.
The flux coming from the innermost cluster region looks morphologically
similar in all cases, with a compact and spherical ''envelope'' of energetic shocks
concentrated inside the virial volume of halos. The differences are
more sizable at the scale of filaments and in the outer region of clusters, where we notice very sharp
shock surfaces even in projection in grid methods, while much
smoother pattern are found in GADGET3, with external accretion shock extending 
at larger distances from the center of clusters. This effect mirrors the corresponding
distribution of gas entropy at larger scales, which we reported in the analysis of
the radial profile of the entropic function, in Sect\ref{subsec:clusters_cp}.
The zoomed images of Fig.~\ref{fig:fth_shocks_map} 
additionally shows that complex intersections of merger
shocks are modeled inside the over-dense regions in grid codes,
 while very smooth distribution appears in the projected GADGET
 maps. 
Taking as a reference the ENZO run with $256^{3}$, we see that the above differences are not trivially
due to resolution 
effects, since the
large scale shock patterns in the grid code do not significantly get smoother or shift in position even if the
resolution of the simulation is made coarser.
The differential distributions of $f_{\rm th}$ for all runs is
reported in Fig.~\ref{fig:fth_shocks}. 
In this case the contribution coming from the low density regions is fairly negligible and results are found to be in an overall good agreement. 
As in the case of number distributions, the EJ method presents the largest
degree of self-convergence, and the VJ presents the slowest degree
of self-convergence.

The grid codes present the clear trend of processing less thermal flux
at $M>>10$ shocks when resolution is increased, while in SPH slightly
more energy flux is processed at strong shocks when resolution is
increased (although this amount is negligible compared to the peak
of thermalization in the box).
In the bottom panels of the same Figure, we also show the cumulative
distribution for the same run, normalized to the total flux
inside the cosmic volume for each run.

At their best resolution, all codes agree in several important findings:
a) the peak of thermalization is found at $M \sim 2$, consistent with
most of previous works in the literature (e.g. Ryu et al.  2003; 
Pfrommer et al. 2006,2007; Vazza et al. 2009,2010; Skillman et 
al. 2008); b) the general shape of the distributions is quite similar, with a steep power-law behaviour, $dlog f_{\rm th}/dlog M \sim M^{-\alpha}$ . The slope is $\alpha \sim 3$ in grid codes and $\alpha \sim 2.5$ in GADGET3 runs; this is steeper compared to the findings
in the literature, because we are not modelling here the re-ionization background.
c) the cumulative distributions for $M<10$ shocks are very similar in all codes, and 
only $\sim 1$ per cent of the total thermal flux inside the cosmic volume belongs to shocks with  $M>10$.  
These findings suggest that, despite sizable differences in the
shapes and statistics of strong external shocks in the 
accretion regions of large scale structures, the bulk of the
energetic properties of shocks within the cosmic volume
is a rather well converged answer from cosmological
simulations.

\begin{figure*}
\begin{center}
\includegraphics[width=0.3\textwidth]{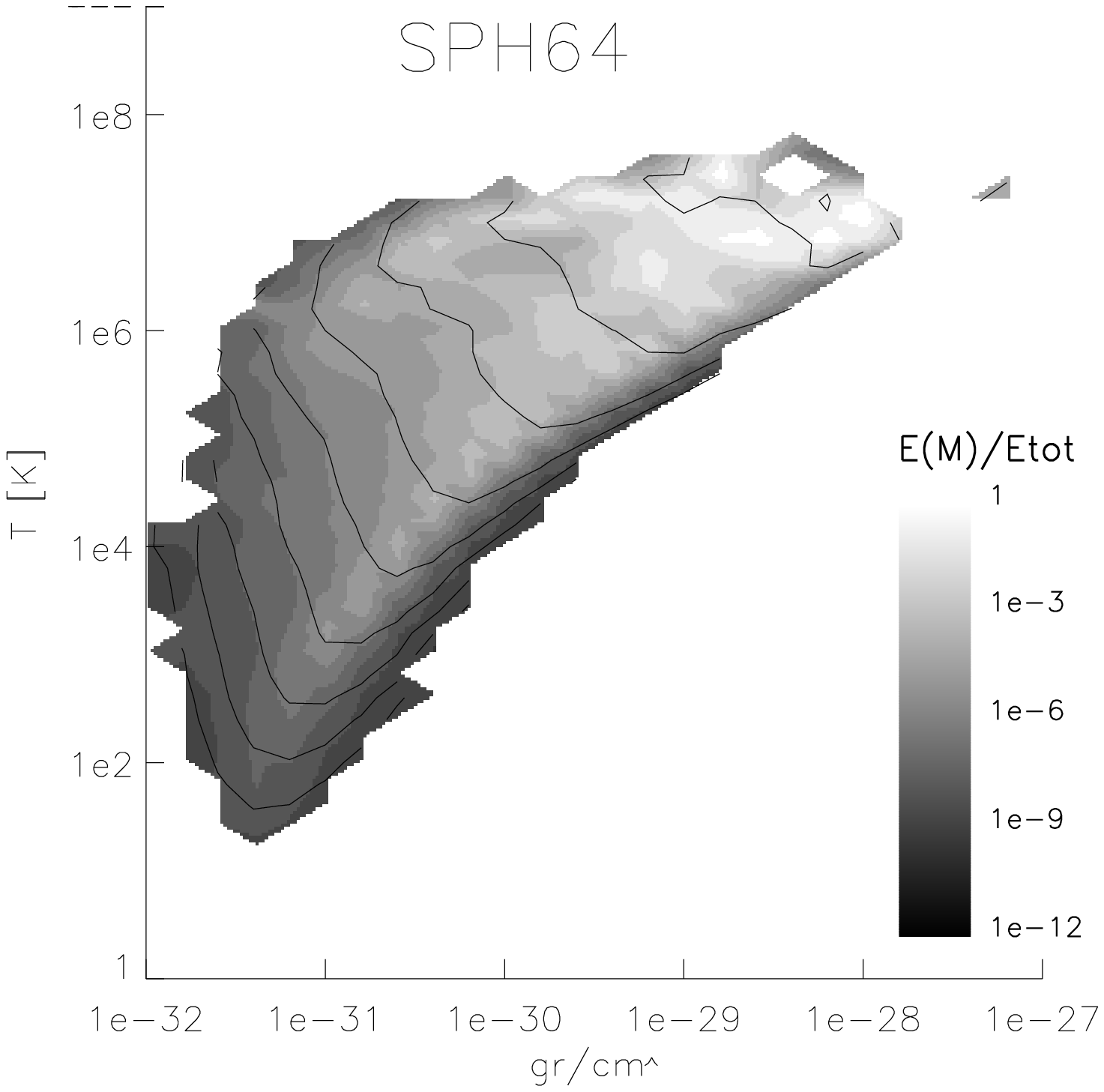}
\includegraphics[width=0.3\textwidth]{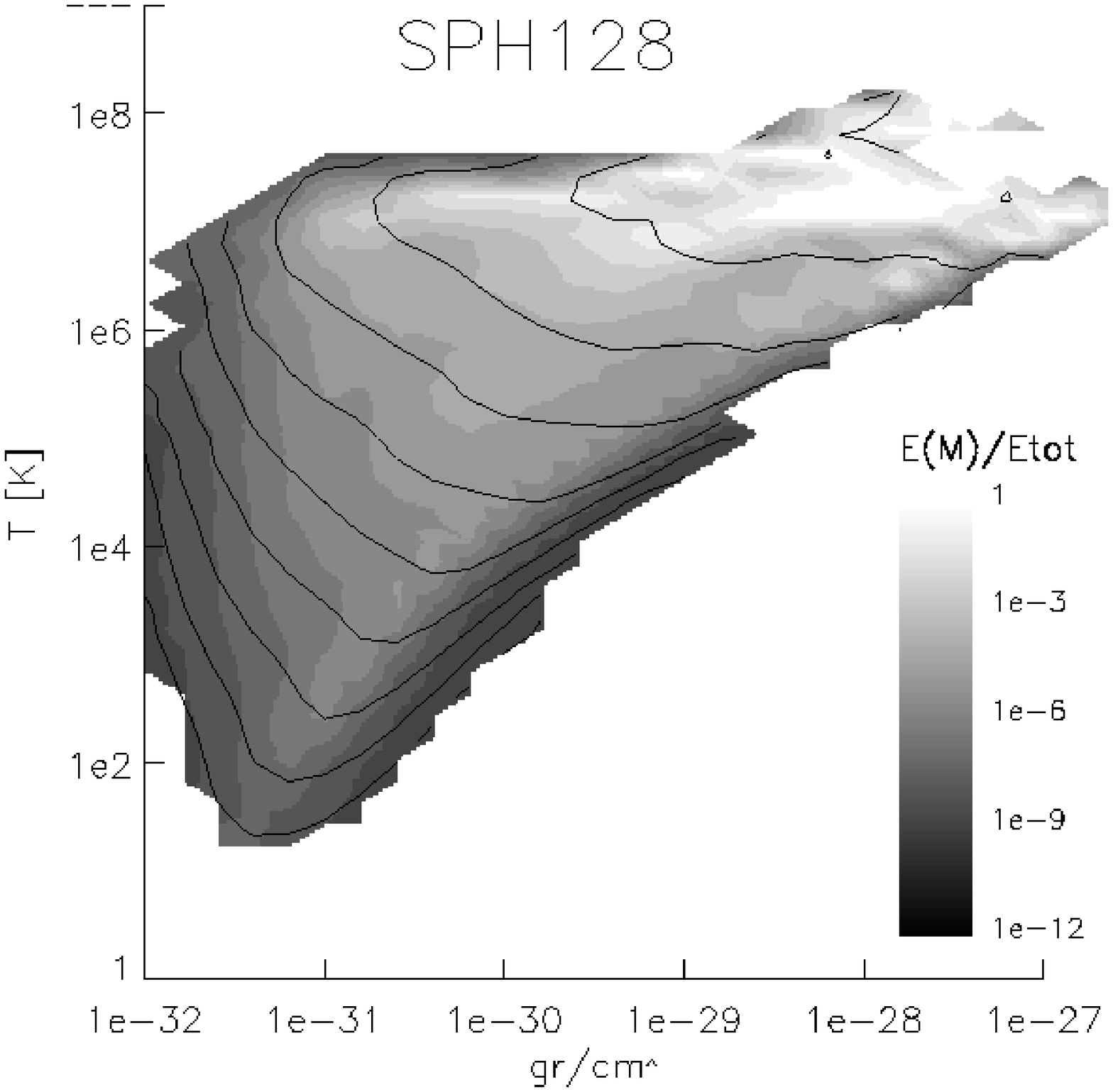}
\includegraphics[width=0.3\textwidth]{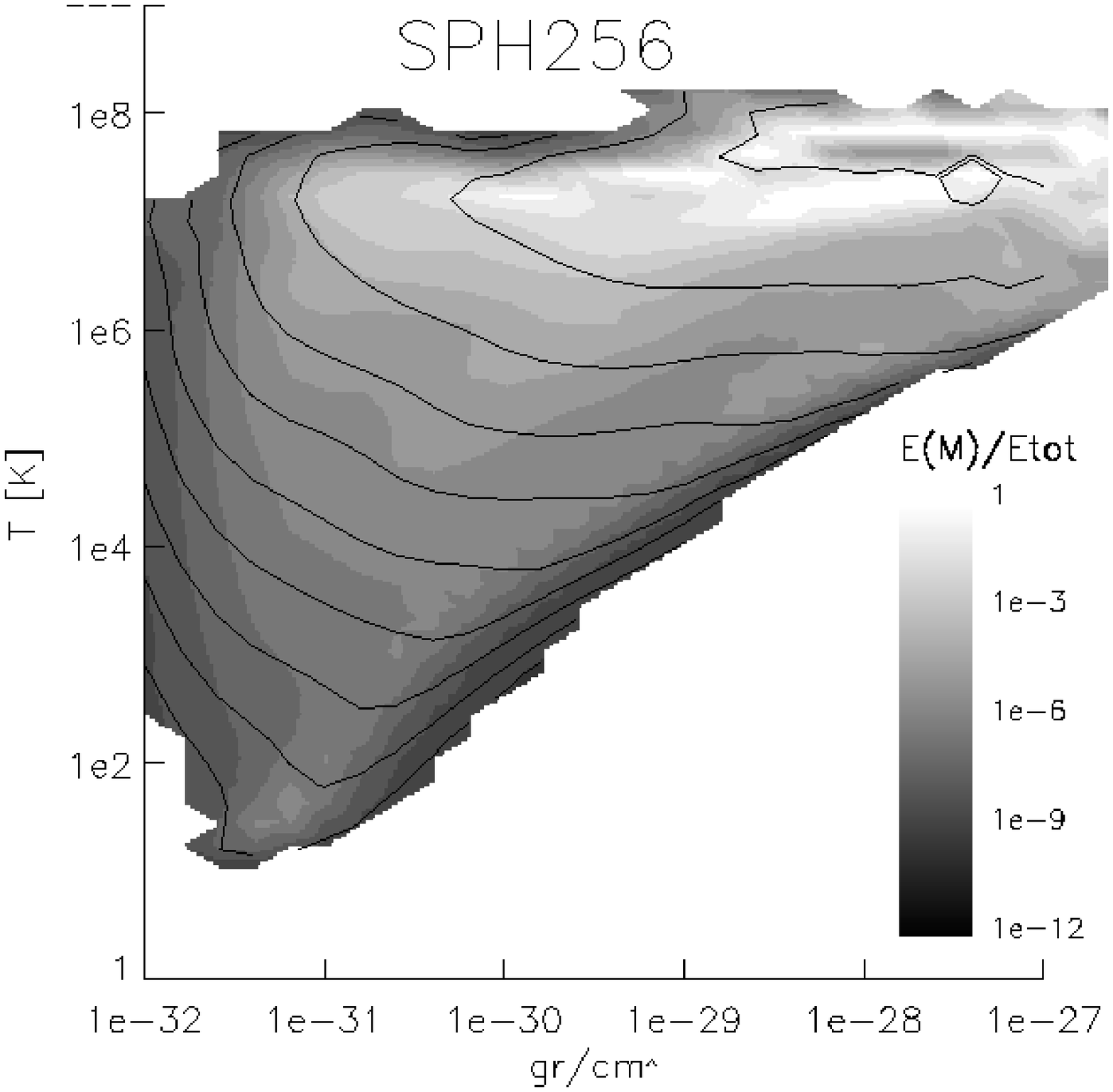}
\includegraphics[width=0.3\textwidth]{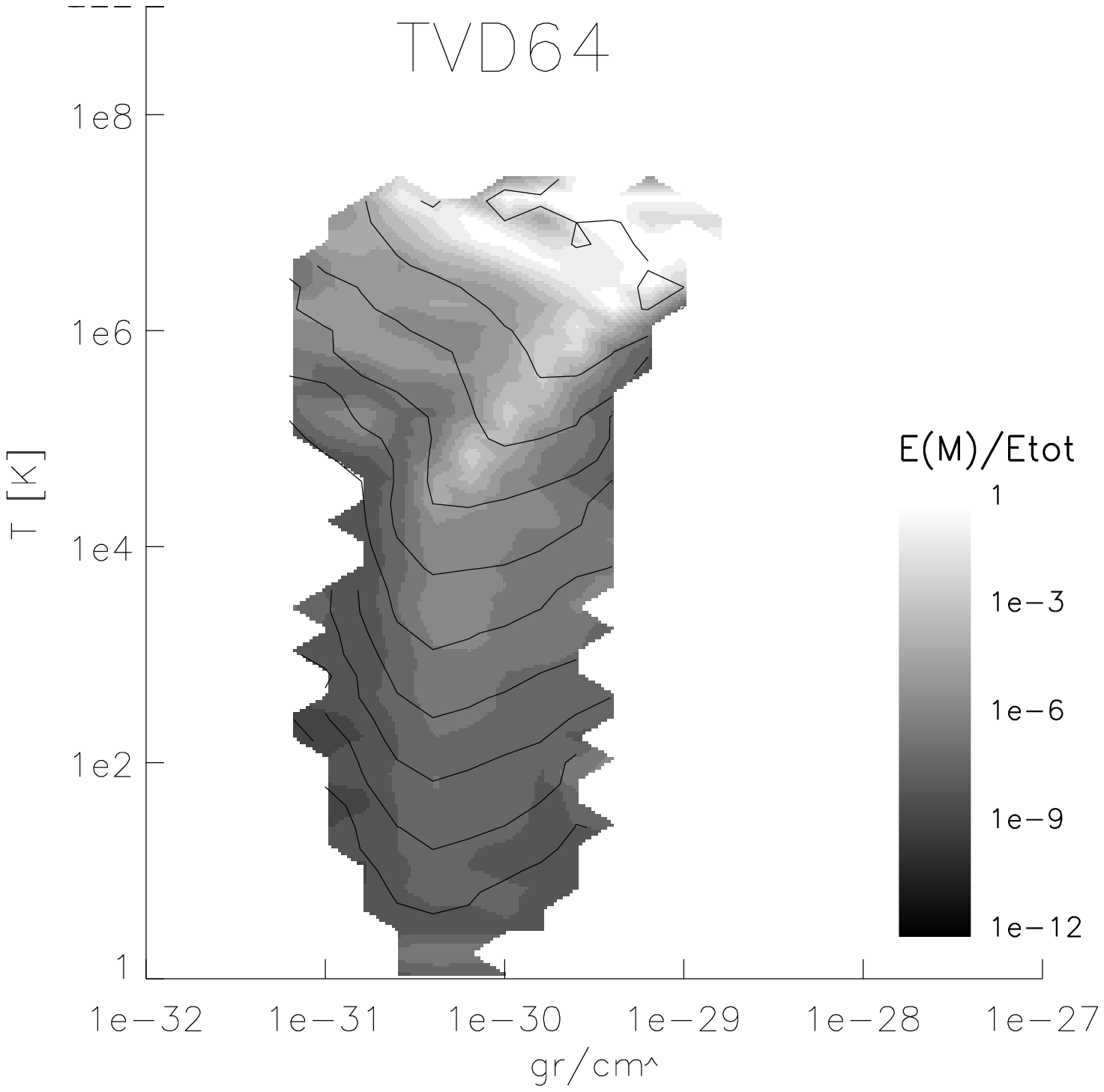}
\includegraphics[width=0.3\textwidth]{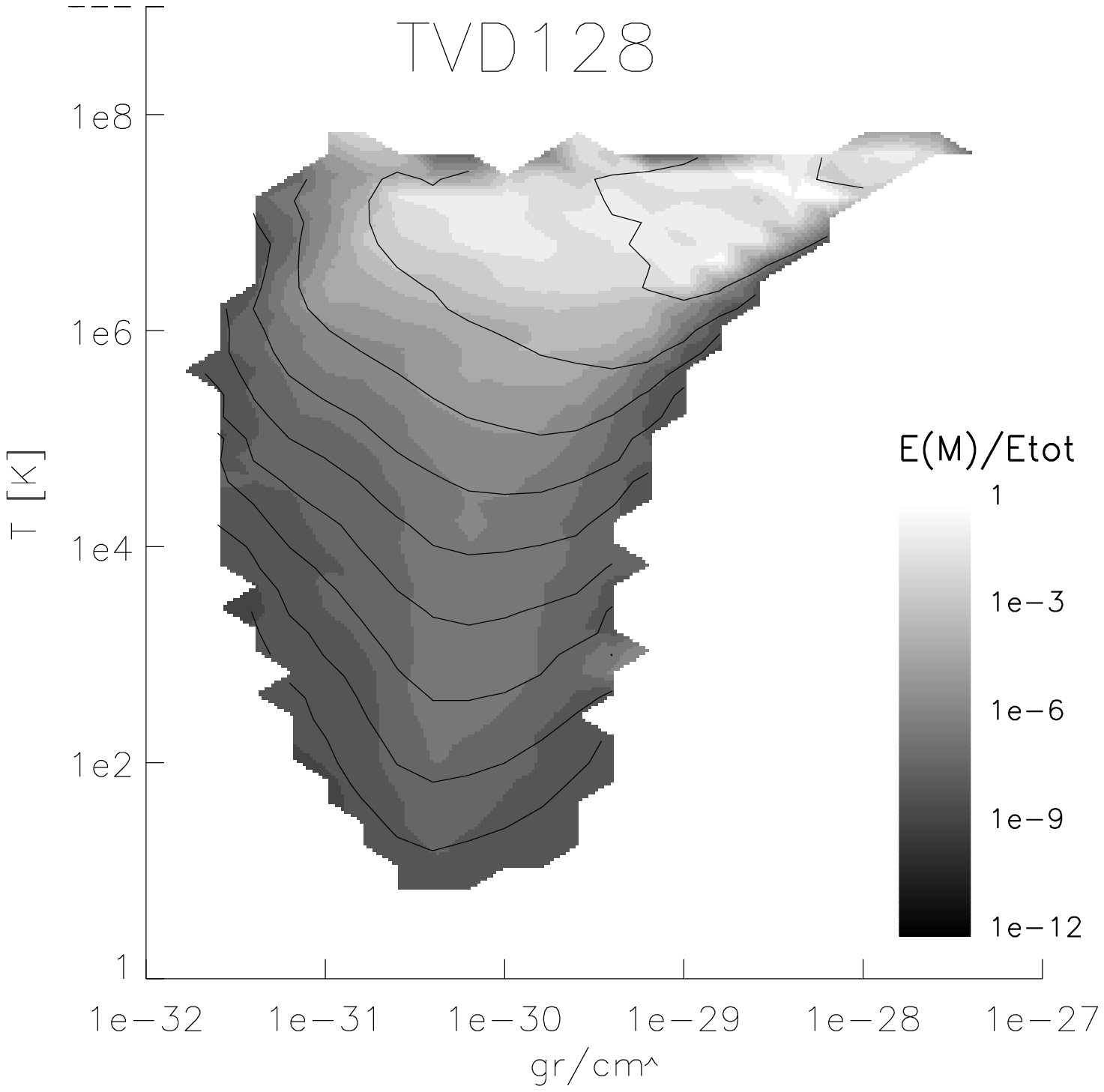}
\includegraphics[width=0.3\textwidth]{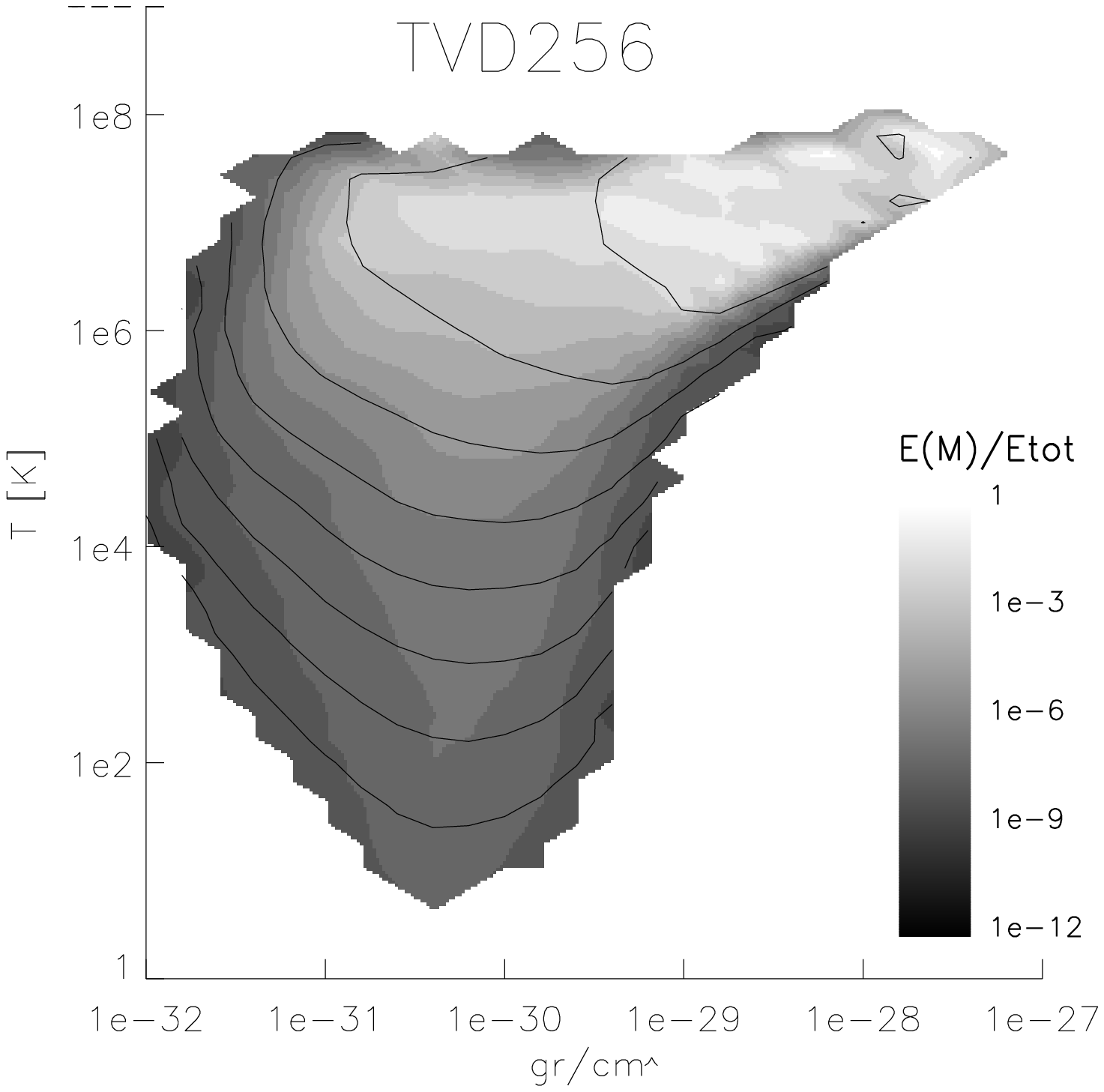}
\includegraphics[width=0.3\textwidth]{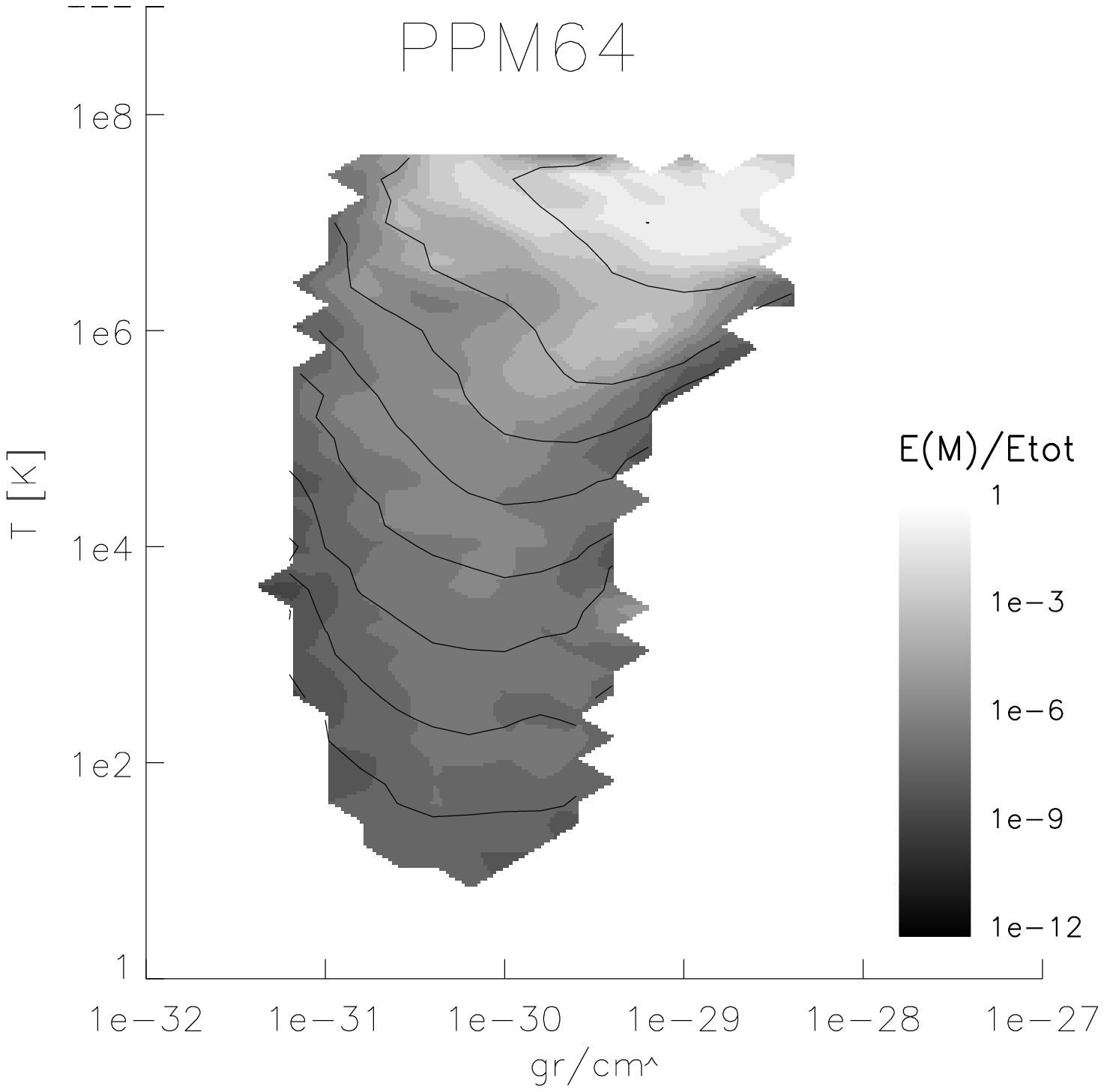}
\includegraphics[width=0.3\textwidth]{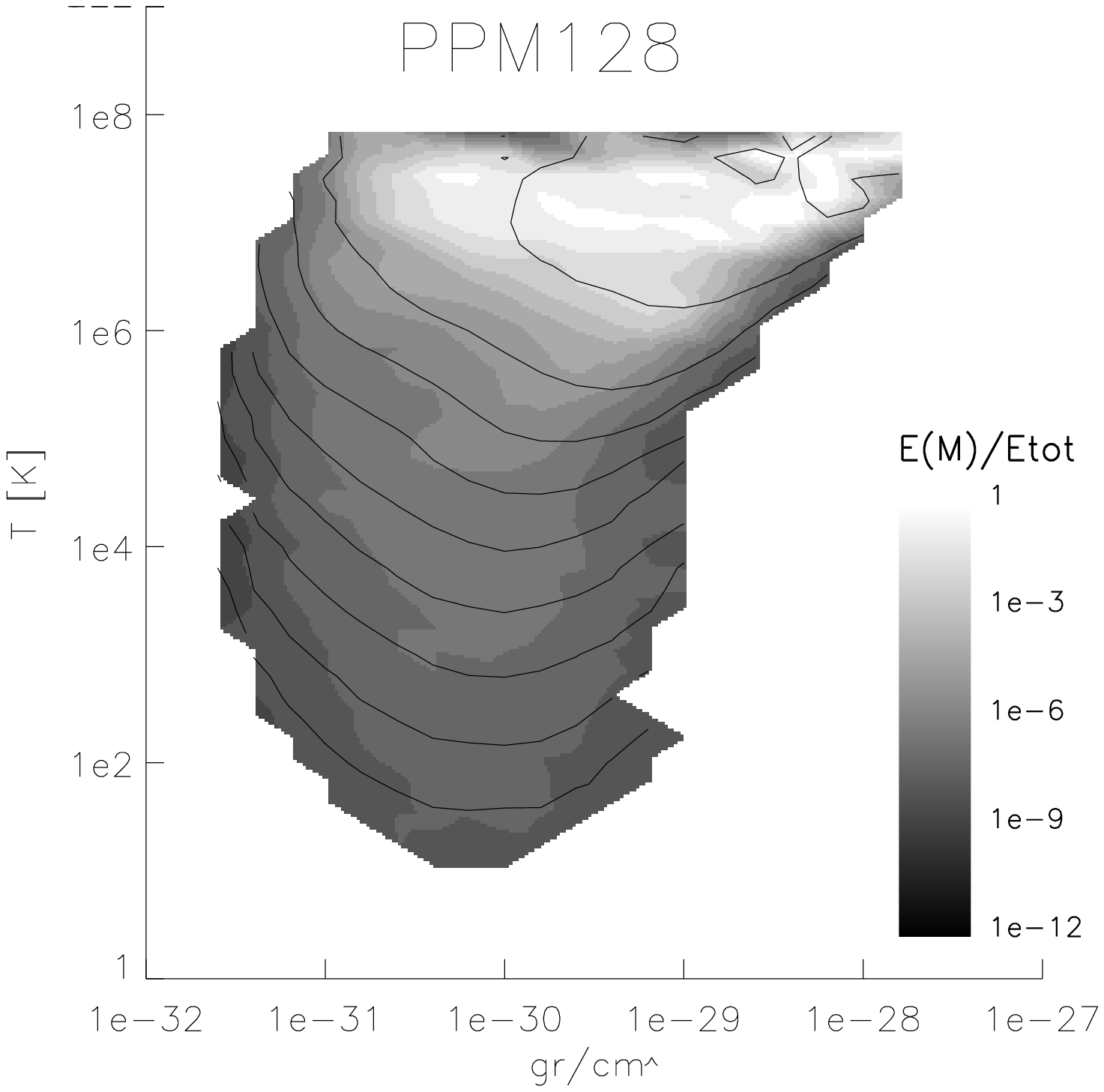}
\includegraphics[width=0.3\textwidth]{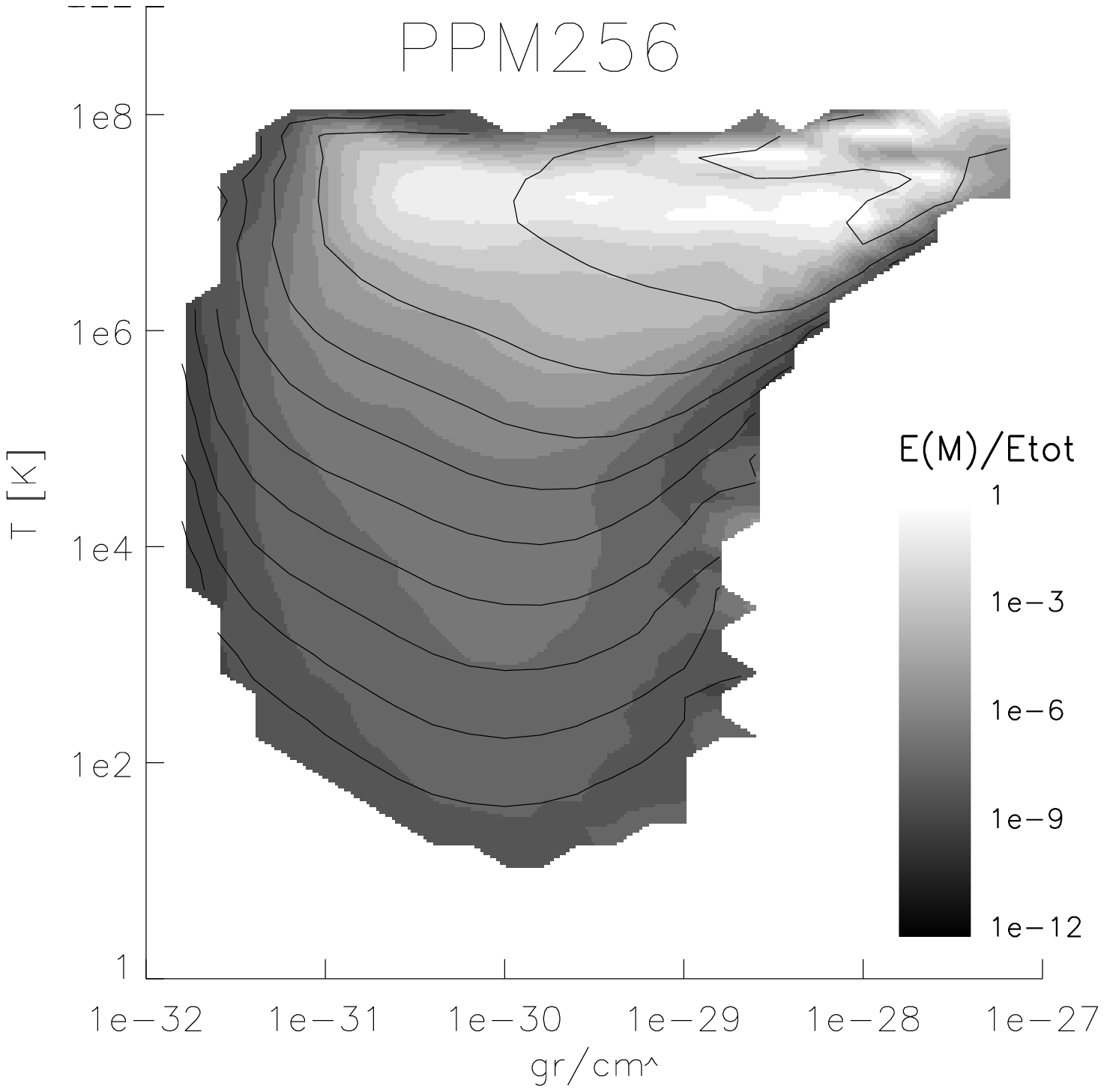}
\caption{Phase diagrams for shocked cells in the simulations, color 
coding shows the ratio the thermal flux, normalized to the total
flux within the simulations. Additional 
isocontours with a coarse binning in $(E(M)/E_{\rm tot})^{1/2}$ 
space are shown for clarity.\label{fig:phase_shocks_flux}}
\end{center}
\end{figure*}

\begin{figure*}
\includegraphics[width=0.33\textwidth]{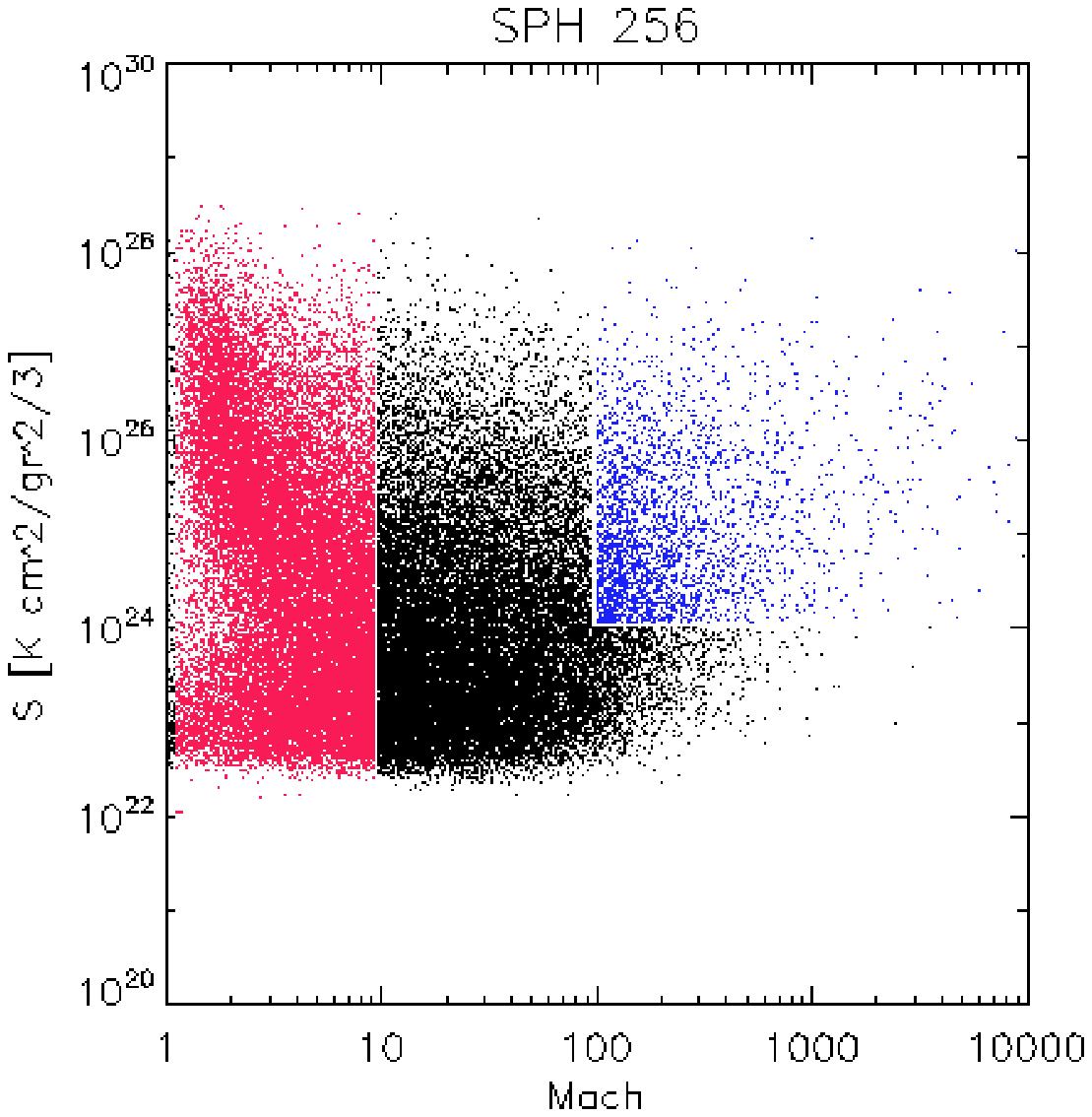}
\includegraphics[width=0.33\textwidth]{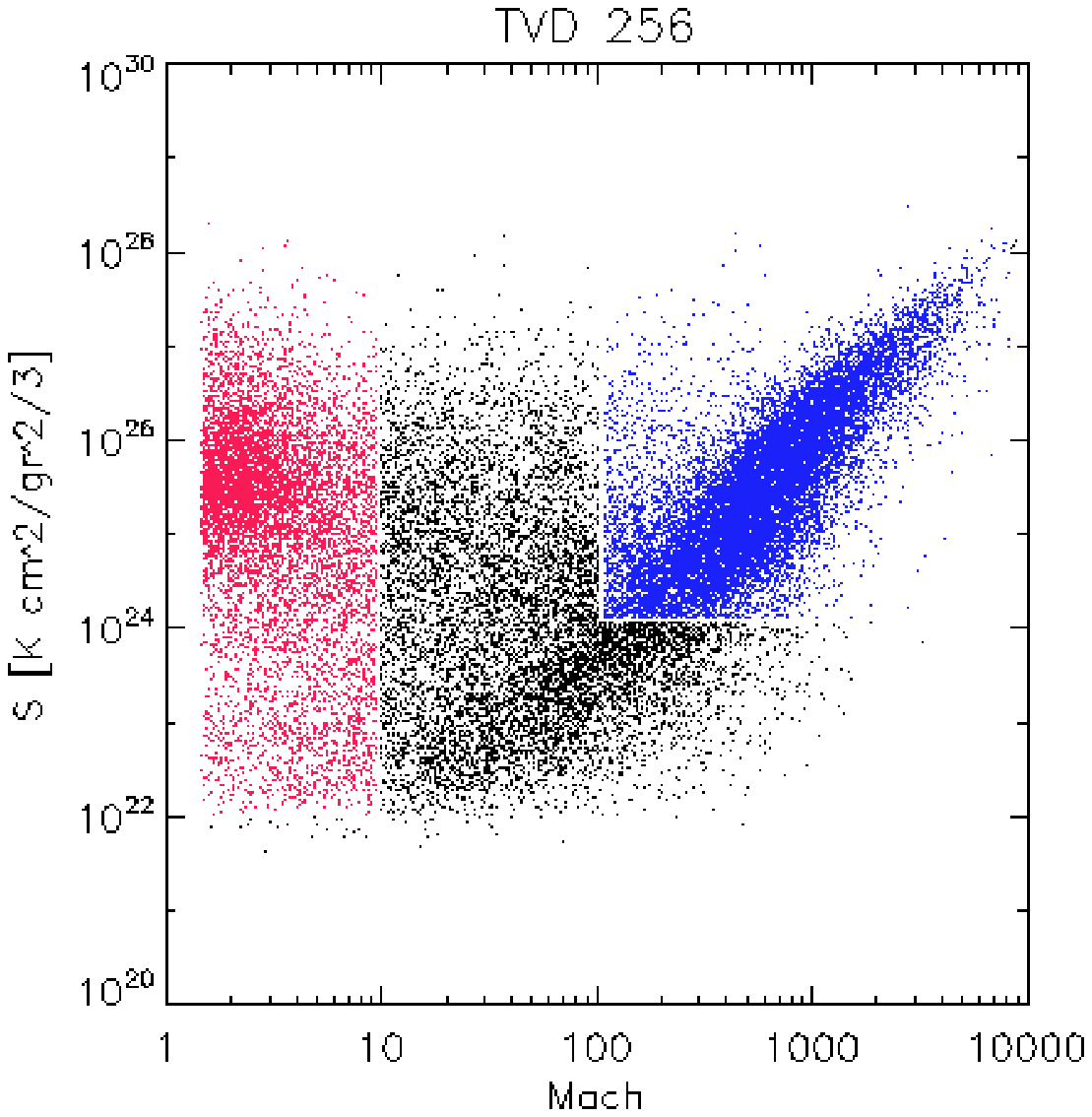}
\includegraphics[width=0.33\textwidth]{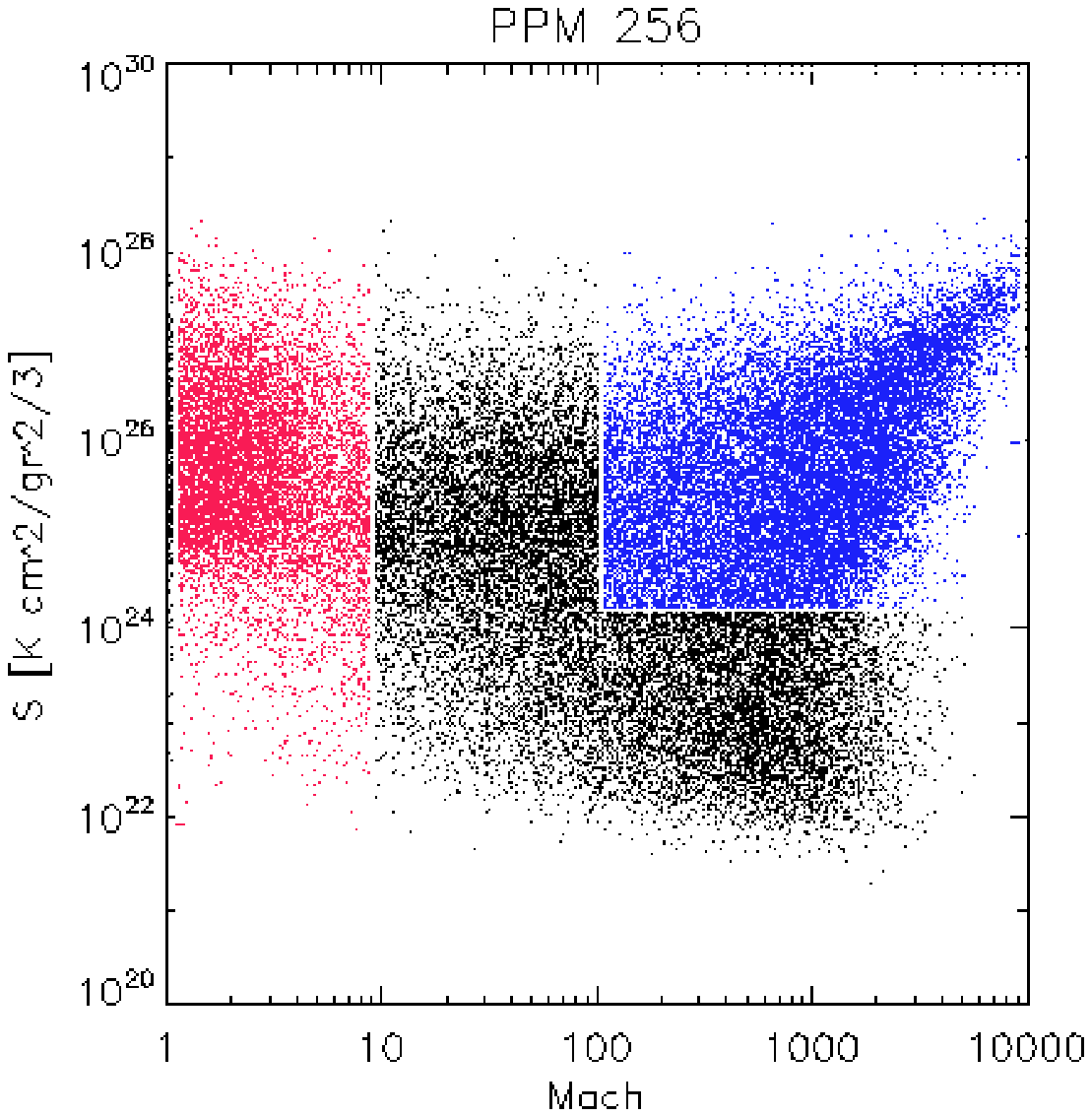}
\caption{Mach versus Entropy diagrams for shocked 
regions of the $256^{3}$ runs.}
\label{fig:entro_phases}
\end{figure*}

\subsection{Phase Diagrams for Shocked Regions.}
\label{subsec:phas_comp}

To pinpoint the differences between the codes, we find it
useful to  
extract the phase diagram of shocked cells for the
various runs within the total cosmic volume.
Panels in Fig.~\ref{fig:phase_shocks} 
and \ref{fig:phase_shocks_flux} show the 
flux-weighted mean Mach number, $\hat{M}$,  and thermal flux 
(normalized to the total thermal flux in
the cosmic volume) for the shocked cells of runs $64^{3}$, $128^{3}$ and $256^{3}$.

In grid codes, as soon as the spatial resolution is large enough to model
the innermost region of collapsed halos, a compact ``group'' of cells at
$\hat{M} \leq 10$ is formed in the upper right corner of the phase diagram,
while a much broader region of strong shocks is found at lower densities
at across a wide range of temperatures.
In GADGET3, a similar ``group'' of points corresponding to halos is 
formed, but it has less sharp contours and it smoothly extends to lower
densities, where strong outer shocks from a concentration which
is much narrower compared to grid codes.

If the dissipated energy flux is concerned
(Fig.~\ref{fig:phase_shocks_flux}), again less disagreement is found
among codes. At all resolutions, about the $\sim 90$ per cent of
total dissipated energy in the box is found at cells with
$\rho/\rho_{\rm cr} \geq 10^{2}$ and  $T \geq 10^{7}$K.

\bigskip

One should expect a high degree of convergence in the statistics and morphologies of energy dissipating structures in the three codes: indeed the main 
sources of heating in these adiabatic runs are shocks, and the
cross comparisons in the previous Sections (Sec.\ref{subsec:map_cp}-Sec.\ref{subsec:clusters_cp}) have shown that most of the thermal properties of halos are in good
agreement.

On the other hand, shocks are also the main source of entropy generation in 
these simulations, and we showed 
that the halos in the different codes present noticeable
differences both in the inner and outer entropy distributions (Sec.\ref{subsec:clusters_cp}--\ref{subsec:tracers}), are likely 
related to details of shocks dynamics away from the most dissipative structures in simulations.

Fig.~\ref{fig:entro_phases} shows the illustrative case of the scatter
plot for the post shock entropy versus $\hat{M}$ diagram. We restrict to 
$T > 100$ K regions in order to avoid any artifacts 
due to different low temperature floors adopted in the various codes
(see Sect.\ref{subsec:pdf}. 

A concentration of high entropy and 
weak shocks (in red color, in the Figure) is common to all simulated data, 
and marks the shock energy dissipation in innermost region
of galaxy clusters. 

However, in grid codes a concentration of points
is also present for $\hat{M}>10^{2}$, as diagonal stripe in the
plane ($\hat{M}$,$S$).
The points in this region (in blue color) trace {\it external} shocks,
for which the post shock entropy
is tightly correlated with $\hat{M}$ (Eqn.~\ref{eq:entropy}) for
strong $M>10$ shocks, leading to 
a $S_{2} \propto \hat{M}^{2}$.

This "phase" of shocked gas is almost completely missing in SPH runs.

%The spatial location of the two concentration 
%is highlighted in Fig.~~\ref{fig:map_phases}
%by color coding in red cells with $10^{22} K cm^{2}/gr^{2/3}<S<10^{28} K %cm^{2}/gr^{2/3}$ and $M<10$ and in blue cells with 
%$S>10^{24} K cm^{2}/gr^{2/3}$ and $M>100$.

We verified that in the grid codes, 
the strong shocks following the $S_{2} \propto \hat{M}^{2}$ correlation are systematically located at the outskirts of galaxy clusters and
filaments, while the concentration at $\hat{M}<10$ shocks comes from cells within collapsed halos. In this second case, energetic and weak shocks
are unable to change the post-shock entropy in a relevant
way, and no strong relation is found between $S$ and $\hat{M}$.
Therefore, in GADGET entropy is released in the simulation at the 
same location of the most dissipative structures in the universe,
whereas in both grid codes a sizable amount of entropy is also released at outer accretion shocks, which are not responsible
for sizable energy dissipation.

This suggests the important point that, although thermalized energy is processed
in the various codes in a rather consistent way, the gas entropy 
in grid codes and in SPH
is increased in shock structures with rather different morphologies and thermodynamical properties. 
Considering that the production of entropy at outer shocks is also 
responsible for the innermost entropy profile in clusters (Sec.\ref{subsec:tracers}),
we suggest that this findings is also relevant to understand the 
detailed properties of advection of matter (and possibly CR) inside
galaxy clusters, over cosmic time.

One possibility is that the absence of strong entropy generation at outer shocks in GADGET3
is due to pre-shock entropy generation by to artificial
viscosity (e.g. O'Shea et al. 2005), which would also be consistent
with the trend reported in the temperature distributions
of Sec.\ref{subsec:pdf}. An additional effect here is likely
the smearing of shocks  at low densities in SPH, which makes difficult to the shock solver in GADGET3 to update the particles entropy in a fully consistent way, if several smeared shocks merge together in the accretion regions.

\begin{figure*}
\begin{center}
\includegraphics[width=0.45\textwidth,height=0.38\textwidth]{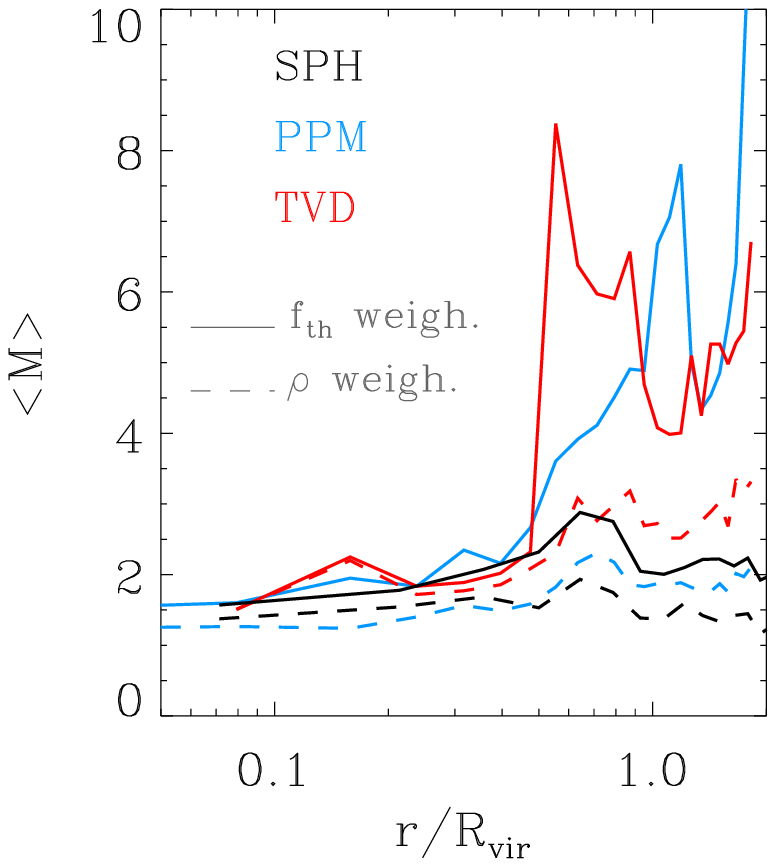}
\includegraphics[width=0.45\textwidth,height=0.38\textwidth]{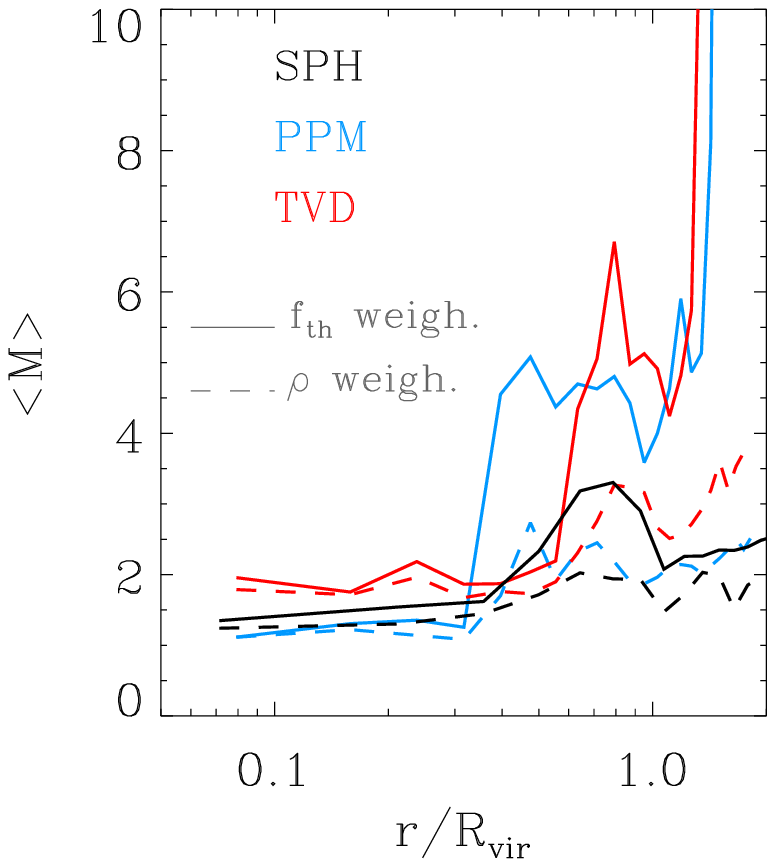}
\includegraphics[width=0.45\textwidth,height=0.38\textwidth]{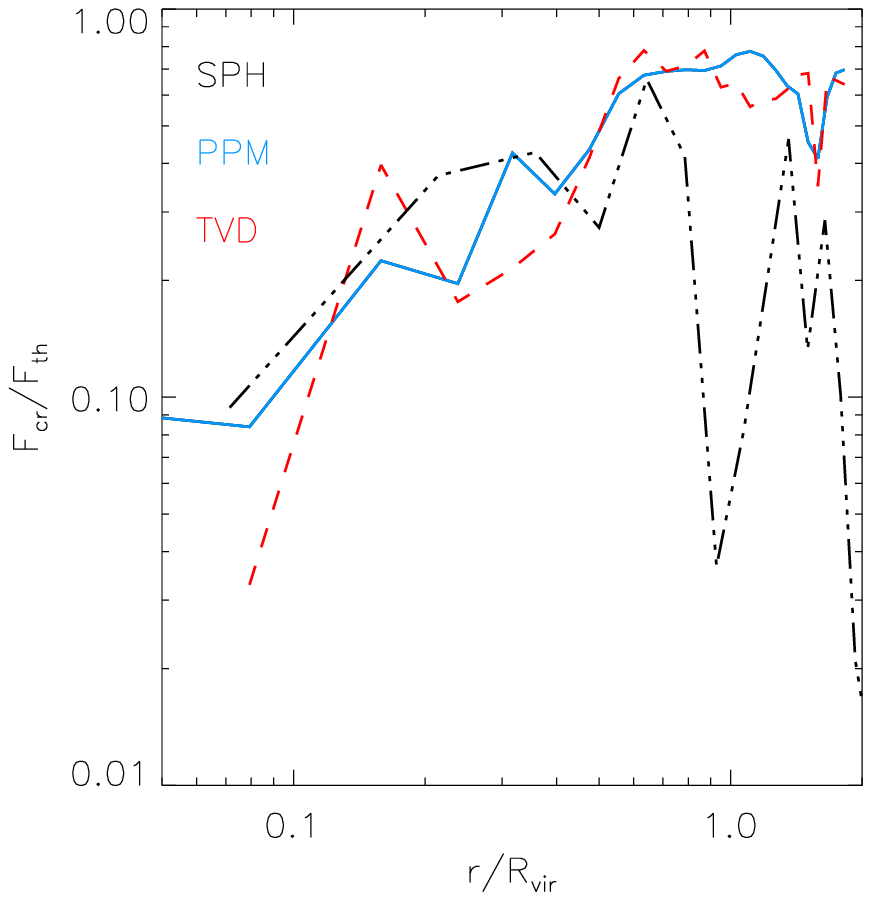}
\includegraphics[width=0.45\textwidth,height=0.38\textwidth]{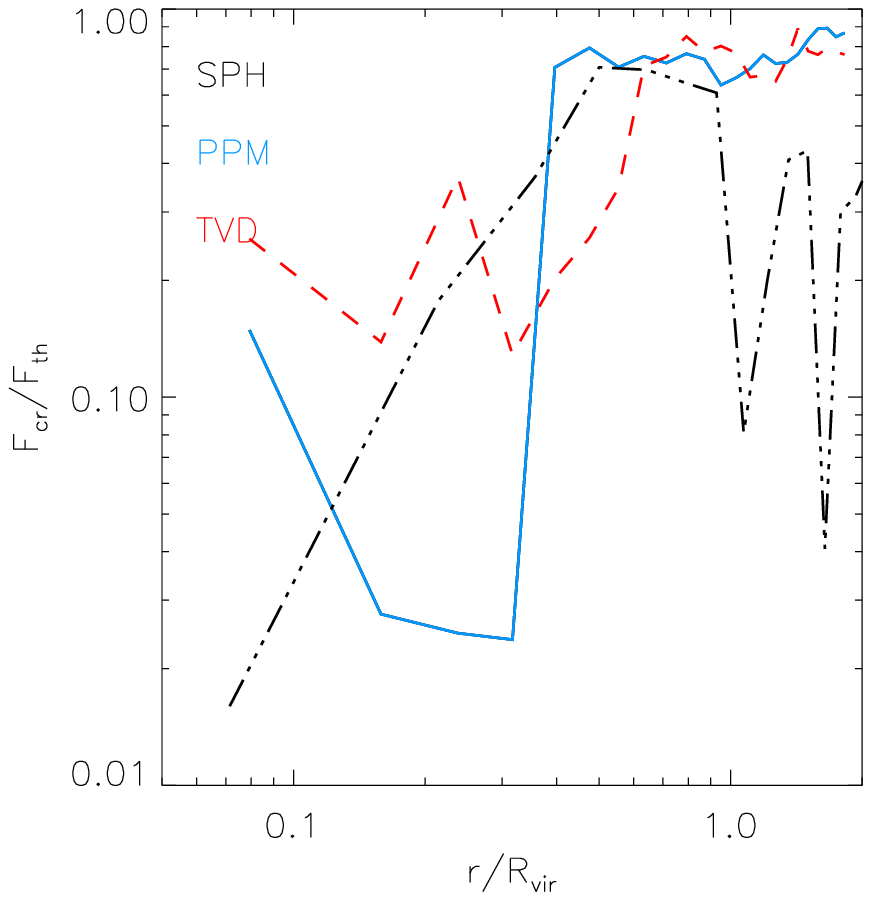}
\caption{{\it Top panels:} profiles of density-weighted and energy flux-weighted average Mach number
for cluster A ({\it left} panel) and cluster B ({\it right} panel).
{\it Bottom panels:} profiles of the CR acceleration efficiency, $f_{\rm CR}/f_{\rm th}$, for cluster A ({\it left}) and cluster B ({\it right}) at the best available resolutions in all codes.}
\label{fig:prof_mach_cp}
\end{center}
\end{figure*}

\subsection{Shocks in Clusters and Cosmic Rays Acceleration}
\label{subsec:comp_CR_clust}

Galaxy clusters are expected to be the most important
accelerators of CR in the universe (e.g. Miniati et al.  2001; Ryu et al. 2003; Pfrommer
et al.  2007); it is therefore important to analyse in detail also the 
estimated properties of CR acceleration at shocks,  in the most massive galaxy clusters of our simulated volume. 

 In Fig.~\ref{fig:prof_mach_cp} we report the average radial profile of mean Mach number, $\hat{M}$, for clusters A and B,
showing both the results of the weighting by gas density (dashed lines), and by the dissipated
energy flux (solid lines). 
Despite the different dynamical state of the two systems, we measure 
$\hat{M} \sim 2$ for $r<0.5 R_{\rm vir}$ in all runs. 
Approaching the cluster virial radius, the grid codes show a sharp increase in the 
mean Mach number (weighted by dissipated flux), which reaches strong shocks, $\hat{M} \sim 10$ at $\sim 2 R_{\rm vir}$. In GADGET3  the increase in the mean shocks strength is smooth, and $\hat{M} < 3$ is 
always found inside $R_{\rm vir}$. The above trends are similar but less evident, if the weighting by gas density is
adopted.  
These two trends mirror the trends in the outer 
entropy profiles (Sec.\ref{subsec:clusters_cp}--\ref{subsec:tracers}), and can be
explained by noting that the medium is more clumpy in GADGET3 
runs, and that the shocks are always thinner and stronger at this
location in grid codes, marking a very sharp the transition between large
scale structures and the rarefied Universe. 
 
\bigskip

In order to explore the possible effect played by the above differences in  the global efficiency of clusters to produce the CR energy flux at shocks, we
applied to all simulations a recipe to estimate the CR acceleration efficiency at shocks, with a standard application of the Diffusive
Shock Acceleration theory (e.g. Kang \& Jones 2002).
According to this model, the CR acceleration at each shocks is 
parametrized as a function of the Mach number:

\begin{equation}
f_{\rm CR} = \eta (M) \cdot \rho_{\rm pre} M^{3} v^{3}_{s}/2;
\label{eq:fcr}
\end{equation}

where $\eta(M)$ is a monotonically increasing function of $M$, whose
numerical approximation can be found instance in Kang et al. (2007).
This prescription for the acceleration of CR particles is 
quite idealized, and that more recent work by the same authors
take also into
account Alfv\'{e}n waves drift and dissipation at the shock
precursor (Kang et al. 2007), causing a lower acceleration efficiency
for shocks with $M<10$. Also, this recipe neglects the role of the re-acceleration of pre-existing CR, which can as well affect in 
a significant way the efficiency of acceleration at weak
shocks  (e.g. Kang \& Ryu 2010).

The bottom panels in Fig.~\ref{fig:prof_mach_cp} 
shows the radial profiles for the mean 
acceleration efficiency at shocks $f_{\rm CR}/f_{\rm th}$, for
cluster A and cluster B at the best available resolutions in all
codes.
In the relaxed cluster A, the agreement is reasonably
good and all codes show a minimum efficiency  $f_{\rm CR}/f_{\rm th} \sim 0.1$,
at the cluster core, with a similar increasing profile up to a maximum
of  $f_{\rm CR}/f_{\rm th} \sim 0.7$ at $R_{\rm vir}$.
Outside of this radius, the trends of grid codes and SPH largely
diverge as in all cases reported before, and the acceleration efficiency 
in GADGET3 run decreases. 

The comparison of the results for cluster B suffers of the 
timing issue reported in Sec.\ref{subsec:clusters_cp}, which 
are further amplified by the
non linearity of Eqn.~\ref{eq:fcr}. This produces a large scatter 
from code to code in $f_{\rm CR}/f_{\rm th}$ inside 
the cluster, but approaches the same values and trend of cluster
A for  $\geq R_{\rm vir}$. 

We stress that the  reported differences for cluster B are representative
of the level of {\it intrinsic} scatter that simulations with different
numerical codes are subject to, which in turn adds a level of unavoidable uncertainty when estimates of CR injections from clusters are estimated using too small number of objects.
\bigskip 

It is worth stressing that the above estimate of CR acceleration efficiency
are already at the edge, if not outside, of the range of permissible
energy ratio between CR and thermal gas from gamma rays 
(e.g. Reimer 2004; Pfrommer \& En{\ss}lin 2004; Aharonian et al. 2008; Ackermann et al.  2010; Aleksic et al.  2010; Donnert et al.  2010; Pinzke et al.  2011), radio
 (Brunetti et al. 2007; Brunetti et al.  2008) and X-ray/optical observations (Churazov et al. 2008).
Given the fairly simple setups of the simulations considered in this
work (e.g. no radiative processes, no re-ionization, idealized recipe for CR acceleration
at shocks, no self-consistent CR feedback,
coarse spatial and mass resolutions, no magnetic fields), this is not
surprising and it suggests that a completely self-consistent treatment
of CR, in presence of other important non-thermal component (such like
magnetic fields) is needed to model observations.

On the other hand, these findings may also imply that the numerical implementation of the complex non-linear physics of
non-thermal phenomena in large scale structures can be subject
to additional uncertainties, because the
basic thermo-dynamical evolution of accreted cosmic baryons in
large scale structures is not yet unambiguously constrained even by
rather simple cosmological simulations.

\bigskip

\section{Discussion}
\label{sec:comp_discussion}

In this work we presented the results of a numerical study
 which compares cosmological simulations at various resolutions, obtained with
with GADGET3 (Springel 2005), ENZO (Norman et al. 2007) 
and TVD (Ryu et al. 1993).

The chosen simulation setup is very simple (only gravity forces and non-radiative hydrodynamics are modeled) and it is therefore 
particularly suitable to study the convergence 
among widely used, complementary numerical approaches. This kind of
comparison my also be helpful
to explore the reasons for differences in the thermal and non thermal
properties of galaxy clusters runs.

We have analyzed in detail the properties of the DM distribution, thermal gas
matter distribution, shock waves and CR acceleration efficiencies within the simulated volume in all codes, and we  highlighted all most 
convergent and least convergent findings of all codes, as
a function of the numerical resolution and of
cosmic environment.

\bigskip

\subsection{Summary of dark matter and thermal gas properties}

An overall satisfactory agreement between the 3 codes 
is found for runs with DM mass resolution better than $m_{\rm dm} < 4 \cdot 10^{10} M_{\odot}/h$, 
in line with previous comparison works.
In particular, we report a good cross convergence of the following
measures:

\begin{itemize}  
\item the mass distribution function and baryon fraction for halos in the simulations are found in agreement within a $\approx 5-10$ per cent, across a range of masses. The rate of convergence with resolution in grid codes is much slower than in GADGET3. These results are in line with the works by O'Shea et al. (2005) and by Heitmann et al. (2008);

\item the profiles of DM matter of halos are well converged in all codes, 
for all the virial volume except for the scales close to the gravitational softening of all codes, consistently with the
literature (e.g. Frenk et al.  1999);

\item the gas density distribution are agreement within $10-20$ per cent,
for densities in the range $1 \leq \rho/\rho_{\rm cr} \leq 100$. High density peaks are found to be located at equal positions within the spatial  resolution of the simulations;

\item the gas temperature distributions are in agreement with a $5-10$ per cent accuracy only for $T>10^{6}K$ regions, which correspond to the typical virial
temperature of the smallest halos produced in the simulations, in agreement with the findings reported by Kang et al. (1994) and O'Shea
et al.  (2005);

\item the gas temperature and the gas density profiles of the most massive clusters in the sample are similar within a $10-20$ cent accuracy, consistently with Frenk et al. (1999). 
Time integration of a 
chaotic system results in slightly different spatial realizations of 
substructure, in particular during mergers. This introduces an episodic 
source of additional uncertainty.

\end{itemize}

On the other hand, noticeable differences are found in the following measures:

\begin{itemize}

\item  the gas density and gas temperature distributions for
$\rho/\rho_{\rm cr} < 1$ and for $T<10^{6}K$ regions are in 
disagreement up to $2-3$ orders of magnitude among simulations, even
at the best available resolutions in the project;

\item  the entropy profiles for clusters simulated with grid codes show a sharp peak located at $\sim 2-3 R_{\rm vir}$, while the  profiles in GADGET3 present a similar shape, but spread across a sizable
lager volume;

\item the inner entropy profile of clusters simulated with ENZO is flat inside
$\sim 0.1 R_{\rm vir}$, while it is steep in GADGET3 (consistently with early
results from Frenk et al. 1999 and more recent ones by Mitchell et al. 2009);

\item the gas clumping within the most massive halos, and
expecially in the outermost cluster regions,  is rather different if grid codes and GADGET3 are compared;

\item the time evolution of the accretion of matter clumps is also found to be radically different when ENZO and GADGET3 are compared: in grid codes their initial entropy is substantially increased by shock heating, while in SPH shock heating mechanisms are more gentle. The accreted material is distributed at larger cluster radii in ENZO than in GADGET3.  

\end{itemize}
\bigskip

\subsection{Summary of shocks properties}

Shocks were identified in all runs according to the shock detecting schemes
specifically conceived for each simulation:
and Entropy based method for GADGET3 (Pfrommer et al. 2006), a temperature based
method for TVD (Ryu et al. 2003) and a velocity based method for ENZO (Vazza,
Brunetti \& Gheller 2009).

\bigskip

The most interesting convergent findings are:

\begin{itemize}

\item the peak of thermal flux in the universe is at $M \sim 2$, and originates
in shocks internal to clusters;

\item the volume distribution and thermal energy flux distribution are very steep, and
are dominated by strong $M \sim 100-1000$ shocks in the external regions of large
scale structures; 

\item $\sim 99$ percent of the total thermal energy flux in the universe is 
processed by shocks with $M<10$;

\item inside the virial radius of the most massive clusters, the density weighted
profile of shocks are very flat, with $\hat{M} \sim 2$;

\item the estimated acceleration efficiency of CR (assuming Kang \& Jones 2002) is small in the innermost cluster
region, $f_{\rm CR}/f_{\rm th} \sim 0.1$, and increases towards the virial radius, with 
$f_{\rm CR}/f_{\rm th} \sim 0.8$ (however, the absolute numbers are  likely to change as this recipe does not account for Alfven wave drift and dissipation at the shock precursor).

 .
\end{itemize}

On the other hand, the findings where we do not find agreement at the investigated
resolutions are:

\begin{itemize}

\item shocks in grid codes are morphologically similar at all resolutions, 
while shocks in GADGET3 show substantial difference at external shocks;

\item the volume-weighted mean Mach number for $\rho/\rho_{\rm cr}<10$ presents different trends in each code;

\item in the vast majority of the simulated volume (outside halos), shocks
in grid codes show rather different properties in the phase diagrams ($\rho$ versus $T$ and $S$ versus $M$) compared to shocks in GADGET3. In particular, 
strong accretion shocks in 
grid codes are associated with large entropy jumps, while accretion shocks in GADGET3 are not characterized by large values of entropy;

\item  in massive clusters, grid codes produce a sharp increase of the shock strength outside $R_{\rm vir}$, while a continuous transition to weaker shocks is 
is found in GADGET3 runs;

\item  the CR injection efficiencies outside the virial
radius show different radial trends when grid runs and GADGET3
 runs are compared. 

\end{itemize}

\subsection{Conclusions}

Overall, when cosmological numerical simulations with GADGET3, ENZO and cosmological TVD are compared within similar range of DM mass resolution, we report
agreement better than $\sim 10$ per cent level in many statistics concerning
hot, over-dense regions of the universe (i.e. halos, filaments). This is reassuring and it is in line with a number of previous works dealing with similar topics
(Kang et al. 1994; Frenk et al. 1999; Heitmann et al. 2008). 
The statistical distributions of halos masses, halos baryon fraction, density 
distributions, thermal profiles and internal shocks are 
characterized by a high 
rate of convergence with resolution in GADGET.
In the most over-dense regions, ENZO and TVD converge at a rather small rate, but produce very similar estimates at the end for most of investigated cases, despite the radically different hydro method
they use to solve baryon gas dynamics.
The application of Adaptive Mesh Refinement techniques
is expected to further reduce the discrepancy between grid methods
and SPH, at least in some cases (e.g. O'Shea et al. 2005; Tasker et al. 2008; Robertson et al.  2010). In the case of lower density regions (i.e. outer accretion
regions of clusters, voids) the 
temperature distributions, entropy distributions, shock morphologies and Mach number distributions converge to rather
different estimates when SPH and grid codes are compared. The role
played by the effective viscosity and diffusivity of each method away from shocks may be partially responsible for the above differences.

\bigskip
One interesting finding is the 
substantially different
characterization of external shocks and entropy profiles in the grid and SPH methods, a feature
that has a number of important consequences in both thermal and
non thermal issues.
The different dynamics felt by accreted clumps (Sec.\ref{subsec:tracers})
show that the prediction of mixing and gas matter deposition rates 
in cluster cosmological simulations is still an open problem.
Given the rather simple setup employed in these simulations (no radiative
processes, no heating mechanism other than shocks, no CR feedback, small
turbulent motions due to lack of resolution and artificial viscosity in SPH)
shocks dynamics has to be regarded as the leading player in setting
the entropy profiles in clusters.
These results conclusively suggest that the differences in shocks morphologies
and shock dynamics across the clusters evolution 
leave major imprints also in substructures distributions and entropy
distributions in the ICM, which is a rather new evidence 
provided by this work.

Tightly connected to this, is the high degree of non-linearity which is 
present in all CR acceleration recipes. However, to date it is not clear 
whether these non-linearities would amplify any of the above differences 
at shocks and potentially lead to a different CR pressure distribution 
in galaxy clusters, or whether the average CR pressure support results 
from a combination of an average shock acceleration efficiency at the 
strongest shocks and successive CR transport.

\bigskip
Based on the results of this project, we notice that  performing of "high-precision" cosmology
(e.g. relating cosmological observables and theoretical models, based on scaling relations affected
by less than $\sim 1$ percent scatter in simulations)  may still be a challenge for many applications, 
since some very important measurements related to the volume filling properties of galaxy clusters simulated
with some of the best available numerical codes 
appear to be still affected by uncertainties of $\sim 10$  percent (or more). This is found even in the case
of the very similar physical setup analyzed here (only gravity and hydrodynamical forces) and 
the reason for this appear to be mostly of numerical nature, meaning that some important details
concerning the production and transport of  entropy in these simulations can be very different from
code to code. This results are based on rather low or moderate resolution simulations presented in this paper, and it
 is likely that going to much higher resolution levels off most of the above differences; however in this paper
we have shown that not all significant differences are related to resolution only (e.g. differences in accretion
shocks in SPH or grid methods).

\bigskip

It is unclear if the application of more physical ingredients
which are not accounted in this work (e.g. magnetic
fields, thermal conductions, feedback of relativistic particles)
may be able to soften any of the reported above reported differences.

The suggestion of this work is that, together with the design
of more sophisticated physical recipes to model the thermal
and non thermal components of the real Universe, our
theoretical understanding of cosmic structures would also
greatly benefit from other detailed comparative studies 
of different numerical recipes, since the convergence of
simulated estimates of a sizable fraction of the cosmic
volume is presently yet to be reached.

 \section{acknowledgements}

F.V. acknowledges support
from the grant FOR1254 from Deutschen Forschungsgemeinschaft. 
K.D. acknowledges the support by the DFG Priority Programme 1177 and additional 
support by the DFG Cluster of Excellence "Origin and Structure of the Universe". 
The work of D. R. and H. K. was supported by the National Research Foundation of Korea through grant 2007-0093860.
F.V. and G.B. acknowledge partial 
support through grant PRIN INAF 20082009/2010.
We acknowledge the 
usage of computational resources under the CINECA-INAF 2008-2010 agreement
and the 2009 Key Project ``Turbulence, shocks and cosmic rays electrons 
in massive galaxy clusters at high resolution''. 
We acknowledge R.Brunino, M. Hoeft,  T. Jones, B. O'Shea and V. Springel of fruitful scientific discussions. 
The computation of cosmological distances in this paper are done thanks to the Cosmology Calculator by E. Wright (Wright 2006).

\bigskip

\bigskip

\end{document}